\documentclass[reprint,aps,prc,twocolumn,longbibliography,floatfix,10pt]{revtex4-1}

\usepackage{bm}
\usepackage{dcolumn}
\usepackage{amsthm}
\usepackage{amssymb}
\usepackage{amsmath}
\usepackage{graphicx}
\usepackage{bm}
\usepackage{xcolor}
\usepackage{fix-cm}
\usepackage{mathptmx} 
\usepackage[T1]{fontenc}
\usepackage[colorlinks,allcolors=blue]{hyperref}
\makeatletter
\def\NAT@def@citea{\def\@citea{\NAT@separator}}
\makeatother

\begin{document}

\title{Deformed shell effects in $^{48}$Ca+$^{249}$Bk quasifission fragments}

\author{K. Godbey}\email{kyle.s.godbey@vanderbilt.edu}
\author{A.S. Umar}\email{umar@compsci.cas.vanderbilt.edu}
\affiliation{Department of Physics and Astronomy, Vanderbilt University, Nashville, Tennessee 37235, USA}
\author{C. Simenel}\email{cedric.simenel@anu.edu.au}
\affiliation{Department of Theoretical Physics and Department of Nuclear Physics, Research School of Physics and Engineering, The Australian National University, Canberra ACT  2601, Australia}
\date{\today}


\begin{abstract}
\edef\oldrightskip{\the\rightskip}
\begin{description}
\rightskip\oldrightskip\relax
\setlength{\parskip}{0pt} 
\item[Background] Quasifission is the main reaction channel hindering the formation of superheavy nuclei (SHN). Its understanding will help to optimize entrance channels for SHN studies. Quasifission also provides a probe to understand the influence of shell effects in the formation of the fragments.
\item[Purpose] Investigate the role of shell effects in quasifission and their interplay with the orientation of the deformed target in the entrance channel.
\item[Methods] $^{48}$Ca$+^{249}$Bk collisions are studied with the time-dependent Hartree-Fock approach for a range of angular momenta and orientations.
\item[Results] Unlike similar reactions with a $^{238}$U target, no significant shell effects which could be attributed to $^{208}$Pb ``doubly-magic'' nucleus are found. However, the octupole deformed shell gap at $N=56$ seems to strongly influence quasifission in the most central collisions.
\item[Conclusions] Shell effects similar to those observed in fission affect the formation of quasifission fragments. Mass-angle correlations could be used to experimentally isolate the fragments influenced by $N=56$ octupole shell gaps.
\end{description}
\end{abstract}
\maketitle


\section{Introduction}

Quasifission occurs when the collision of two heavy nuclei produces two fragments with similar characteristics to fusion-fission fragments, but without the intermediate formation of a fully equilibrated compound nucleus~\cite{heusch1978,back1981,back1983,bock1982}.
It is the main mechanism that hinders fusion of heavy nuclei and consequently the formation of
superheavy elements~\cite{sahm1984,gaggeler1984,schmidt1991,back2014,khuyagbaatar2018,banerjee2019}.
It is thus crucial to achieve a deeper insight of quasifission in order to minimize its impact and maximize the formation of compound nuclei for heavy and superheavy nuclei searches.

Quasifission also provides a unique probe to quantum many-body dynamics of out-of-equilibrium nuclear systems.
For instance, quasifission studies bring information on mass equilibration time-scales~\cite{toke1985,shen1987,durietz2011}, on shell effects in the exit channels~\cite{itkis2004,nishio2008,kozulin2014,wakhle2014,morjean2017}, as well as on the nuclear equation of state~\cite{veselsky2016,zheng2018}.
In fusion-fission, the exit channel is essentially determined by the properties of the compound nucleus, and does not depend a priori on the specificity of the entrance channel.
This is not the case in quasifission which is known to preserve a strong memory of the entrance channel properties.
As a result, understanding the interplay between the entrance and exit channels requires a significant amount of experimental systematic studies.
These include investigations of the role of beam energy~\cite{back1996,nishio2008,nishio2012}, dissipation~\cite{williams2018}, fissility of the compound nucleus~\cite{lin2012,durietz2013},
deformation of the target~\cite{hinde1995,hinde1996,knyazheva2007,hinde2008,nishio2008},
spherical shells of the collision partners~\cite{simenel2012b,mohanto2018}, and the neutron-to-proton ratio
$N/Z$ of the compound nucleus~\cite{hammerton2015,hammerton2019}.

On the theory side, quasifission has been studied with various approaches.
This includes classical methods such as
a transport model~\cite{diaz-torres2001}, the dinuclear system model \cite{adamian2003,huang2010,bao2015,guo2018c}, and models based on the Langevin equation~\cite{zagrebaev2005,aritomo2009,aritomo2012,karpov2017,sekizawa2019b}.
Microscopic approaches such as quantum molecular dynamics~\cite{wen2013,wang2016,zhao2016} and the
time-dependent Hartree-Fock (TDHF) theory \cite{golabek2009,kedziora2010,wakhle2014,oberacker2014,hammerton2015,umar2015a,umar2016,sekizawa2016,yu2017,ayik2017,ayik2018,sekizawa2017a,wakhle2018,morjean2017,sekizawa2019b} have also been used. See~\cite{simenel2012,simenel2018,sekizawa2019,stevenson2019} for recent reviews on TDHF.

An advantage of microscopic calculations is that their only inputs are the parameters of the energy density functional describing the interaction between the nucleons.
Since these parameters are usually fitted on nuclear structure properties only, such calculations do not require additional parameters determined from reaction mechanisms, such as nucleus-nucleus potentials.
In addition, TDHF calculations treat both reaction mechanisms and structure properties on the same footing.
This is important for reactions with actinide targets which exhibit a strong quadrupole deformation.

Indeed, the outcome of the calculations strongly depend on the orientation of the nuclei.
For instance, TDHF calculations of $^{40}$Ca$+^{238}$U reaction showed that only collisions with the side of the $^{238}$U could lead to configurations which are compact enough to enable fusion~\cite{wakhle2014}.
This is contrary to the collisions with the tip of $^{238}$U which seem to always lead to a fast quasifission (after $\sim5-10$ zeptoseconds (zs) of contact time) as long as contact between collision partners is achieved.
A remarkable observation of this work was the systematic production of lead nuclei ($Z=82$), known to possess a strong spherical proton shell gap, in tip collisions only, showing a strong influence of orientation dependent shell effects in the production of the fragments.
Such influence of shell effects was proposed to explain peaks in fragment mass distributions~\cite{itkis2004,nishio2008,wakhle2014}, but experimental confirmation came only recently with the observation of a peak of quasifission fragments at $Z=82$ protons from x-ray measurements~\cite{morjean2017}.

Deformed shell effects in the region of $^{100}$Zr have also been invoked to interpret the outcome of TDHF simulations of $^{40,48}$Ca+$^{238}$U, $^{249}$Bk collisions~\cite{oberacker2014,umar2016}.
It is then natural to wonder if other shell effects, spherical or deformed, could be driving the dinuclear system out of its compact shape, into quasifission.
Potential candidates are shell effects known to influence the outcome of fission reactions.
It has recently been proposed that octupole deformed shell effects, in particular with $Z$ or $N=52-56$, are the main driver to asymmetric fission~\cite{scamps2018,scamps2019}.
The fact that $^{208}$Pb can easily acquire an octupole deformation (its first excited state is a $3^-$ octupole vibration) is compatible with this interpretation.
Note also that some superheavy nuclei like $^{294}$Og are expected to encounter super-asymmetric fission and produce a heavy fragment around $^{208}$Pb~\cite{poenaru2018,warda2018,matheson2019,zhang2018b}, confronting the idea that quasifission valleys could match fission ones.

In this work we study the $^{48}$Ca+$^{249}$Bk reaction with the TDHF approach.
The choice of this reaction is motivated by its success in
forming the element $Z=117$~\cite{oganessian2010,oganessian2011,oganessian2012,oganessian2013,khuyagbaatar2014}.
Previous TDHF studies of quasifission with actinide targets were restricted to one or two orientations of the target to limit computational time.
However, to allow possible comparison with experimental data, it is important to simulate a range of orientations in addition to the usual tip and side configurations.
We therefore performed systematic simulations, spanning both a range of orientations and a range of angular momenta.
This allow us to study correlations between, e.g.,  mass, angle, kinetic energy, as well as to predict distributions of neutron and proton numbers at the mean-field level.
These distributions are used to identify potential shell gaps driving quasifission.
The method is described in Sec.~\ref{sec:method}.
The results are discussed in Sec.~\ref{sec:results}.
We then conclude in Sec.~\ref{sec:conclusion}.


\section{Method}\label{sec:method}

The TDHF theory provides a microscopic approach to investigate
a large selection of phenomena
observed in low energy nuclear physics~\cite{negele1982,simenel2012,simenel2018}.
In particular, TDHF provides a dynamic quantum many-body description of
nuclear reactions in the
vicinity of the Coulomb barrier, such as fusion~\cite{bonche1978,flocard1978,simenel2001,umar2008a,umar2006d,
washiyama2008,umar2010a,guo2012,keser2012,simenel2013a,oberacker2012,oberacker2010,umar2012a,simenel2013b,umar2014a,jiang2014},
deep-inelastic reactions and transfer~\cite{koonin1977,simenel2010,simenel2011,umar2008a,
sekizawa2013,scamps2013a,sekizawa2014,bourgin2016,umar2017,sekizawa2019},
and dynamics of (quasi)fission fragments~\cite{umar2010a,wakhle2014,oberacker2014,simenel2014a,
umar2015a,umar2015c,scamps2015a,goddard2015,bulgac2016,sekizawa2016,umar2016}.
The classification of various reaction types in TDHF is done by calculating the
time evolution of expectation values of one-body observables:
fragments' centers of masses, mass and charges on each side of the neck, kinetic energy,
orbital angular momentum, among others. Quasifission is characterized by two final
state fragments that emerge after a long lived composite system (typically longer
than 5\,zs) and final fragment masses $A_{f} = A_{\rm CN}/2 \pm 20$ or more. In
addition, final TKEs distinguish quasifission from highly damped deep-inelastic
collisions, which have a smaller mass and charge difference between initial and
final fragments. In TDHF the mass and charge difference between the initial nuclei
and the final fragments measure the number of nucleons transferred. As discussed
above fusion corresponds to the case where the final product remains as a single composite
for a reasonably long time, chosen here to be 35~zs.

The TDHF equations for the single-particle wave functions
\begin{equation}
h(\{\phi_{\mu}\}) \ \phi_{\lambda} (r,t) = i \hbar \frac{\partial}{\partial t} \phi_{\lambda} (r,t)
            \ \ \ \ (\lambda = 1,...,A) \ ,
\label{eq:TDHF}
\end{equation}
can be derived from a variational principle. The main approximation in TDHF is
that the many-body wave function $\Phi(t)$  is assumed to be a single time-dependent
Slater determinant at all times. It describes the time-evolution of the single-particle
wave functions in a mean-field corresponding to the dominant reaction channel.
During the past decade it has become numerically feasible to perform TDHF calculations on a
3D Cartesian grid without any symmetry restrictions
and with much more accurate numerical methods~\cite{bottcher1989,umar2006c,sekizawa2013,maruhn2014}.
\begin{figure}[!htb]
\centerline{\includegraphics*[width=8.6cm]{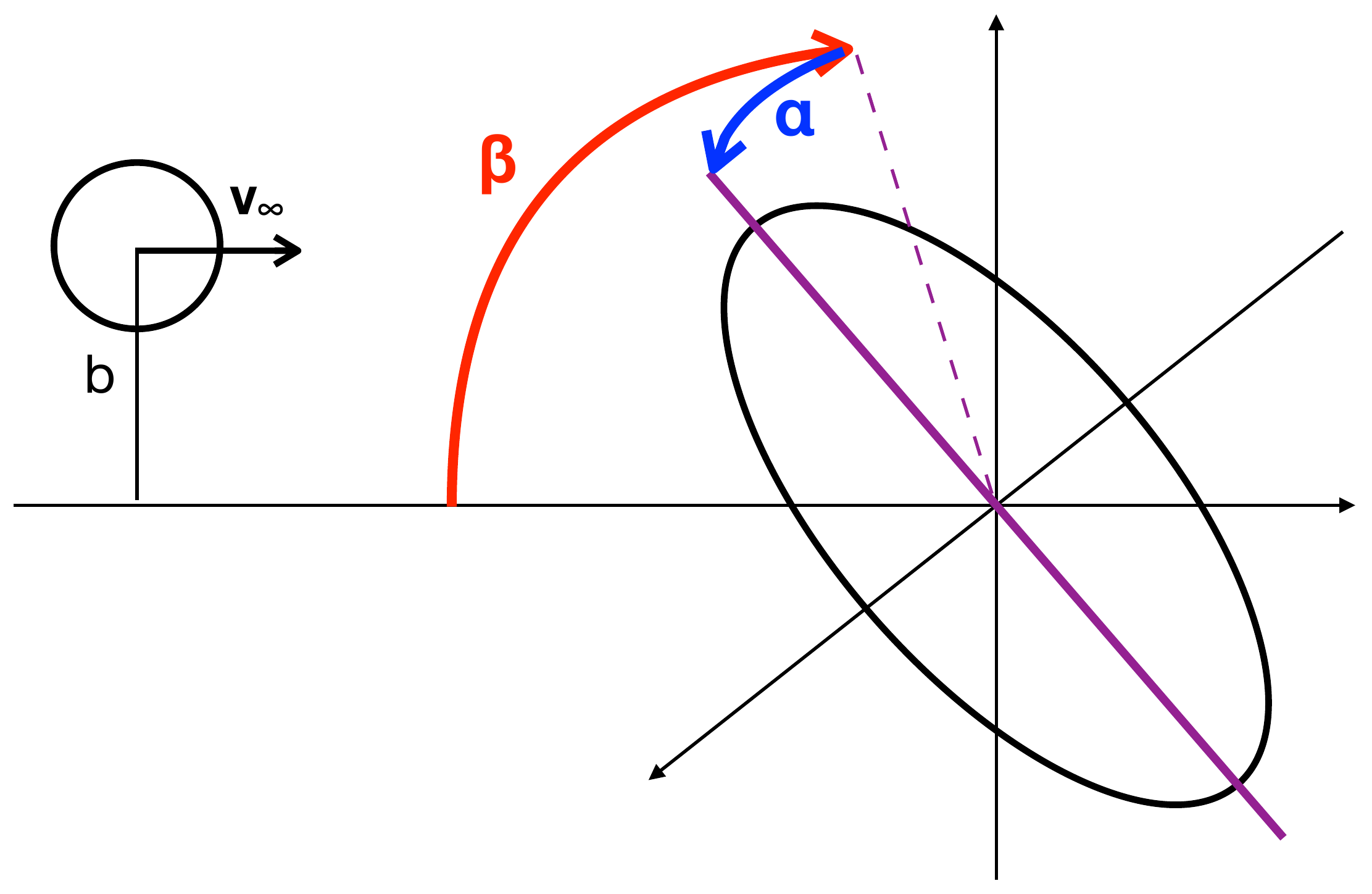}}
\caption{\protect(Color online) Schematic representation of the initial configuration for an impact parameter $b$ and  a velocity vector $\mathbf{v}_\infty$ defining the collision plane and the collision axis. The orientation of the target is defined by the angles $\beta$ (rotation around the axis perpendicular to the reaction plane) and $\alpha$ (rotation around the collision axis).}
\label{fig:geometry}
\end{figure}

In this paper, we focus on fusion and quasifission in the reaction
$^{48}\mathrm{Ca}+^{249}\mathrm{Bk}$. In our TDHF calculations
we use the Skyrme SLy4d energy density functionals~\cite{kim1997}
including all of the relevant time-odd terms in the mean-field Hamiltonian.
Static Hartree-Fock (HF) calculations without pairing predict a spherical
density distribution for $^{48}$Ca while $^{249}$Bk shows prolate quadrupole
and hexadecupole deformation, in agreement with experimental observations.
Numerically, we proceed as follows: First we generate very well-converged static
HF wave functions for the two nuclei on the 3D grid.
Three-dimensional TDHF initialization of the deformed $^{249}$Bk nucleus, with a particular alignment of its symmetry
axis with respect to the collision axis, can be most easily achieved by
evaluating the initial guess for HF calculations
on mesh values rotated with respect to the code axes. Subsequent HF iterations do not change this
orientation thus resulting in the desired HF solution. This procedure involves no interpolation
procedure and is the most straightforward method to implement in TDHF codes~\cite{pigg2014}.
Otherwise, static solutions obtained for extreme angles ($0^{\circ}$ or $90^{\circ}$ with respect to collision
axis) can be very accurately interpolated to arbitrary angles~\cite{pigg2014} followed by a few additional static iterations for extra accuracy.
\begin{figure}[!thb]
\includegraphics*[width=8.6cm]{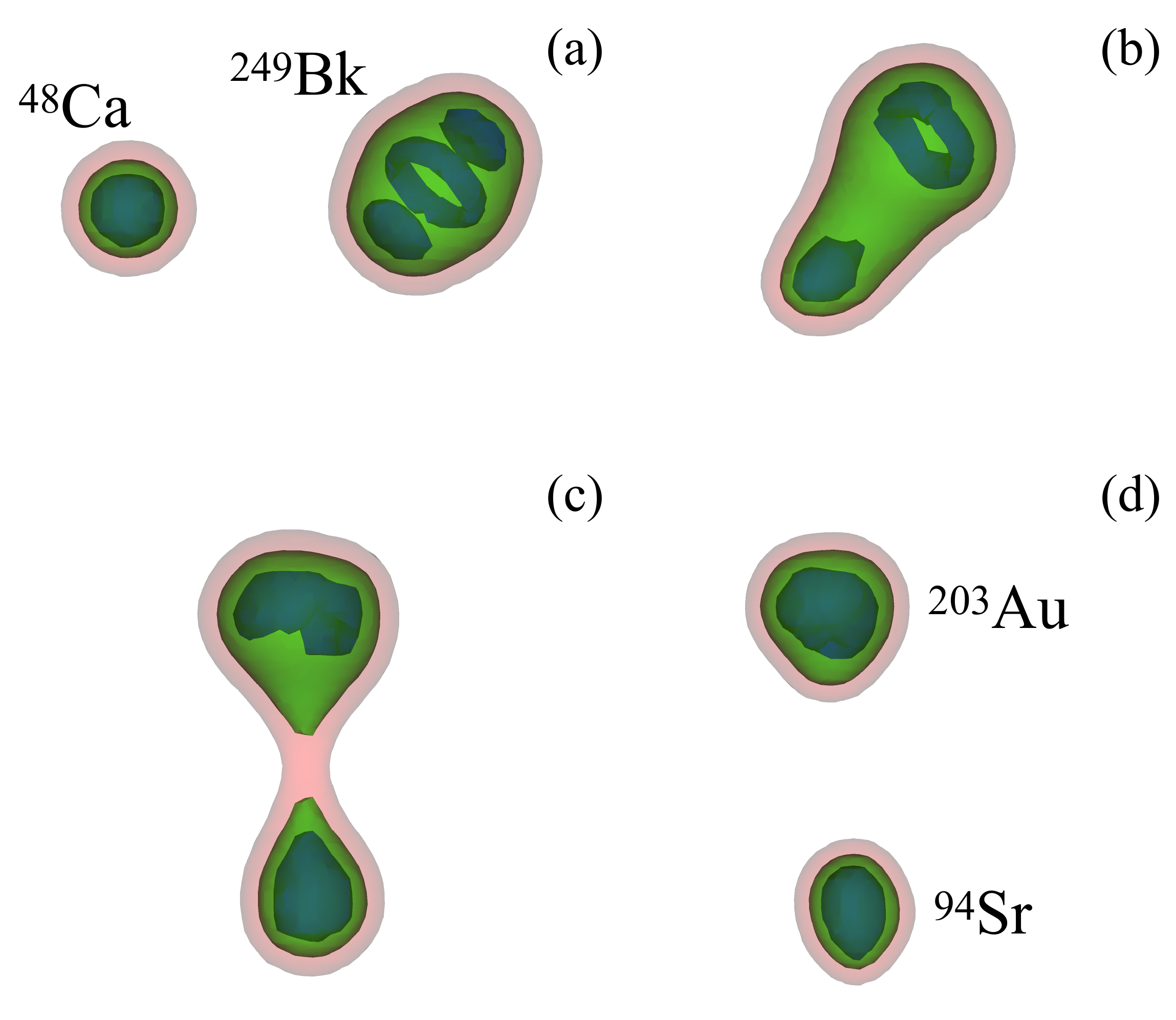}
\caption{\protect(Color online) Isodensity surfaces at $\rho=0.145$, $0.1$, and $0.02$~fm$^{-3}$ in blue, green, and pink, respectively, shown at times $t\simeq0$ (a), $2.1$ (b), $5.8$ (c), and $6.4$~zs (d) for an initial orientation $\beta=135^{\circ}$ and an angular momentum $L=60\hbar$. For visualization purposes, the reaction plane  is $37^{\circ}$ off the plane of the page.}
\label{fig:density}
\end{figure}

The initial separation is chosen to be $30$~fm
with nuclei in their ground states. The nuclei are assumed to arrive
to this separation on a Coulomb trajectory for the purpose of initializing
the proper boosts.
In the second
step, we apply a boost operator to the single-particle wave functions.
The calculations end when the fragments are well separated (or after 35~zs if they are still in contact). Outgoing Coulomb trajectories are then assumed to determine the scattering angle.

The time-propagation
is carried out using a Taylor series expansion (up to orders $10-12$) of the
unitary mean-field propagator, with a time step $\Delta t = 0.4$~fm/c.
For reactions leading to superheavy dinuclear systems, the TDHF calculations
require very long CPU times: a single TDHF run at fixed $E_\mathrm{c.m.}$ energy
for a fixed impact parameter $b$ and orientation angle $\beta$ takes about 2-3 days of CPU time
on a 16-processor LINUX workstation.

Assuming the $^{249}$Bk nucleus to be axially symmetric with no octupole deformation, the cross-section or yield for a specific reaction channel $\xi$ is proportional to
\begin{equation}
 \sigma_\xi\propto \sum_L (2L+1) \int_0^{\frac{\pi}{2}} d\beta \,\sin\beta \int_0^\pi d\alpha \,P_L^{(\xi)}(\beta,\alpha). \label{eq:sigma}
\end{equation}
Here, $P_L^{(\xi)}(\beta,\alpha)$ is the probability for the reaction channel $\xi$ and an orientation of the target defined by the rotation angles $\beta$ and $\alpha$ (see Fig.~\ref{fig:geometry}).
The orientation of the deformation axis is obtained by applying first a rotation of an angle $\beta$ around the axis perpendicular to the reaction plane, and then a rotation of an angle $\alpha$ around the collision axis.

The TDHF calculations are performed for a range of orbital angular momenta $L_i\hbar$ with $\{L_i\}=\{0,10,20\cdots N_L\}$ and $N_L=12$ or 13, depending on the orientation (some orientations lead to quasi-elastic collisions at $L=120$, in which case $L=130$ is not computed).
The first term is then replaced by $$\sum_L (2L+1) \rightarrow \sum_{i=1}^{N_L}K_i \,\,\,\mbox{~with~} \,\,\,
K_i=\sum_{L=L_i-\Delta_i^-}^{L_i+\Delta^+}(2L+1),$$
where $\Delta^+=5$, $\Delta_1^-=0$ and $\Delta_{i\ne 1}=4$.

The double integral in Eq.~\eqref{eq:sigma} is computationally too demanding.
The integral over $\alpha$ is then replaced by a sum over probabilities for $\alpha=0$ and $\pi$.
Equivalently, we can ignore $\alpha$ and extend the integral over $\beta$ up to $\pi$.
We then define the probability
$$\tilde{P}_{L_i}^{(\xi)}(\beta) = \begin{cases} P_{L_i}^{(\xi)}(\beta,0)\mbox{~if~}\beta\le\frac{\pi}{2}\\
P_{L_i}^{(\xi)}(\pi-\beta,\pi)\mbox{~if~}\beta > \frac{\pi}{2}\end{cases}.$$\\

The remaining integral over $\beta$ is discretized with $N_\beta=12$ angles $\{\beta_n\}=\{0^{\circ},15^{\circ},30^{\circ},\cdots ,165^{\circ}\}$.
We can finally write the approximate cross-section as
\begin{equation}
\sigma_\xi\simeq \sum_{i=1}^{N_L}K_i \sum_{n=1}^{N_\beta} C_n\, \tilde{P}^{(\xi)}_{L_i}(\beta_n)\;,
\end{equation}
where we have defined
$$C_n=\begin{cases}
2(1-\cos\delta) \mbox{~if~} n=1\\
\cos(\beta_n-\delta)-\cos(\beta_n+\delta) \mbox{~if~} n>1 \;,
      \end{cases}
$$
with $\delta=7.5^{\circ}$.
Note that, because of its semi-classical behavior, the TDHF theory leads to probabilities $\tilde{P}^{(\xi)}_{L_i}(\beta_n)=0$ or 1 for the reaction channel $\xi$ for a given orientation and angular momentum.

\section{Results}\label{sec:results}

The $^{48}$Ca$+^{249}$Bk at $E_{c.m.}=234$~MeV has been studied as a function of the orientation $\beta$ of the target (see Fig.~\ref{fig:geometry}) and as a function of orbital angular momentum $L$, given in units of $\hbar$, totaling 148 collisions.

\subsection{Quasifission characteristics}
\begin{figure}[!htb]
\includegraphics*[width=8.6cm]{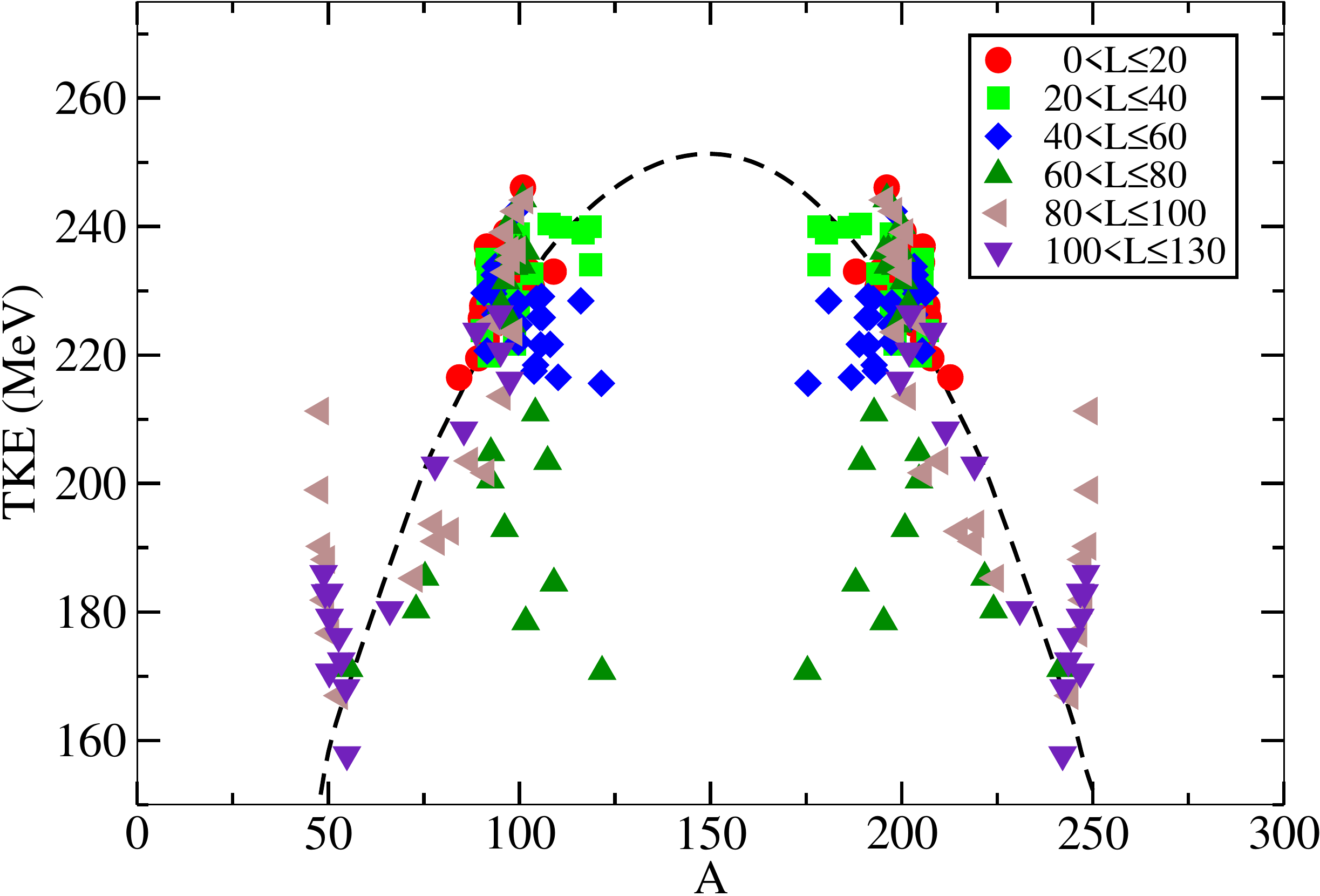}
\caption{\protect(Color online) Total kinetic energy of the fragments as a function of their mass ratio. The curve corresponds to the Viola systematics~\cite{viola1985,hinde1987}.}
\label{fig:MED}
\end{figure}

Figure~\ref{fig:density} shows a typical example of density evolution for a non-central collision.
Different isodensity surfaces are represented.
The rings observed at highest density in panels (a) and (b) are coming from shell structure effects \cite{simenel2012}.
After contact, the nuclei are trapped in a potential pocket, forming a dinuclear system (panel (b)) which, unlike in fusion, does not reach an equilibrated compound nucleus.
When the dinuclear system fissions (panel (c)), it forms two fragments (panel (d)) which preserve a memory of the entrance channel.
\begin{figure}[!htb]
\includegraphics*[width=8.6cm]{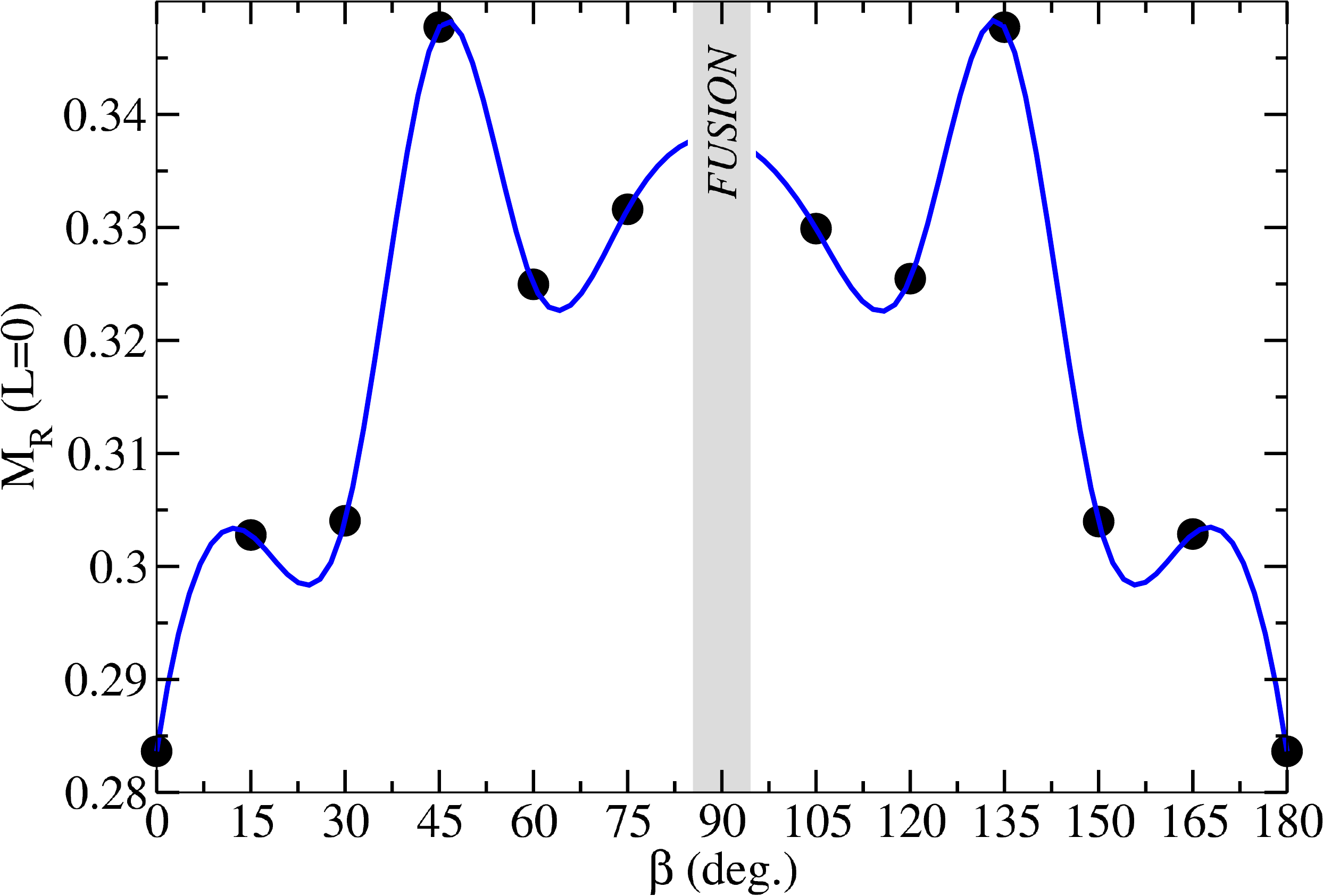}
\caption{\protect(Color online). Mass ratio $M_R$ as a function of orientation angle $\beta$ for central collisions. Fusion is indicated by the shaded area. }
\label{fig:central_beta}
\end{figure}

The outgoing fragments for this reaction are $^{94}$Sr and $^{203}$Au.
Such a significant mass transfer towards a more mass symmetric configuration is one of the characteristics of quasifission.
A second characteristic is the rotation of the dinuclear system before scission.
This rotation is due to the initial angular momentum for non-central collisions.
For contact times $\tau<20$~zs, the dinuclear system usually does not undergo a full rotation before scission, resulting in so-called fast quasifission~\cite{durietz2013,hinde2018}.
Such times are also too short for the system to achieve full mass equilibration and form two fragments with similar masses.
Fast quasifission then results in correlations between masses and angles which can be used to infer the time scale of the reaction~\cite{toke1985,durietz2011}.
The density evolution represented in Fig.~\ref{fig:density} is an example of fast quasifission reaction as the fragments are
in contact for $\sim6$~zs and the dinuclear system rotates by only $\sim90$~degrees.
In fact, all quasifissions observed in our calculations for this system correspond to fast quasifission, producing fragment mass-angle correlations which will be studied in Section~\ref{sec:MAD}.

Another characteristic of quasifission is that the reaction is fully damped.
In quasifission, the outgoing fragments have a total kinetic energy (TKE) essentially determined by their Coulomb repulsion at scission.
As a first approximation, this TKE does not depend on the beam energy.
Figure~\ref{fig:MED} shows the mass-energy distribution (MED), i.e., the distribution of TKE as a function of the number of nucleons $A$ in the fragments.
Except for quasi-elastic reactions in which the masses of the fragments are very close to the projectile and target masses, the TKE are generally distributed around the Viola systematics~\cite{viola1985,hinde1987} (dashed line) which gives an empirical estimate of fully damped fission fragments.

Each color in Fig.~\ref{fig:MED} shows the location in the MED that is expected for a given range of orbital angular momenta. In each case, two or three values of L and thirteen angles $\beta$  are included.
The more central collisions ($L\le80 \hbar$) all lead to quasifission, while more peripheral collisions ($L>80\hbar$) lead to both quasi-elastic and quasifission reactions.
This indicates a strong influence of  orientation on the reaction outcome.

\subsection{Effect of target orientation in central collisions}
\begin{figure}[!htb]
\includegraphics*[scale=0.9]{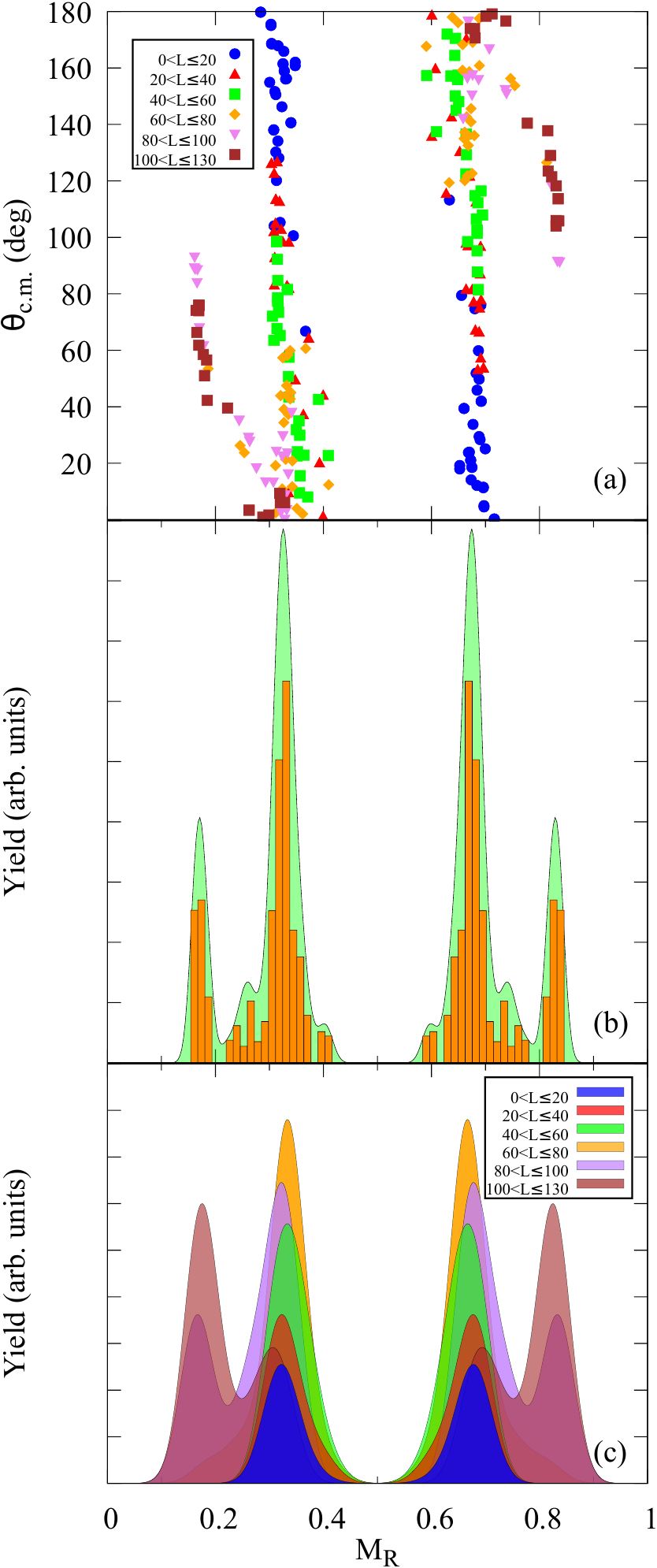}
\caption{\protect(Color online). (a) Distribution of scattering angle $\theta_{c.m.}$ versus mass ratio $M_R$ (MAD). The colors correspond to different ranges of angular momenta. (b) Fragment mass yield (histogram). The solid line gives a smooth representation of the histogram using the kernel density estimation with bandwidth 0.012. (c) Mass yields obtained for different ranges of angular momenta.}
\label{fig:MAD}
\end{figure}

Different orientations of the target lead to different compactness of the dinuclear system.
A clear relation between orientation and compactness is obtained in the case of central collisions ($L=0$) in which case less compact configurations are obtained for $\beta=0$ and $180$~degrees, leading to collisions with the tips of the target, while the most compact configurations are obtained for $\beta=90$~degrees, leading to collisions with the side.
For non-central collisions, the relationship between orientation and compactness is less straightforward and can be estimated assuming Coulomb trajectories until the distance of closest approach~\cite{wakhle2014}.

Figure~\ref{fig:central_beta} shows the mass ratio of the fragments, defined as the ratio between the mass of the fragment and the total mass of the system, as a function of the orientation for central collisions.
A slight asymmetry between $\beta$ and $\pi-\beta$ is observed due to a small violation of symmetry under reflection across the plane orthogonal to the main deformation axis of  $^{249}$Bk HF ground-state.

Fusion is only observed for side collisions, in agreement with previous TDHF studies \cite{wakhle2014,oberacker2014,umar2016}.
Overall, a small increase of the mass ratio from $M_R\approx0.28$ to $0.35$ is observed when going from tip orientations to more compact configurations.
There is, however, no clear transition associated with an eventual critical angle $\beta_{crit}$ when going from tip to side orientation in this system (except for when fusion is achieved).
This shows the importance of considering a full range of intermediate orientations in order to realize quantitative predictions.

\subsection{Correlations between fragment masses and scattering angles}\label{sec:MAD}

Experimental studies of correlations between fragment masses and scattering angles have led to considerable insights into quasifission mechanisms in the past \cite{toke1985,shen1987,hinde2008,simenel2012b,durietz2013,wakhle2014,hammerton2015,morjean2017,mohanto2018,hinde2018}.
TDHF calculations have been used recently to help interpret qualitatively these correlations \cite{wakhle2014,hammerton2015,umar2016,sekizawa2016}.
However, these theoretical studies were somewhat limited by the restriction of initial orientations.

The mass-angle distribution (MAD) of the fragments is shown in Fig.~\ref{fig:MAD}(a).
The horizontal axis gives the mass ratio $M_R=\frac{m_1}{m_1+m_2}$ where $m_1$ and $m_2$ are the masses of the fragments.
These masses are for primary fragments, i.e., before nucleon emission takes place.
This is also what is measured experimentally using 2-body kinematics techniques \cite{toke1985,hinde1996}.
The colors represent different angular momentum ranges, as in Fig.~\ref{fig:MED}.

Most calculations lead to quasifission with fragment mass ratios $0.28<M_R<0.72$, while projectile and target mass ratios are at $M_R\simeq0.16$ and $0.84$, respectively.
This indicates significant mass transfer towards more symmetric mass repartitions. However, full symmetry is never achieved in these TDHF calculations, unlike in $^{40}$Ca$+^{238}$U~\cite{wakhle2014}.
Most peripheral collisions with $L\ge70\hbar$ lead to larger mass asymmetries and a transition from quasifission to deep-inelastic and quasi-elastic reactions.
Note that fragments from elastic scattering are not shown.

We also see that quasifission fragments are distributed among the full range of scattering angles, from $\theta_{c.m.}=0$ (forward angles) to $180$ degrees (backward angles).
This wide angular distribution motivates the development of larger angular acceptance detectors \cite{khuyagbaatar2018,banerjee2019}.
Note that each angular momentum range leads itself to a broad distribution of angles.
For instance, results from $L\le20 \hbar$ are found all the way from backward angles to $\theta_{c.m.}\simeq70$ degrees, while $L\le40 \hbar$ spans all angles.
This is a manifestation of the impact of orientation on the angular distribution: for a given angular momentum, the scattering angle strongly depends on the orientation of the target.
However, there is much less dependence of the mass on the orientation, as each orientation leads to approximately similar mass ratio for quasifission outcomes in this system.

Interestingly, the correlation between quasifission fragment masses and angles shows a narrow mass distribution for the light fragment around $M_R\simeq0.3$ at more backward angles with $\theta_{c.m.}>70$~degrees.
At more forward angles ($\theta_{c.m.}<70$~degrees), the light fragment mass distribution  broadens and slightly shifts towards larger masses ($M_R\sim0.34$).
For symmetry reasons, a similar narrow (respectively broad) mass distribution is found in the heavy fragment at $M_R\simeq0.7$ (resp. $M_R\sim0.66$) for $\theta_{c.m.}<110$ (resp. $\theta_{c.m.}>110$) degrees.
The origin of these features will be discussed using neutron and proton distributions in Sec.~\ref{sec:shell}.

\subsection{Fragment mass distributions}

The theoretical MAD in Fig.~\ref{fig:MAD}(a) is useful to investigate correlations between mass and angle. However it is not directly related to yields and cross-sections as it does not account for the $2L+1$ and $\sin\beta$ terms in Eq.~(\ref{eq:sigma}).
Yields are better represented in one-dimensional spectra.
Figure~\ref{fig:MAD}(b) shows a histogram of the mass ratio yield obtained from Eq.~(\ref{eq:sigma}).
The solid line curve gives a smooth representation of the histogram.
As these are more illustrative, we will only use these smooth representations of yields in later figures.

The quasifission mass yields in Fig.~\ref{fig:MAD}(b) are strongly peaked at $M_R\sim0.33$ and $0.67$,
with a full width half maximum FWHM~$\simeq0.1$ corresponding to a standard deviation $\sigma_{M_R}\simeq0.042$.
Note that the present TDHF calculations neglect mass distributions associated with each single TDHF calculation outcome.
The latter can be computed using particle-number projection techniques~\cite{simenel2010,sekizawa2013,scamps2013a,scamps2017b}.
However, the width of the resulting distributions are known to be underestimated in dissipative collisions~\cite{dasso1979}.
Beyond mean-field calculations incorporating one-body
fluctuations could also be used~\cite{simenel2011,williams2018,lacroix2014,ayik2015,ayik2015a,ayik2016,ayik2018,tanimura2017}.
However, these approaches are not used here as they would significantly increase computing time and would become prohibitive with large ranges of orientations and angular momenta.

We can nevertheless attempt a comparison with typical experimental mass width for quasifission distributions, keeping in mind that our theoretical prediction is a lower bound.
Experimental spread $\sigma_{M_R}$ can roughly be parameterized as a linear function from $\sigma_{M_R}^{(DIC)}\approx0.025$ typical for deep-inelastic collisions (DIC) at the mass ratio of the projectile and target, to $\sigma_{M_R}^{(FF)}=0.07$ in fusion-fission at $M_R=0.5$~\cite{durietz2011}.
We then get an estimate of $\sigma_{M_R}^{(QF)}\approx 0.047$ at $M_R=0.33$, which is only $\sim10\%$ higher than the TDHF prediction.
The present calculations, to a large extent, account for the expected fluctuations of the mass of the quasifission fragments.
These fluctuations are essentially coming from the various orientations of the deformed target nucleus.

Figure~\ref{fig:MAD}(c) shows the expected mass ratio yield distributions for various ranges of angular momenta $L$.
The purpose of this figure is to compare quantitatively the relative contributions to the yields when going from central to peripheral collisions.
For instance, we see that, because of the $2L+1$ weighting factor in Eq.~(\ref{eq:sigma}), the most central collisions with $L\le20\hbar$, which are found at backward angles in Fig.~\ref{fig:MAD}(a), have also the smallest contribution to the total yield.
In order to understand the transition from $M_R\simeq0.30$ to $0.34$ discussed at the end of Sec.~\ref{sec:MAD}, it will then be necessary to fully exploit the correlations between  masses and angles of the quasifission fragments.

\subsection{Identification of shell effects in quasifission fragments}\label{sec:shell}

Experimental indications of the role of shell effects in the production of quasifission fragments initially came from mass-yield measurements \cite{itkis2004,nishio2008,wakhle2014}.
Theoretical predictions from TDHF calculations then supported these views \cite{wakhle2014,oberacker2014,umar2016}.
However, to unambiguously confirm the role of shell effects, proton or neutron numbers distributions have to be measured.
Only recently this was done for quasifission for the $^{48}$Ti$+^{238}$U reaction using  x-ray detectors to identify proton numbers in the fragments \cite{morjean2017}, thus confirming the role of $Z=82$ ``magic'' shell in this reaction.

\begin{figure}[!htb]
\centerline{\includegraphics*[width=8.6cm]{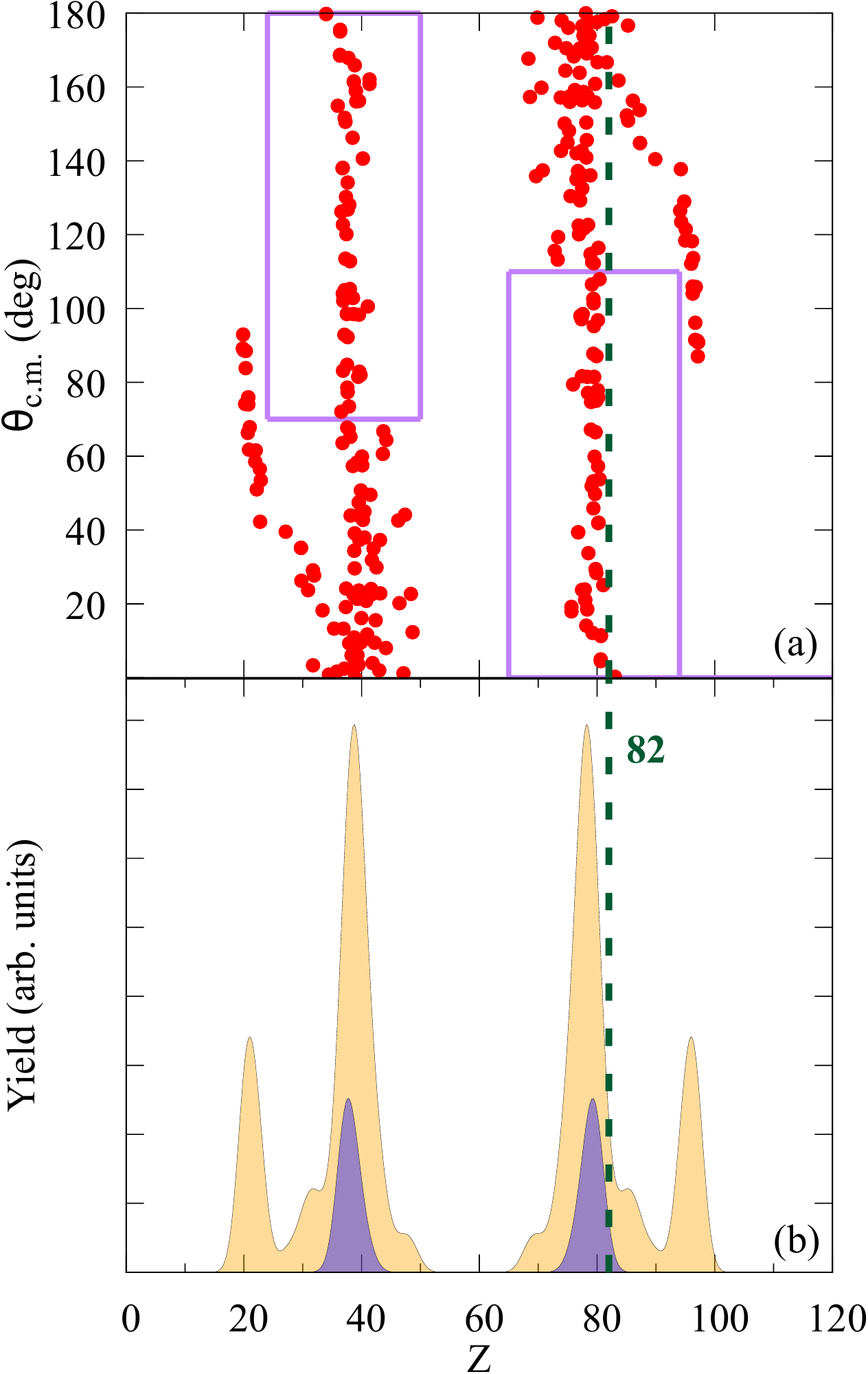}}
\caption{\protect(Color online). (a) Distribution of scattering angle $\theta_{c.m.}$ versus proton number $Z$ (ZAD). (b) Fragment proton number yield without (lighter shade) and with angular cut $\theta_{c.m.}>70$~degrees (darker shade). The vertical line represents potential proton shell gap.}
\label{fig:ZAD}
\end{figure}

\begin{figure}[!htb]
\centerline{\includegraphics*[width=8.6cm]{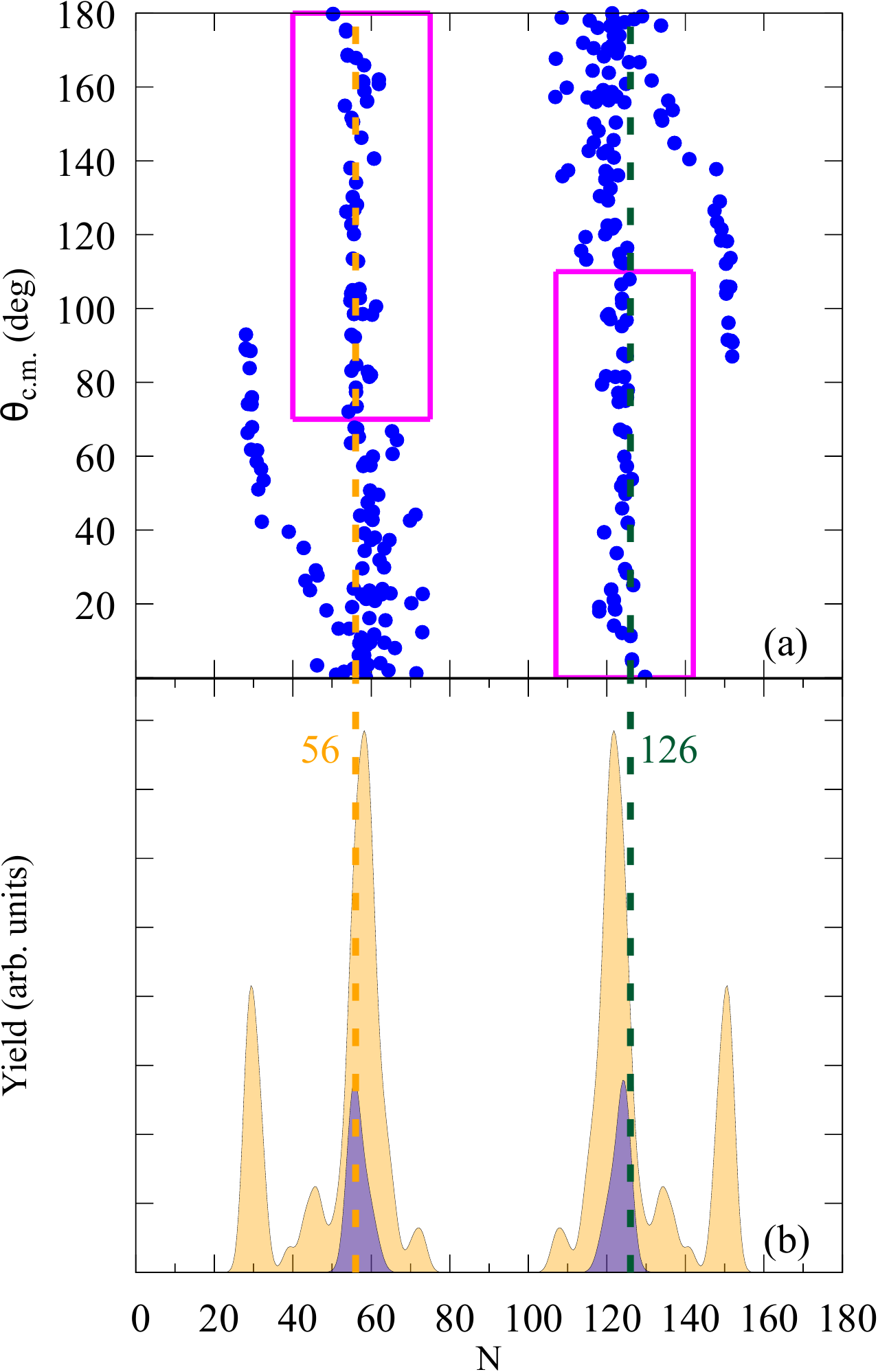}}
\caption{\protect(Color online). (a) Distribution of scattering angle $\theta_{c.m.}$ versus neutron number $N$ (NAD). (b) Fragment neutron number yield without (lighter shade) and with angular cut $\theta_{c.m.}>70$~degrees (darker shade). The vertical lines represent potential neutron shell gaps.}
\label{fig:NAD}
\end{figure}

To investigate the role of potential shell effects in $^{48}$Ca$+^{249}$Bk quasifission, the correlations between proton and neutron numbers with scattering angles have been plotted in Figs.~\ref{fig:ZAD}(a) and \ref{fig:NAD}(a), respectively.
Proton and neutron numbers yields are also shown in  Figs.~\ref{fig:ZAD}(b) and \ref{fig:NAD}(b), respectively.
In addition to the total yields obtained without restriction on scattering angles and nucleon numbers (orange spectra), gates on quasifission fragments have also been used (rectangles in Figs.~\ref{fig:ZAD}(a) and \ref{fig:NAD}(a)) with $\theta_{c.m.}>70$ degrees for the light fragments and $\theta_{c.m.}<110$ degrees for the heavy ones.
The resulting gated spectra are shown in purple in Figs.~\ref{fig:ZAD}(b) and \ref{fig:NAD}(b).

The vertical dotted line in Fig.~\ref{fig:ZAD} shows the expected position of fragments affected by $Z=82$ shell effects.
The heavy fragments seem to be systematically lighter, indicating that $Z=82$ may not play a significant role in this reaction.
This is surprising as TDHF studies have shown the importance of this shell gap in quasifission for $^{40,48}$Ca,$^{48}$Ti+$^{238}$U \cite{wakhle2014,oberacker2014,morjean2017}.

A similar comparison is made with the ``magic'' number $N=126$ in Fig.~\ref{fig:NAD}.
Here, we see that some fragments are indeed formed with $N=126$.
However, both the centroids of the ungated and gated distributions are shifted towards smaller neutron numbers.
For the gated spectrum, the shift is relatively small as the peak is centered at $\overline{N}_{gated}\simeq124$. Nevertheless, spherical shell effects are known to be quite localized in the nuclear chart and this ``proximity'' may as well be coincidental.
Other spherical shell effects are also excluded for both protons and neutrons.
In particular, the quasifission peaks are far from $Z=50$ or $N=50$.

This leaves us with potential deformed shell effects.
For instance, the importance of octupole deformed shell gaps at $Z=52-56$ \cite{scamps2018} and $N=52-56$ \cite{scamps2019} have recently been shown to have an important role in driving heavy systems towards asymmetric fission.
As a results of these gaps, the nuclei can easily acquire octupole deformations for a small cost (and sometimes even a gain) in energy.
This is why their production as fission fragments is naturally favored, as the fissioning system has no choice but to go through a shape with a neck just before scission, imposing strong octupole deformations in the fragments.
Despite its strong spherical shell effects which are expected to energetically favor its production, the formation of $^{132}$Sn as a fission fragment is hindered by its strong resistance to octupole deformations.
This is not the case, however, of $^{208}$Pb which can easily acquire octupole deformations thanks to its low-lying octupole $3^-$ state.

The orange vertical dotted line in Figure~\ref{fig:NAD} indicates the expected location of fragments affected by the $N=56$ octupole deformed shell gap.
It matches very well the position of the gated peak, providing a plausible explanation for the origin of this narrow distribution of quasifission fragments at backward angles, corresponding to more central collisions.

As discussed in Sec.~\ref{sec:MAD}, however, more peripheral collisions ($\theta_{c.m.}<70$ degrees for the light fragment) lead to the production of slightly more symmetric quasifission fragments.
For the light fragment, the $Z$ and $N$ distributions of these more peripheral quasifission events [see Figs.~\ref{fig:ZAD}(a) and \ref{fig:NAD}(b)] seem to be centered around $\overline{N}_{periph}\approx60$ and $\overline{Z}_{periph}\approx40$, respectively, indicating the production of fragments in the $^{100}$Zr region.
Similar observations were already made in $^{40,48}$Ca+$^{238}$U systems \cite{oberacker2014}.
\begin{figure}[!htb]
\centerline{\includegraphics*[width=8.6cm]{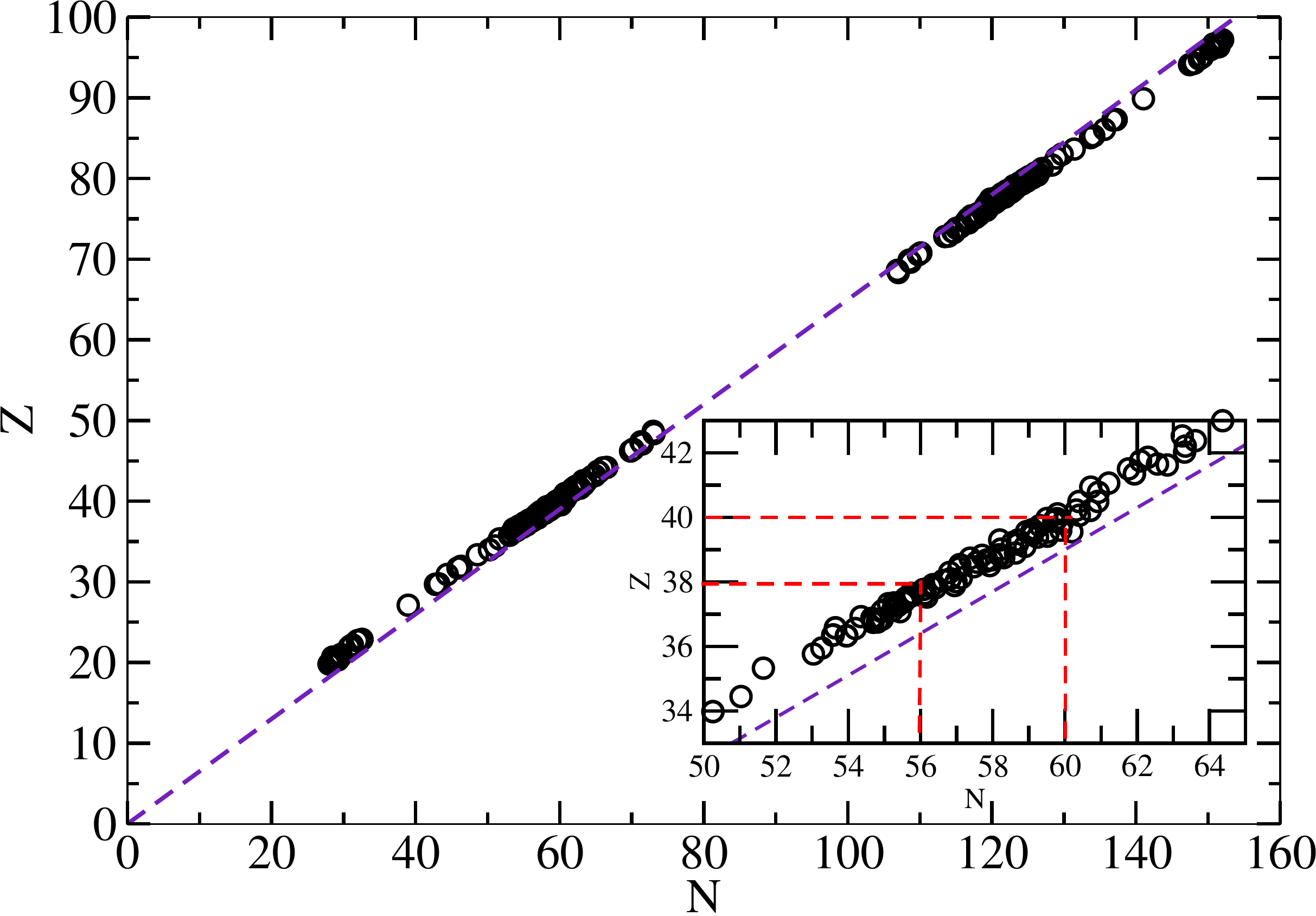}}
\caption{\protect(Color online). Distribution of proton number $Z$ versus neutron number $N$ of the fragments. The dashed line represents the $N/Z$ ratio of the compound nucleus.
The inset is a zoom around the light fragment. Thin dashed lines indicate the positions of $^{94}$Sr ($Z=38$, $N=56$) and $^{100}$Zr ($Z=40$, $N=60$).}
\label{fig:NZ}
\end{figure}
\begin{figure}[!htb]
\centerline{\includegraphics*[width=8.6cm]{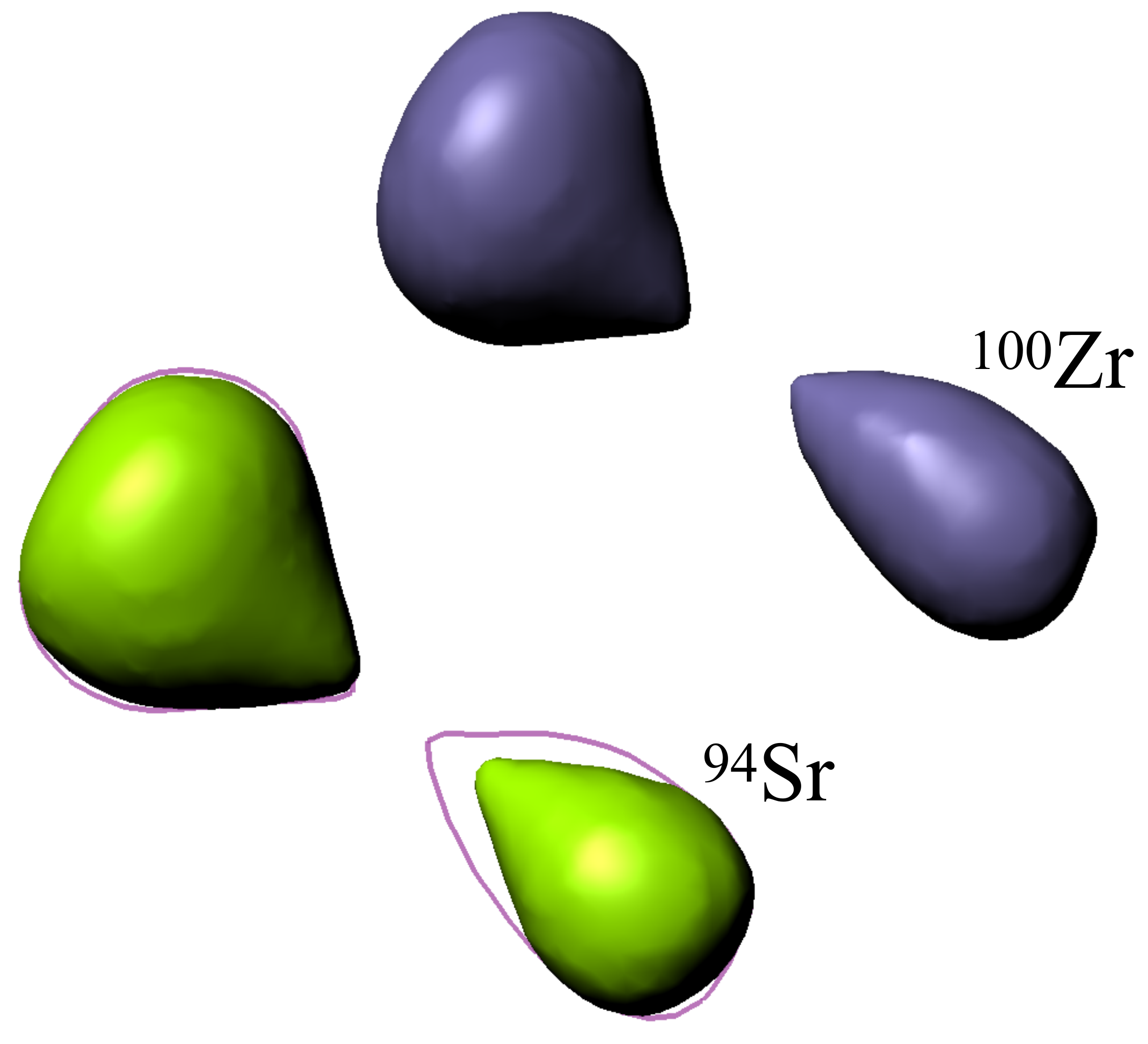}}
\caption{\protect(Color online). Isodensity surfaces at $\rho=0.1$~fm$^{-3}$ for $L=90\hbar$ and $\beta=120^{\circ}$ (top), and $L=60\hbar$ and $\beta=135^{\circ}$ (bottom), just after the breaking of the neck. The light fragment (right) in the top is a $^{94}$Sr ($Z=38$, $N=56$) and a $^{100}$Zr ($Z=40$, $N=60$) in the bottom. The contour line in the bottom represents the same density as in the top. }
\label{fig:shapes}
\end{figure}

Figure~\ref{fig:NZ}  shows the distribution of fragments in the $N$ and $Z$ plane.
We see that, due to a strong symmetry energy, the fragments have $N/Z$ ratios very close to the one of the compound nucleus.
Nevertheless, the light fragments are slightly more proton rich, and the heavy fragments more proton deficient, due to the stronger Coulomb repulsion in the latter.
The production of fragments in the $^{100}$Zr region is confirmed in the inset of Fig.~\ref{fig:NZ}. We also see that the fragments with $N=56$ neutrons correspond essentially to $^{94}$Sr, as also illustrated in Fig.~\ref{fig:density}

Shell effects are known to evolve with the deformation of the nucleus.
To confirm the presence of shell effects, it is then necessary to verify that the deformation is the one expected to exhibit a shell gap.
Typical isosurface densities for reactions just after scission leading to the production of a $^{100}$Zr (top) and of a $^{94}$Sr (bottom) fragment are shown in Fig.~\ref{fig:shapes}.
In particular, the $^{94}$Sr fragment is quite compact with a strong octupole shape, similar to what is observed in fission of mercury isotopes producing $N=56$ fragments with to octupole shell gaps \cite{scamps2019}.
The $^{100}$Zr fragment is also octupole deformed (as the density is shown just after breaking of the neck) but with a much more elongated shape.
Neutron rich zirconium isotopes are indeed expected to exhibit strong quadrupole deformations \cite{lalazissis1999,blazkiewicz2005,hwang2006}.

\section{Conclusions}\label{sec:conclusion}

The $^{48}$Ca$+^{249}$Bk reaction, used experimentally to produce Tennessine ($Z=117$), has been studied at a center of mass energy of $234$~MeV with time-dependent Hartree-Fock simulations.
Properties of quasifission fragments, such as mass, numbers of protons and neutrons, kinetic energy, and scattering angles have been studied systematically.

Unlike previous TDHF studies of quasifission, a broad distribution of orientations of the target has been considered for the first time, allowing for the prediction of, e.g.,  mass yield characteristics that can be directly compared with experiment.
Except for a few collisions compatible with fusion or long-time quasifission, the largely dominant outcome is fast quasifission.
It is shown that the orientation has also a strong influence on the scattering angle.

Fast quasifission produces peaks in the mass yield distribution for the projectile-like and target-like fragments with a width in good agreement with empirical estimates, despite the fact that the TDHF approach does not account for beyond mean-field fluctuations.
Here, the observed fluctuations come mainly from the various orientations of the target in the entrance channel.

The influence of shell effects on the formation of the fragments has been investigated.
Unlike similar reactions with $^{238}$U targets, no influence of $^{208}$Pb is observed unambiguously.
However, elongated fragments in the $^{100}$Zr region are produced in the more peripheral quasifission reactions.
More central collisions consistently produce fragments with $N=56$ nucleons for all orientations.
This is interpreted as an effect of octupole deformed shells favoring the production of fragments with pear shapes at scission.
A similar effect has recently been discussed in the case of fission.

This is the first indication of a potential influence of octupole shell gaps in quasifission.
Its experimental confirmation would be particularly interesting as it would point towards strong similarities in how shell effects affect both fission and quasifission.
These shell effects in the light fragments will be more easily investigated experimentally at backward angles.

\begin{acknowledgments}
We thank D. J. Hinde for useful discussions.
This work has been supported by the U.S. Department of Energy under grant No.
DE-SC0013847 with Vanderbilt University and by the
Australian Research Council Discovery Project (project numbers DP160101254 and DP190100256) funding schemes.
\end{acknowledgments}

\bibliography{VU_bibtex_master.bib}

\begin{thebibliography}{120}%
\makeatletter
\providecommand \@ifxundefined [1]{%
 \@ifx{#1\undefined}
}%
\providecommand \@ifnum [1]{%
 \ifnum #1\expandafter \@firstoftwo
 \else \expandafter \@secondoftwo
 \fi
}%
\providecommand \@ifx [1]{%
 \ifx #1\expandafter \@firstoftwo
 \else \expandafter \@secondoftwo
 \fi
}%
\providecommand \natexlab [1]{#1}%
\providecommand \enquote  [1]{``#1''}%
\providecommand \bibnamefont  [1]{#1}%
\providecommand \bibfnamefont [1]{#1}%
\providecommand \citenamefont [1]{#1}%
\providecommand \href@noop [0]{\@secondoftwo}%
\providecommand \href [0]{\begingroup \@sanitize@url \@href}%
\providecommand \@href[1]{\@@startlink{#1}\@@href}%
\providecommand \@@href[1]{\endgroup#1\@@endlink}%
\providecommand \@sanitize@url [0]{\catcode `\\12\catcode `\$12\catcode
  `\&12\catcode `\#12\catcode `\^12\catcode `\_12\catcode `\%12\relax}%
\providecommand \@@startlink[1]{}%
\providecommand \@@endlink[0]{}%
\providecommand \url  [0]{\begingroup\@sanitize@url \@url }%
\providecommand \@url [1]{\endgroup\@href {#1}{\urlprefix }}%
\providecommand \urlprefix  [0]{URL }%
\providecommand \Eprint [0]{\href }%
\providecommand \doibase [0]{http://dx.doi.org/}%
\providecommand \selectlanguage [0]{\@gobble}%
\providecommand \bibinfo  [0]{\@secondoftwo}%
\providecommand \bibfield  [0]{\@secondoftwo}%
\providecommand \translation [1]{[#1]}%
\providecommand \BibitemOpen [0]{}%
\providecommand \bibitemStop [0]{}%
\providecommand \bibitemNoStop [0]{.\EOS\space}%
\providecommand \EOS [0]{\spacefactor3000\relax}%
\providecommand \BibitemShut  [1]{\csname bibitem#1\endcsname}%
\let\auto@bib@innerbib\@empty
\bibitem [{\citenamefont {Heusch}\ \emph {et~al.}(1978)\citenamefont {Heusch},
  \citenamefont {Volant}, \citenamefont {Freiesleben}, \citenamefont
  {Chestnut}, \citenamefont {Hildenbrand}, \citenamefont {P\"uhlhofer},
  \citenamefont {Schneider}, \citenamefont {Kohlmeyer},\ and\ \citenamefont
  {Pfeffer}}]{heusch1978}%
  \BibitemOpen
  \bibfield  {author} {\bibinfo {author} {\bibfnamefont {B.}~\bibnamefont
  {Heusch}}, \bibinfo {author} {\bibfnamefont {C.}~\bibnamefont {Volant}},
  \bibinfo {author} {\bibfnamefont {H.}~\bibnamefont {Freiesleben}}, \bibinfo
  {author} {\bibfnamefont {R.~P.}\ \bibnamefont {Chestnut}}, \bibinfo {author}
  {\bibfnamefont {K.~D.}\ \bibnamefont {Hildenbrand}}, \bibinfo {author}
  {\bibfnamefont {F.}~\bibnamefont {P\"uhlhofer}}, \bibinfo {author}
  {\bibfnamefont {W.~F.~W.}\ \bibnamefont {Schneider}}, \bibinfo {author}
  {\bibfnamefont {B.}~\bibnamefont {Kohlmeyer}}, \ and\ \bibinfo {author}
  {\bibfnamefont {W.}~\bibnamefont {Pfeffer}},\ }\bibfield  {title} {\enquote
  {\bibinfo {title} {{T}he reaction mechanism in the system
  $^{132}${X}e+$^{56}${F}e at 5.73 {M}e{V}/u: {E}vidence for a new type of
  strongly damped collisions},}\ }\href {\doibase 10.1007/BF01417723}
  {\bibfield  {journal} {\bibinfo  {journal} {Z. Phys. A}\ }\textbf {\bibinfo
  {volume} {288}},\ \bibinfo {pages} {391--400} (\bibinfo {year}
  {1978})}\BibitemShut {NoStop}%
\bibitem [{\citenamefont {Back}\ \emph {et~al.}(1981)\citenamefont {Back},
  \citenamefont {Clerc}, \citenamefont {Betts}, \citenamefont {Glagola},\ and\
  \citenamefont {Wilkins}}]{back1981}%
  \BibitemOpen
  \bibfield  {author} {\bibinfo {author} {\bibfnamefont {B.~B.}\ \bibnamefont
  {Back}}, \bibinfo {author} {\bibfnamefont {H.-G.}\ \bibnamefont {Clerc}},
  \bibinfo {author} {\bibfnamefont {R.~R.}\ \bibnamefont {Betts}}, \bibinfo
  {author} {\bibfnamefont {B.~G.}\ \bibnamefont {Glagola}}, \ and\ \bibinfo
  {author} {\bibfnamefont {B.~D.}\ \bibnamefont {Wilkins}},\ }\bibfield
  {title} {\enquote {\bibinfo {title} {{O}bservation of {A}nisotropy in the
  {F}ission {D}ecay of {N}uclei with {V}anishing {F}ission {B}arrier},}\ }\href
  {\doibase 10.1103/PhysRevLett.46.1068} {\bibfield  {journal} {\bibinfo
  {journal} {Phys. Rev. Lett.}\ }\textbf {\bibinfo {volume} {46}},\ \bibinfo
  {pages} {1068--1071} (\bibinfo {year} {1981})}\BibitemShut {NoStop}%
\bibitem [{\citenamefont {Back}\ \emph {et~al.}(1983)\citenamefont {Back},
  \citenamefont {Betts}, \citenamefont {Cassidy}, \citenamefont {Glagola},
  \citenamefont {Gindler}, \citenamefont {Glendenin},\ and\ \citenamefont
  {Wilkins}}]{back1983}%
  \BibitemOpen
  \bibfield  {author} {\bibinfo {author} {\bibfnamefont {B.~B.}\ \bibnamefont
  {Back}}, \bibinfo {author} {\bibfnamefont {R.~R.}\ \bibnamefont {Betts}},
  \bibinfo {author} {\bibfnamefont {K.}~\bibnamefont {Cassidy}}, \bibinfo
  {author} {\bibfnamefont {B.~G.}\ \bibnamefont {Glagola}}, \bibinfo {author}
  {\bibfnamefont {J.~E.}\ \bibnamefont {Gindler}}, \bibinfo {author}
  {\bibfnamefont {L.~E.}\ \bibnamefont {Glendenin}}, \ and\ \bibinfo {author}
  {\bibfnamefont {B.~D.}\ \bibnamefont {Wilkins}},\ }\bibfield  {title}
  {\enquote {\bibinfo {title} {{E}xperimental {S}ignatures of {Q}uasifission
  {R}eactions},}\ }\href {\doibase 10.1103/physrevlett.50.818} {\bibfield
  {journal} {\bibinfo  {journal} {Phys. Rev. Lett.}\ }\textbf {\bibinfo
  {volume} {50}},\ \bibinfo {pages} {818--821} (\bibinfo {year}
  {1983})}\BibitemShut {NoStop}%
\bibitem [{\citenamefont {Bock}\ \emph {et~al.}(1982)\citenamefont {Bock},
  \citenamefont {Chu}, \citenamefont {Dakowski}, \citenamefont {Gobbi},
  \citenamefont {Grosse}, \citenamefont {Olmi}, \citenamefont {Sann},
  \citenamefont {Schwalm}, \citenamefont {Lynen}, \citenamefont {M\"uller},
  \citenamefont {Bj\o{}rnholm}, \citenamefont {Esbensen}, \citenamefont
  {W\"olfli},\ and\ \citenamefont {Morenzoni}}]{bock1982}%
  \BibitemOpen
  \bibfield  {author} {\bibinfo {author} {\bibfnamefont {R.}~\bibnamefont
  {Bock}}, \bibinfo {author} {\bibfnamefont {Y.~T.}\ \bibnamefont {Chu}},
  \bibinfo {author} {\bibfnamefont {M.}~\bibnamefont {Dakowski}}, \bibinfo
  {author} {\bibfnamefont {A.}~\bibnamefont {Gobbi}}, \bibinfo {author}
  {\bibfnamefont {E.}~\bibnamefont {Grosse}}, \bibinfo {author} {\bibfnamefont
  {A.}~\bibnamefont {Olmi}}, \bibinfo {author} {\bibfnamefont {H.}~\bibnamefont
  {Sann}}, \bibinfo {author} {\bibfnamefont {D.}~\bibnamefont {Schwalm}},
  \bibinfo {author} {\bibfnamefont {U.}~\bibnamefont {Lynen}}, \bibinfo
  {author} {\bibfnamefont {W.}~\bibnamefont {M\"uller}}, \bibinfo {author}
  {\bibfnamefont {S.}~\bibnamefont {Bj\o{}rnholm}}, \bibinfo {author}
  {\bibfnamefont {H.}~\bibnamefont {Esbensen}}, \bibinfo {author}
  {\bibfnamefont {W.}~\bibnamefont {W\"olfli}}, \ and\ \bibinfo {author}
  {\bibfnamefont {E.}~\bibnamefont {Morenzoni}},\ }\bibfield  {title} {\enquote
  {\bibinfo {title} {{D}ynamics of the fusion process},}\ }\href {\doibase
  10.1016/0375-9474(82)90420-1} {\bibfield  {journal} {\bibinfo  {journal}
  {Nucl. Phys. A}\ }\textbf {\bibinfo {volume} {388}},\ \bibinfo {pages}
  {334--380} (\bibinfo {year} {1982})}\BibitemShut {NoStop}%
\bibitem [{\citenamefont {Sahm}\ \emph {et~al.}(1984)\citenamefont {Sahm},
  \citenamefont {Clerc}, \citenamefont {Schmidt}, \citenamefont {Reisdorf},
  \citenamefont {Armbruster}, \citenamefont {He\ss{}berger}, \citenamefont
  {Keller}, \citenamefont {M\"unzenberg},\ and\ \citenamefont
  {Vermeulen}}]{sahm1984}%
  \BibitemOpen
  \bibfield  {author} {\bibinfo {author} {\bibfnamefont {C.-C.}\ \bibnamefont
  {Sahm}}, \bibinfo {author} {\bibfnamefont {H.-G.}\ \bibnamefont {Clerc}},
  \bibinfo {author} {\bibfnamefont {K.-H.}\ \bibnamefont {Schmidt}}, \bibinfo
  {author} {\bibfnamefont {W.}~\bibnamefont {Reisdorf}}, \bibinfo {author}
  {\bibfnamefont {P.}~\bibnamefont {Armbruster}}, \bibinfo {author}
  {\bibfnamefont {F.~P.}\ \bibnamefont {He\ss{}berger}}, \bibinfo {author}
  {\bibfnamefont {J.~G.}\ \bibnamefont {Keller}}, \bibinfo {author}
  {\bibfnamefont {G.}~\bibnamefont {M\"unzenberg}}, \ and\ \bibinfo {author}
  {\bibfnamefont {D.}~\bibnamefont {Vermeulen}},\ }\bibfield  {title} {\enquote
  {\bibinfo {title} {{H}indrance of fusion in central collisions of heavy
  symmetric nuclear systems},}\ }\href {\doibase 10.1007/BF01415623} {\bibfield
   {journal} {\bibinfo  {journal} {Z. Phys. A}\ }\textbf {\bibinfo {volume}
  {319}},\ \bibinfo {pages} {113--118} (\bibinfo {year} {1984})}\BibitemShut
  {NoStop}%
\bibitem [{\citenamefont {G\"aggeler}\ \emph {et~al.}(1984)\citenamefont
  {G\"aggeler}, \citenamefont {Sikkeland}, \citenamefont {Wirth}, \citenamefont
  {Br\"uchle}, \citenamefont {B\"ogl}, \citenamefont {Franz}, \citenamefont
  {Herrmann}, \citenamefont {Kratz}, \citenamefont {Sch\"adel}, \citenamefont
  {S\"ummerer},\ and\ \citenamefont {Weber}}]{gaggeler1984}%
  \BibitemOpen
  \bibfield  {author} {\bibinfo {author} {\bibfnamefont {H.}~\bibnamefont
  {G\"aggeler}}, \bibinfo {author} {\bibfnamefont {T.}~\bibnamefont
  {Sikkeland}}, \bibinfo {author} {\bibfnamefont {G.}~\bibnamefont {Wirth}},
  \bibinfo {author} {\bibfnamefont {W.}~\bibnamefont {Br\"uchle}}, \bibinfo
  {author} {\bibfnamefont {W.}~\bibnamefont {B\"ogl}}, \bibinfo {author}
  {\bibfnamefont {G.}~\bibnamefont {Franz}}, \bibinfo {author} {\bibfnamefont
  {G.}~\bibnamefont {Herrmann}}, \bibinfo {author} {\bibfnamefont {J.~V.}\
  \bibnamefont {Kratz}}, \bibinfo {author} {\bibfnamefont {M.}~\bibnamefont
  {Sch\"adel}}, \bibinfo {author} {\bibfnamefont {K.}~\bibnamefont
  {S\"ummerer}}, \ and\ \bibinfo {author} {\bibfnamefont {W.}~\bibnamefont
  {Weber}},\ }\bibfield  {title} {\enquote {\bibinfo {title} {{P}robing
  sub-barrier fusion and extra-push by measuring fermium evaporation residues
  in different heavy ion reactions},}\ }\href {\doibase 10.1007/BF01439902}
  {\bibfield  {journal} {\bibinfo  {journal} {Z. Phys. A}\ }\textbf {\bibinfo
  {volume} {316}},\ \bibinfo {pages} {291--307} (\bibinfo {year}
  {1984})}\BibitemShut {NoStop}%
\bibitem [{\citenamefont {Schmidt}\ and\ \citenamefont
  {Morawek}(1991)}]{schmidt1991}%
  \BibitemOpen
  \bibfield  {author} {\bibinfo {author} {\bibfnamefont {K.-H.}\ \bibnamefont
  {Schmidt}}\ and\ \bibinfo {author} {\bibfnamefont {W.}~\bibnamefont
  {Morawek}},\ }\bibfield  {title} {\enquote {\bibinfo {title} {{T}he
  conditions for the synthesis of heavy nuclei},}\ }\href {\doibase
  10.1088/0034-4885/54/7/002} {\bibfield  {journal} {\bibinfo  {journal} {Rep.
  Prog. Phys.}\ }\textbf {\bibinfo {volume} {54}},\ \bibinfo {pages} {949}
  (\bibinfo {year} {1991})}\BibitemShut {NoStop}%
\bibitem [{\citenamefont {Back}\ \emph {et~al.}(2014)\citenamefont {Back},
  \citenamefont {Esbensen}, \citenamefont {Jiang},\ and\ \citenamefont
  {Rehm}}]{back2014}%
  \BibitemOpen
  \bibfield  {author} {\bibinfo {author} {\bibfnamefont {B.~B.}\ \bibnamefont
  {Back}}, \bibinfo {author} {\bibfnamefont {H.}~\bibnamefont {Esbensen}},
  \bibinfo {author} {\bibfnamefont {C.~L.}\ \bibnamefont {Jiang}}, \ and\
  \bibinfo {author} {\bibfnamefont {K.~E.}\ \bibnamefont {Rehm}},\ }\bibfield
  {title} {\enquote {\bibinfo {title} {{R}ecent developments in heavy-ion
  fusion reactions},}\ }\href {\doibase 10.1103/RevModPhys.86.317} {\bibfield
  {journal} {\bibinfo  {journal} {Rev. Mod. Phys.}\ }\textbf {\bibinfo {volume}
  {86}},\ \bibinfo {pages} {317--360} (\bibinfo {year} {2014})}\BibitemShut
  {NoStop}%
\bibitem [{\citenamefont {Khuyagbaatar}\ \emph {et~al.}(2018)\citenamefont
  {Khuyagbaatar}, \citenamefont {David}, \citenamefont {Hinde}, \citenamefont
  {Carter}, \citenamefont {Cook}, \citenamefont {Dasgupta}, \citenamefont
  {D\"ullmann}, \citenamefont {Jeung}, \citenamefont {Kindler}, \citenamefont
  {Lommel}, \citenamefont {Luong}, \citenamefont {Prasad}, \citenamefont
  {Rafferty}, \citenamefont {Sengupta}, \citenamefont {Simenel}, \citenamefont
  {Simpson}, \citenamefont {Smith}, \citenamefont {Vo-Phuoc}, \citenamefont
  {Walshe}, \citenamefont {Wakhle}, \citenamefont {Williams},\ and\
  \citenamefont {Yakushev}}]{khuyagbaatar2018}%
  \BibitemOpen
  \bibfield  {author} {\bibinfo {author} {\bibfnamefont {J.}~\bibnamefont
  {Khuyagbaatar}}, \bibinfo {author} {\bibfnamefont {H.~M.}\ \bibnamefont
  {David}}, \bibinfo {author} {\bibfnamefont {D.~J.}\ \bibnamefont {Hinde}},
  \bibinfo {author} {\bibfnamefont {I.~P.}\ \bibnamefont {Carter}}, \bibinfo
  {author} {\bibfnamefont {K.~J.}\ \bibnamefont {Cook}}, \bibinfo {author}
  {\bibfnamefont {M.}~\bibnamefont {Dasgupta}}, \bibinfo {author}
  {\bibfnamefont {{\relax Ch. E}.}~\bibnamefont {D\"ullmann}}, \bibinfo
  {author} {\bibfnamefont {D.~Y.}\ \bibnamefont {Jeung}}, \bibinfo {author}
  {\bibfnamefont {B.}~\bibnamefont {Kindler}}, \bibinfo {author} {\bibfnamefont
  {B.}~\bibnamefont {Lommel}}, \bibinfo {author} {\bibfnamefont {D.~H.}\
  \bibnamefont {Luong}}, \bibinfo {author} {\bibfnamefont {E.}~\bibnamefont
  {Prasad}}, \bibinfo {author} {\bibfnamefont {D.~C.}\ \bibnamefont
  {Rafferty}}, \bibinfo {author} {\bibfnamefont {C.}~\bibnamefont {Sengupta}},
  \bibinfo {author} {\bibfnamefont {C.}~\bibnamefont {Simenel}}, \bibinfo
  {author} {\bibfnamefont {E.~C.}\ \bibnamefont {Simpson}}, \bibinfo {author}
  {\bibfnamefont {J.~F.}\ \bibnamefont {Smith}}, \bibinfo {author}
  {\bibfnamefont {K.}~\bibnamefont {Vo-Phuoc}}, \bibinfo {author}
  {\bibfnamefont {J.}~\bibnamefont {Walshe}}, \bibinfo {author} {\bibfnamefont
  {A.}~\bibnamefont {Wakhle}}, \bibinfo {author} {\bibfnamefont
  {E.}~\bibnamefont {Williams}}, \ and\ \bibinfo {author} {\bibfnamefont
  {A.}~\bibnamefont {Yakushev}},\ }\bibfield  {title} {\enquote {\bibinfo
  {title} {Nuclear structure dependence of fusion hindrance in heavy element
  synthesis},}\ }\href {\doibase 10.1103/PhysRevC.97.064618} {\bibfield
  {journal} {\bibinfo  {journal} {Phys. Rev. C}\ }\textbf {\bibinfo {volume}
  {97}},\ \bibinfo {pages} {064618} (\bibinfo {year} {2018})}\BibitemShut
  {NoStop}%
\bibitem [{\citenamefont {Banerjee}\ \emph {et~al.}(2019)\citenamefont
  {Banerjee}, \citenamefont {Hinde}, \citenamefont {Dasgupta}, \citenamefont
  {Simpson}, \citenamefont {Jeung}, \citenamefont {Simenel}, \citenamefont
  {{Swinton-Bland}}, \citenamefont {Williams}, \citenamefont {Carter},
  \citenamefont {Cook}, \citenamefont {David}, \citenamefont {D\"ullmann},
  \citenamefont {Khuyagbaatar}, \citenamefont {Kindler}, \citenamefont
  {Lommel}, \citenamefont {Prasad}, \citenamefont {Sengupta}, \citenamefont
  {Smith}, \citenamefont {{Vo-Phuoc}}, \citenamefont {Walshe},\ and\
  \citenamefont {Yakushev}}]{banerjee2019}%
  \BibitemOpen
  \bibfield  {author} {\bibinfo {author} {\bibfnamefont {K.}~\bibnamefont
  {Banerjee}}, \bibinfo {author} {\bibfnamefont {D.~J.}\ \bibnamefont {Hinde}},
  \bibinfo {author} {\bibfnamefont {M.}~\bibnamefont {Dasgupta}}, \bibinfo
  {author} {\bibfnamefont {E.~C.}\ \bibnamefont {Simpson}}, \bibinfo {author}
  {\bibfnamefont {D.~Y.}\ \bibnamefont {Jeung}}, \bibinfo {author}
  {\bibfnamefont {C.}~\bibnamefont {Simenel}}, \bibinfo {author} {\bibfnamefont
  {B.~M.~A.}\ \bibnamefont {{Swinton-Bland}}}, \bibinfo {author} {\bibfnamefont
  {E.}~\bibnamefont {Williams}}, \bibinfo {author} {\bibfnamefont {I.~P.}\
  \bibnamefont {Carter}}, \bibinfo {author} {\bibfnamefont {K.~J.}\
  \bibnamefont {Cook}}, \bibinfo {author} {\bibfnamefont {H.~M.}\ \bibnamefont
  {David}}, \bibinfo {author} {\bibfnamefont {{Ch. E.}}\ \bibnamefont
  {D\"ullmann}}, \bibinfo {author} {\bibfnamefont {J.}~\bibnamefont
  {Khuyagbaatar}}, \bibinfo {author} {\bibfnamefont {B.}~\bibnamefont
  {Kindler}}, \bibinfo {author} {\bibfnamefont {B.}~\bibnamefont {Lommel}},
  \bibinfo {author} {\bibfnamefont {E.}~\bibnamefont {Prasad}}, \bibinfo
  {author} {\bibfnamefont {C.}~\bibnamefont {Sengupta}}, \bibinfo {author}
  {\bibfnamefont {J.~F.}\ \bibnamefont {Smith}}, \bibinfo {author}
  {\bibfnamefont {K.}~\bibnamefont {{Vo-Phuoc}}}, \bibinfo {author}
  {\bibfnamefont {J.}~\bibnamefont {Walshe}}, \ and\ \bibinfo {author}
  {\bibfnamefont {A.}~\bibnamefont {Yakushev}},\ }\bibfield  {title} {\enquote
  {\bibinfo {title} {Mechanisms {S}uppressing {S}uperheavy {E}lement {Y}ields
  in {C}old {F}usion {R}eactions},}\ }\href {\doibase
  10.1103/PhysRevLett.122.232503} {\bibfield  {journal} {\bibinfo  {journal}
  {Phys. Rev. Lett.}\ }\textbf {\bibinfo {volume} {122}},\ \bibinfo {pages}
  {232503} (\bibinfo {year} {2019})}\BibitemShut {NoStop}%
\bibitem [{\citenamefont {T{\~{o}}ke}\ \emph {et~al.}(1985)\citenamefont
  {T{\~{o}}ke}, \citenamefont {Bock}, \citenamefont {Dai}, \citenamefont
  {Gobbi}, \citenamefont {Gralla}, \citenamefont {Hildenbrand}, \citenamefont
  {Kuzminski}, \citenamefont {M\"uller}, \citenamefont {Olmi}, \citenamefont
  {Stelzer}, \citenamefont {Back},\ and\ \citenamefont
  {Bj\o{}rnholm}}]{toke1985}%
  \BibitemOpen
  \bibfield  {author} {\bibinfo {author} {\bibfnamefont {J.}~\bibnamefont
  {T{\~{o}}ke}}, \bibinfo {author} {\bibfnamefont {R.}~\bibnamefont {Bock}},
  \bibinfo {author} {\bibfnamefont {G.~X.}\ \bibnamefont {Dai}}, \bibinfo
  {author} {\bibfnamefont {A.}~\bibnamefont {Gobbi}}, \bibinfo {author}
  {\bibfnamefont {S.}~\bibnamefont {Gralla}}, \bibinfo {author} {\bibfnamefont
  {K.~D.}\ \bibnamefont {Hildenbrand}}, \bibinfo {author} {\bibfnamefont
  {J.}~\bibnamefont {Kuzminski}}, \bibinfo {author} {\bibfnamefont {W.~F.~J.}\
  \bibnamefont {M\"uller}}, \bibinfo {author} {\bibfnamefont {A.}~\bibnamefont
  {Olmi}}, \bibinfo {author} {\bibfnamefont {H.}~\bibnamefont {Stelzer}},
  \bibinfo {author} {\bibfnamefont {B.~B.}\ \bibnamefont {Back}}, \ and\
  \bibinfo {author} {\bibfnamefont {S.}~\bibnamefont {Bj\o{}rnholm}},\
  }\bibfield  {title} {\enquote {\bibinfo {title} {{Q}uasi-fission: {T}he
  mass-drift mode in heavy-ion reactions},}\ }\href {\doibase
  10.1016/0375-9474(85)90344-6} {\bibfield  {journal} {\bibinfo  {journal}
  {Nucl. Phys. A}\ }\textbf {\bibinfo {volume} {440}},\ \bibinfo {pages}
  {327--365} (\bibinfo {year} {1985})}\BibitemShut {NoStop}%
\bibitem [{\citenamefont {Shen}\ \emph {et~al.}(1987)\citenamefont {Shen},
  \citenamefont {Albinski}, \citenamefont {Gobbi}, \citenamefont {Gralla},
  \citenamefont {Hildenbrand}, \citenamefont {Herrmann}, \citenamefont
  {Kuzminski}, \citenamefont {M\"uller}, \citenamefont {Stelzer}, \citenamefont
  {T{\~{o}}ke}, \citenamefont {Back}, \citenamefont {Bj\o{}rnholm},\ and\
  \citenamefont {S\o{}rensen}}]{shen1987}%
  \BibitemOpen
  \bibfield  {author} {\bibinfo {author} {\bibfnamefont {W.~Q.}\ \bibnamefont
  {Shen}}, \bibinfo {author} {\bibfnamefont {J.}~\bibnamefont {Albinski}},
  \bibinfo {author} {\bibfnamefont {A.}~\bibnamefont {Gobbi}}, \bibinfo
  {author} {\bibfnamefont {S.}~\bibnamefont {Gralla}}, \bibinfo {author}
  {\bibfnamefont {K.~D.}\ \bibnamefont {Hildenbrand}}, \bibinfo {author}
  {\bibfnamefont {N.}~\bibnamefont {Herrmann}}, \bibinfo {author}
  {\bibfnamefont {J.}~\bibnamefont {Kuzminski}}, \bibinfo {author}
  {\bibfnamefont {W.~F.~J.}\ \bibnamefont {M\"uller}}, \bibinfo {author}
  {\bibfnamefont {H.}~\bibnamefont {Stelzer}}, \bibinfo {author} {\bibfnamefont
  {J.}~\bibnamefont {T{\~{o}}ke}}, \bibinfo {author} {\bibfnamefont {B.~B.}\
  \bibnamefont {Back}}, \bibinfo {author} {\bibfnamefont {S.}~\bibnamefont
  {Bj\o{}rnholm}}, \ and\ \bibinfo {author} {\bibfnamefont {S.~P.}\
  \bibnamefont {S\o{}rensen}},\ }\bibfield  {title} {\enquote {\bibinfo {title}
  {{F}ission and quasifission in {U}-induced reactions},}\ }\href {\doibase
  10.1103/PhysRevC.36.115} {\bibfield  {journal} {\bibinfo  {journal} {Phys.
  Rev. C}\ }\textbf {\bibinfo {volume} {36}},\ \bibinfo {pages} {115--142}
  (\bibinfo {year} {1987})}\BibitemShut {NoStop}%
\bibitem [{\citenamefont {{du Rietz}}\ \emph {et~al.}(2011)\citenamefont {{du
  Rietz}}, \citenamefont {Hinde}, \citenamefont {Dasgupta}, \citenamefont
  {Thomas}, \citenamefont {Gasques}, \citenamefont {Evers}, \citenamefont
  {Lobanov},\ and\ \citenamefont {Wakhle}}]{durietz2011}%
  \BibitemOpen
  \bibfield  {author} {\bibinfo {author} {\bibfnamefont {R.}~\bibnamefont {{du
  Rietz}}}, \bibinfo {author} {\bibfnamefont {D.~J.}\ \bibnamefont {Hinde}},
  \bibinfo {author} {\bibfnamefont {M.}~\bibnamefont {Dasgupta}}, \bibinfo
  {author} {\bibfnamefont {R.~G.}\ \bibnamefont {Thomas}}, \bibinfo {author}
  {\bibfnamefont {L.~R.}\ \bibnamefont {Gasques}}, \bibinfo {author}
  {\bibfnamefont {M.}~\bibnamefont {Evers}}, \bibinfo {author} {\bibfnamefont
  {N.}~\bibnamefont {Lobanov}}, \ and\ \bibinfo {author} {\bibfnamefont
  {A.}~\bibnamefont {Wakhle}},\ }\bibfield  {title} {\enquote {\bibinfo {title}
  {Predominant {T}ime {S}cales in {F}ission {P}rocesses in {R}eactions of {S},
  {T}i and {N}i with {W}: {Z}eptosecond versus {A}ttosecond},}\ }\href
  {\doibase 10.1103/PhysRevLett.106.052701} {\bibfield  {journal} {\bibinfo
  {journal} {Phys. Rev. Lett.}\ }\textbf {\bibinfo {volume} {106}},\ \bibinfo
  {pages} {052701} (\bibinfo {year} {2011})}\BibitemShut {NoStop}%
\bibitem [{\citenamefont {Itkis}\ \emph {et~al.}(2004)\citenamefont {Itkis},
  \citenamefont {\"Ayst\"o}, \citenamefont {Beghini}, \citenamefont {Bogachev},
  \citenamefont {Corradi}, \citenamefont {Dorvaux}, \citenamefont {Gadea},
  \citenamefont {Giardina}, \citenamefont {Hanappe}, \citenamefont {Itkis},
  \citenamefont {Jandel}, \citenamefont {Kliman}, \citenamefont {Khlebnikov},
  \citenamefont {Kniajeva}, \citenamefont {Kondratiev}, \citenamefont
  {Kozulin}, \citenamefont {Krupa}, \citenamefont {Latina}, \citenamefont
  {Materna}, \citenamefont {Montagnoli}, \citenamefont {Oganessian},
  \citenamefont {Pokrovsky}, \citenamefont {Prokhorova}, \citenamefont
  {Rowley}, \citenamefont {Rubchenya}, \citenamefont {Rusanov}, \citenamefont
  {Sagaidak}, \citenamefont {Scarlassara}, \citenamefont {Stefanini},
  \citenamefont {Stuttge}, \citenamefont {Szilner}, \citenamefont {Trotta},
  \citenamefont {Trzaska}, \citenamefont {Vakhtin}, \citenamefont {Vinodkumar},
  \citenamefont {Voskressenski},\ and\ \citenamefont {Zagrebaev}}]{itkis2004}%
  \BibitemOpen
  \bibfield  {author} {\bibinfo {author} {\bibfnamefont {M.~G.}\ \bibnamefont
  {Itkis}}, \bibinfo {author} {\bibfnamefont {J.}~\bibnamefont {\"Ayst\"o}},
  \bibinfo {author} {\bibfnamefont {S.}~\bibnamefont {Beghini}}, \bibinfo
  {author} {\bibfnamefont {A.~A.}\ \bibnamefont {Bogachev}}, \bibinfo {author}
  {\bibfnamefont {L.}~\bibnamefont {Corradi}}, \bibinfo {author} {\bibfnamefont
  {O.}~\bibnamefont {Dorvaux}}, \bibinfo {author} {\bibfnamefont
  {A.}~\bibnamefont {Gadea}}, \bibinfo {author} {\bibfnamefont
  {G.}~\bibnamefont {Giardina}}, \bibinfo {author} {\bibfnamefont
  {F.}~\bibnamefont {Hanappe}}, \bibinfo {author} {\bibfnamefont {I.~M.}\
  \bibnamefont {Itkis}}, \bibinfo {author} {\bibfnamefont {M.}~\bibnamefont
  {Jandel}}, \bibinfo {author} {\bibfnamefont {J.}~\bibnamefont {Kliman}},
  \bibinfo {author} {\bibfnamefont {S.~V.}\ \bibnamefont {Khlebnikov}},
  \bibinfo {author} {\bibfnamefont {G.~N.}\ \bibnamefont {Kniajeva}}, \bibinfo
  {author} {\bibfnamefont {N.~A.}\ \bibnamefont {Kondratiev}}, \bibinfo
  {author} {\bibfnamefont {E.~M.}\ \bibnamefont {Kozulin}}, \bibinfo {author}
  {\bibfnamefont {L.}~\bibnamefont {Krupa}}, \bibinfo {author} {\bibfnamefont
  {A.}~\bibnamefont {Latina}}, \bibinfo {author} {\bibfnamefont
  {T.}~\bibnamefont {Materna}}, \bibinfo {author} {\bibfnamefont
  {G.}~\bibnamefont {Montagnoli}}, \bibinfo {author} {\bibfnamefont {{\relax
  Yu. Ts}.}~\bibnamefont {Oganessian}}, \bibinfo {author} {\bibfnamefont
  {I.~V.}\ \bibnamefont {Pokrovsky}}, \bibinfo {author} {\bibfnamefont {E.~V.}\
  \bibnamefont {Prokhorova}}, \bibinfo {author} {\bibfnamefont
  {N.}~\bibnamefont {Rowley}}, \bibinfo {author} {\bibfnamefont {V.~A.}\
  \bibnamefont {Rubchenya}}, \bibinfo {author} {\bibfnamefont {{\relax A.
  Ya}.}~\bibnamefont {Rusanov}}, \bibinfo {author} {\bibfnamefont {R.~N.}\
  \bibnamefont {Sagaidak}}, \bibinfo {author} {\bibfnamefont {F.}~\bibnamefont
  {Scarlassara}}, \bibinfo {author} {\bibfnamefont {A.~M.}\ \bibnamefont
  {Stefanini}}, \bibinfo {author} {\bibfnamefont {L.}~\bibnamefont {Stuttge}},
  \bibinfo {author} {\bibfnamefont {S.}~\bibnamefont {Szilner}}, \bibinfo
  {author} {\bibfnamefont {M.}~\bibnamefont {Trotta}}, \bibinfo {author}
  {\bibfnamefont {W.~H.}\ \bibnamefont {Trzaska}}, \bibinfo {author}
  {\bibfnamefont {D.~N.}\ \bibnamefont {Vakhtin}}, \bibinfo {author}
  {\bibfnamefont {A.~M.}\ \bibnamefont {Vinodkumar}}, \bibinfo {author}
  {\bibfnamefont {V.~M.}\ \bibnamefont {Voskressenski}}, \ and\ \bibinfo
  {author} {\bibfnamefont {V.~I.}\ \bibnamefont {Zagrebaev}},\ }\bibfield
  {title} {\enquote {\bibinfo {title} {{S}hell effects in fission and
  quasi-fission of heavy and superheavy nuclei},}\ }\href {\doibase
  10.1016/j.nuclphysa.2004.01.022} {\bibfield  {journal} {\bibinfo  {journal}
  {Nucl. Phys. A}\ }\textbf {\bibinfo {volume} {734}},\ \bibinfo {pages}
  {136--147} (\bibinfo {year} {2004})}\BibitemShut {NoStop}%
\bibitem [{\citenamefont {Nishio}\ \emph {et~al.}(2008)\citenamefont {Nishio},
  \citenamefont {Ikezoe}, \citenamefont {Mitsuoka}, \citenamefont {Nishinaka},
  \citenamefont {Nagame}, \citenamefont {Watanabe}, \citenamefont {Ohtsuki},
  \citenamefont {Hirose},\ and\ \citenamefont {Hofmann}}]{nishio2008}%
  \BibitemOpen
  \bibfield  {author} {\bibinfo {author} {\bibfnamefont {K.}~\bibnamefont
  {Nishio}}, \bibinfo {author} {\bibfnamefont {H.}~\bibnamefont {Ikezoe}},
  \bibinfo {author} {\bibfnamefont {S.}~\bibnamefont {Mitsuoka}}, \bibinfo
  {author} {\bibfnamefont {I.}~\bibnamefont {Nishinaka}}, \bibinfo {author}
  {\bibfnamefont {Y.}~\bibnamefont {Nagame}}, \bibinfo {author} {\bibfnamefont
  {Y.}~\bibnamefont {Watanabe}}, \bibinfo {author} {\bibfnamefont
  {T.}~\bibnamefont {Ohtsuki}}, \bibinfo {author} {\bibfnamefont
  {K.}~\bibnamefont {Hirose}}, \ and\ \bibinfo {author} {\bibfnamefont
  {S.}~\bibnamefont {Hofmann}},\ }\bibfield  {title} {\enquote {\bibinfo
  {title} {{E}ffects of nuclear orientation on the mass distribution of fission
  fragments in the reaction of $^{36}\mathrm{S}+{}^{238}\mathrm{U}$},}\ }\href
  {\doibase 10.1103/PhysRevC.77.064607} {\bibfield  {journal} {\bibinfo
  {journal} {Phys. Rev. C}\ }\textbf {\bibinfo {volume} {77}},\ \bibinfo
  {pages} {064607} (\bibinfo {year} {2008})}\BibitemShut {NoStop}%
\bibitem [{\citenamefont {Kozulin}\ \emph {et~al.}(2014)\citenamefont
  {Kozulin}, \citenamefont {Knyazheva}, \citenamefont {Dmitriev}, \citenamefont
  {Itkis}, \citenamefont {Itkis}, \citenamefont {Loktev}, \citenamefont
  {Novikov}, \citenamefont {Baranov}, \citenamefont {Trzaska}, \citenamefont
  {Vardaci}, \citenamefont {Heinz}, \citenamefont {Beliuskina},\ and\
  \citenamefont {Khlebnikov}}]{kozulin2014}%
  \BibitemOpen
  \bibfield  {author} {\bibinfo {author} {\bibfnamefont {E.~M.}\ \bibnamefont
  {Kozulin}}, \bibinfo {author} {\bibfnamefont {G.~N.}\ \bibnamefont
  {Knyazheva}}, \bibinfo {author} {\bibfnamefont {S.~N.}\ \bibnamefont
  {Dmitriev}}, \bibinfo {author} {\bibfnamefont {I.~M.}\ \bibnamefont {Itkis}},
  \bibinfo {author} {\bibfnamefont {M.~G.}\ \bibnamefont {Itkis}}, \bibinfo
  {author} {\bibfnamefont {T.~A.}\ \bibnamefont {Loktev}}, \bibinfo {author}
  {\bibfnamefont {K.~V.}\ \bibnamefont {Novikov}}, \bibinfo {author}
  {\bibfnamefont {A.~N.}\ \bibnamefont {Baranov}}, \bibinfo {author}
  {\bibfnamefont {W.~H.}\ \bibnamefont {Trzaska}}, \bibinfo {author}
  {\bibfnamefont {E.}~\bibnamefont {Vardaci}}, \bibinfo {author} {\bibfnamefont
  {S.}~\bibnamefont {Heinz}}, \bibinfo {author} {\bibfnamefont
  {O.}~\bibnamefont {Beliuskina}}, \ and\ \bibinfo {author} {\bibfnamefont
  {S.~V.}\ \bibnamefont {Khlebnikov}},\ }\bibfield  {title} {\enquote {\bibinfo
  {title} {{S}hell effects in damped collisions of $^{88}\mathrm{Sr}$ with
  $^{176}\mathrm{Yb}$ at the {C}oulomb barrier energy},}\ }\href {\doibase
  10.1103/PhysRevC.89.014614} {\bibfield  {journal} {\bibinfo  {journal} {Phys.
  Rev. C}\ }\textbf {\bibinfo {volume} {89}},\ \bibinfo {pages} {014614}
  (\bibinfo {year} {2014})}\BibitemShut {NoStop}%
\bibitem [{\citenamefont {Wakhle}\ \emph {et~al.}(2014)\citenamefont {Wakhle},
  \citenamefont {Simenel}, \citenamefont {Hinde}, \citenamefont {Dasgupta},
  \citenamefont {Evers}, \citenamefont {Luong}, \citenamefont {du~Rietz},\ and\
  \citenamefont {Williams}}]{wakhle2014}%
  \BibitemOpen
  \bibfield  {author} {\bibinfo {author} {\bibfnamefont {A.}~\bibnamefont
  {Wakhle}}, \bibinfo {author} {\bibfnamefont {C.}~\bibnamefont {Simenel}},
  \bibinfo {author} {\bibfnamefont {D.~J.}\ \bibnamefont {Hinde}}, \bibinfo
  {author} {\bibfnamefont {M.}~\bibnamefont {Dasgupta}}, \bibinfo {author}
  {\bibfnamefont {M.}~\bibnamefont {Evers}}, \bibinfo {author} {\bibfnamefont
  {D.~H.}\ \bibnamefont {Luong}}, \bibinfo {author} {\bibfnamefont
  {R.}~\bibnamefont {du~Rietz}}, \ and\ \bibinfo {author} {\bibfnamefont
  {E.}~\bibnamefont {Williams}},\ }\bibfield  {title} {\enquote {\bibinfo
  {title} {{I}nterplay between {Q}uantum {S}hells and {O}rientation in
  {Q}uasifission},}\ }\href {\doibase 10.1103/PhysRevLett.113.182502}
  {\bibfield  {journal} {\bibinfo  {journal} {Phys. Rev. Lett.}\ }\textbf
  {\bibinfo {volume} {113}},\ \bibinfo {pages} {182502} (\bibinfo {year}
  {2014})}\BibitemShut {NoStop}%
\bibitem [{\citenamefont {Morjean}\ \emph {et~al.}(2017)\citenamefont
  {Morjean}, \citenamefont {Hinde}, \citenamefont {Simenel}, \citenamefont
  {Jeung}, \citenamefont {Airiau}, \citenamefont {Cook}, \citenamefont
  {Dasgupta}, \citenamefont {Drouart}, \citenamefont {Jacquet}, \citenamefont
  {Kalkal}, \citenamefont {Palshetkar}, \citenamefont {Prasad}, \citenamefont
  {Rafferty}, \citenamefont {Simpson}, \citenamefont {Tassan-Got},
  \citenamefont {Vo-Phuoc},\ and\ \citenamefont {Williams}}]{morjean2017}%
  \BibitemOpen
  \bibfield  {author} {\bibinfo {author} {\bibfnamefont {M.}~\bibnamefont
  {Morjean}}, \bibinfo {author} {\bibfnamefont {D.~J.}\ \bibnamefont {Hinde}},
  \bibinfo {author} {\bibfnamefont {C.}~\bibnamefont {Simenel}}, \bibinfo
  {author} {\bibfnamefont {D.~Y.}\ \bibnamefont {Jeung}}, \bibinfo {author}
  {\bibfnamefont {M.}~\bibnamefont {Airiau}}, \bibinfo {author} {\bibfnamefont
  {K.~J.}\ \bibnamefont {Cook}}, \bibinfo {author} {\bibfnamefont
  {M.}~\bibnamefont {Dasgupta}}, \bibinfo {author} {\bibfnamefont
  {A.}~\bibnamefont {Drouart}}, \bibinfo {author} {\bibfnamefont
  {D.}~\bibnamefont {Jacquet}}, \bibinfo {author} {\bibfnamefont
  {S.}~\bibnamefont {Kalkal}}, \bibinfo {author} {\bibfnamefont {C.~S.}\
  \bibnamefont {Palshetkar}}, \bibinfo {author} {\bibfnamefont
  {E.}~\bibnamefont {Prasad}}, \bibinfo {author} {\bibfnamefont
  {D.}~\bibnamefont {Rafferty}}, \bibinfo {author} {\bibfnamefont {E.~C.}\
  \bibnamefont {Simpson}}, \bibinfo {author} {\bibfnamefont {L.}~\bibnamefont
  {Tassan-Got}}, \bibinfo {author} {\bibfnamefont {K.}~\bibnamefont
  {Vo-Phuoc}}, \ and\ \bibinfo {author} {\bibfnamefont {E.}~\bibnamefont
  {Williams}},\ }\bibfield  {title} {\enquote {\bibinfo {title} {Evidence for
  the {R}ole of {P}roton {S}hell {C}losure in {Q}uasifission {R}eactions from
  {X--Ray} {F}luorescence of {M}ass--{I}dentified {F}ragments},}\ }\href
  {\doibase 10.1103/PhysRevLett.119.222502} {\bibfield  {journal} {\bibinfo
  {journal} {Phys. Rev. Lett.}\ }\textbf {\bibinfo {volume} {119}},\ \bibinfo
  {pages} {222502} (\bibinfo {year} {2017})}\BibitemShut {NoStop}%
\bibitem [{\citenamefont {Veselsky}\ \emph {et~al.}(2016)\citenamefont
  {Veselsky}, \citenamefont {Klimo}, \citenamefont {Ma},\ and\ \citenamefont
  {Souliotis}}]{veselsky2016}%
  \BibitemOpen
  \bibfield  {author} {\bibinfo {author} {\bibfnamefont {M.}~\bibnamefont
  {Veselsky}}, \bibinfo {author} {\bibfnamefont {J.}~\bibnamefont {Klimo}},
  \bibinfo {author} {\bibfnamefont {Yu-Gang}\ \bibnamefont {Ma}}, \ and\
  \bibinfo {author} {\bibfnamefont {G.~A.}\ \bibnamefont {Souliotis}},\
  }\bibfield  {title} {\enquote {\bibinfo {title} {Constraining the equation of
  state of nuclear matter from fusion hindrance in reactions leading to the
  production of superheavy elements},}\ }\href {\doibase
  10.1103/PhysRevC.94.064608} {\bibfield  {journal} {\bibinfo  {journal} {Phys.
  Rev. C}\ }\textbf {\bibinfo {volume} {94}},\ \bibinfo {pages} {064608}
  (\bibinfo {year} {2016})}\BibitemShut {NoStop}%
\bibitem [{\citenamefont {Zheng}\ \emph {et~al.}(2018)\citenamefont {Zheng},
  \citenamefont {Burrello}, \citenamefont {Colonna}, \citenamefont {Lacroix},\
  and\ \citenamefont {Scamps}}]{zheng2018}%
  \BibitemOpen
  \bibfield  {author} {\bibinfo {author} {\bibfnamefont {H.}~\bibnamefont
  {Zheng}}, \bibinfo {author} {\bibfnamefont {S.}~\bibnamefont {Burrello}},
  \bibinfo {author} {\bibfnamefont {M.}~\bibnamefont {Colonna}}, \bibinfo
  {author} {\bibfnamefont {D.}~\bibnamefont {Lacroix}}, \ and\ \bibinfo
  {author} {\bibfnamefont {G.}~\bibnamefont {Scamps}},\ }\bibfield  {title}
  {\enquote {\bibinfo {title} {Connecting the nuclear equation of state to the
  interplay between fusion and quasifission processes in low-energy nuclear
  reactions},}\ }\href {\doibase 10.1103/PhysRevC.98.024622} {\bibfield
  {journal} {\bibinfo  {journal} {Phys. Rev. C}\ }\textbf {\bibinfo {volume}
  {98}},\ \bibinfo {pages} {024622} (\bibinfo {year} {2018})}\BibitemShut
  {NoStop}%
\bibitem [{\citenamefont {Back}\ \emph {et~al.}(1996)\citenamefont {Back},
  \citenamefont {Fernandez}, \citenamefont {Glagola}, \citenamefont
  {Henderson}, \citenamefont {Kaufman}, \citenamefont {Keller}, \citenamefont
  {Sanders}, \citenamefont {Videb{\ae}k}, \citenamefont {Wang},\ and\
  \citenamefont {Wilkins}}]{back1996}%
  \BibitemOpen
  \bibfield  {author} {\bibinfo {author} {\bibfnamefont {B.~B.}\ \bibnamefont
  {Back}}, \bibinfo {author} {\bibfnamefont {P.~B.}\ \bibnamefont {Fernandez}},
  \bibinfo {author} {\bibfnamefont {B.~G.}\ \bibnamefont {Glagola}}, \bibinfo
  {author} {\bibfnamefont {D.}~\bibnamefont {Henderson}}, \bibinfo {author}
  {\bibfnamefont {S.}~\bibnamefont {Kaufman}}, \bibinfo {author} {\bibfnamefont
  {J.~G.}\ \bibnamefont {Keller}}, \bibinfo {author} {\bibfnamefont {S.~J.}\
  \bibnamefont {Sanders}}, \bibinfo {author} {\bibfnamefont {F.}~\bibnamefont
  {Videb{\ae}k}}, \bibinfo {author} {\bibfnamefont {T.~F.}\ \bibnamefont
  {Wang}}, \ and\ \bibinfo {author} {\bibfnamefont {B.~D.}\ \bibnamefont
  {Wilkins}},\ }\bibfield  {title} {\enquote {\bibinfo {title}
  {{E}ntrance-channel effects in quasifission reactions},}\ }\href {\doibase
  10.1103/physrevc.53.1734} {\bibfield  {journal} {\bibinfo  {journal} {Phys.
  Rev. C}\ }\textbf {\bibinfo {volume} {53}},\ \bibinfo {pages} {1734--1744}
  (\bibinfo {year} {1996})}\BibitemShut {NoStop}%
\bibitem [{\citenamefont {Nishio}\ \emph {et~al.}(2012)\citenamefont {Nishio},
  \citenamefont {Mitsuoka}, \citenamefont {Nishinaka}, \citenamefont {Makii},
  \citenamefont {Wakabayashi}, \citenamefont {Ikezoe}, \citenamefont {Hirose},
  \citenamefont {Ohtsuki}, \citenamefont {Aritomo},\ and\ \citenamefont
  {Hofmann}}]{nishio2012}%
  \BibitemOpen
  \bibfield  {author} {\bibinfo {author} {\bibfnamefont {K.}~\bibnamefont
  {Nishio}}, \bibinfo {author} {\bibfnamefont {S.}~\bibnamefont {Mitsuoka}},
  \bibinfo {author} {\bibfnamefont {I.}~\bibnamefont {Nishinaka}}, \bibinfo
  {author} {\bibfnamefont {H.}~\bibnamefont {Makii}}, \bibinfo {author}
  {\bibfnamefont {Y.}~\bibnamefont {Wakabayashi}}, \bibinfo {author}
  {\bibfnamefont {H.}~\bibnamefont {Ikezoe}}, \bibinfo {author} {\bibfnamefont
  {K.}~\bibnamefont {Hirose}}, \bibinfo {author} {\bibfnamefont
  {T.}~\bibnamefont {Ohtsuki}}, \bibinfo {author} {\bibfnamefont
  {Y.}~\bibnamefont {Aritomo}}, \ and\ \bibinfo {author} {\bibfnamefont
  {S.}~\bibnamefont {Hofmann}},\ }\bibfield  {title} {\enquote {\bibinfo
  {title} {{F}usion probabilities in the reactions $^{40,48}${C}a + $^{238}${U}
  at energies around the {C}oulomb barrier},}\ }\href {\doibase
  10.1103/PhysRevC.86.034608} {\bibfield  {journal} {\bibinfo  {journal} {Phys.
  Rev. C}\ }\textbf {\bibinfo {volume} {86}},\ \bibinfo {pages} {034608}
  (\bibinfo {year} {2012})}\BibitemShut {NoStop}%
\bibitem [{\citenamefont {Williams}\ \emph {et~al.}(2018)\citenamefont
  {Williams}, \citenamefont {Sekizawa}, \citenamefont {Hinde}, \citenamefont
  {Simenel}, \citenamefont {Dasgupta}, \citenamefont {Carter}, \citenamefont
  {Cook}, \citenamefont {Jeung}, \citenamefont {McNeil}, \citenamefont
  {Palshetkar}, \citenamefont {Rafferty}, \citenamefont {Ramachandran},\ and\
  \citenamefont {Wakhle}}]{williams2018}%
  \BibitemOpen
  \bibfield  {author} {\bibinfo {author} {\bibfnamefont {E.}~\bibnamefont
  {Williams}}, \bibinfo {author} {\bibfnamefont {K.}~\bibnamefont {Sekizawa}},
  \bibinfo {author} {\bibfnamefont {D.~J.}\ \bibnamefont {Hinde}}, \bibinfo
  {author} {\bibfnamefont {C.}~\bibnamefont {Simenel}}, \bibinfo {author}
  {\bibfnamefont {M.}~\bibnamefont {Dasgupta}}, \bibinfo {author}
  {\bibfnamefont {I.~P.}\ \bibnamefont {Carter}}, \bibinfo {author}
  {\bibfnamefont {K.~J.}\ \bibnamefont {Cook}}, \bibinfo {author}
  {\bibfnamefont {D.~Y.}\ \bibnamefont {Jeung}}, \bibinfo {author}
  {\bibfnamefont {S.~D.}\ \bibnamefont {McNeil}}, \bibinfo {author}
  {\bibfnamefont {C.~S.}\ \bibnamefont {Palshetkar}}, \bibinfo {author}
  {\bibfnamefont {D.~C.}\ \bibnamefont {Rafferty}}, \bibinfo {author}
  {\bibfnamefont {K.}~\bibnamefont {Ramachandran}}, \ and\ \bibinfo {author}
  {\bibfnamefont {A.}~\bibnamefont {Wakhle}},\ }\bibfield  {title} {\enquote
  {\bibinfo {title} {Exploring {Z}eptosecond {Q}uantum {E}quilibration
  {D}ynamics: {F}rom {D}eep-{I}nelastic to {F}usion-{F}ission {O}utcomes in
  $^{58}\mathrm{Ni}+{}^{60}\mathrm{Ni}$ {R}eactions},}\ }\href {\doibase
  10.1103/PhysRevLett.120.022501} {\bibfield  {journal} {\bibinfo  {journal}
  {Phys. Rev. Lett.}\ }\textbf {\bibinfo {volume} {120}},\ \bibinfo {pages}
  {022501} (\bibinfo {year} {2018})}\BibitemShut {NoStop}%
\bibitem [{\citenamefont {Lin}\ \emph {et~al.}(2012)\citenamefont {Lin},
  \citenamefont {du~Rietz}, \citenamefont {Hinde}, \citenamefont {Dasgupta},
  \citenamefont {Thomas}, \citenamefont {Brown}, \citenamefont {Evers},
  \citenamefont {Gasques},\ and\ \citenamefont {Rodriguez}}]{lin2012}%
  \BibitemOpen
  \bibfield  {author} {\bibinfo {author} {\bibfnamefont {C.~J.}\ \bibnamefont
  {Lin}}, \bibinfo {author} {\bibfnamefont {R.}~\bibnamefont {du~Rietz}},
  \bibinfo {author} {\bibfnamefont {D.~J.}\ \bibnamefont {Hinde}}, \bibinfo
  {author} {\bibfnamefont {M.}~\bibnamefont {Dasgupta}}, \bibinfo {author}
  {\bibfnamefont {R.~G.}\ \bibnamefont {Thomas}}, \bibinfo {author}
  {\bibfnamefont {M.~L.}\ \bibnamefont {Brown}}, \bibinfo {author}
  {\bibfnamefont {M.}~\bibnamefont {Evers}}, \bibinfo {author} {\bibfnamefont
  {L.~R.}\ \bibnamefont {Gasques}}, \ and\ \bibinfo {author} {\bibfnamefont
  {M.~D.}\ \bibnamefont {Rodriguez}},\ }\bibfield  {title} {\enquote {\bibinfo
  {title} {{S}ystematic behavior of mass distributions in $^{48}${T}i-induced
  fission at near-barrier energies},}\ }\href {\doibase
  10.1103/PhysRevC.85.014611} {\bibfield  {journal} {\bibinfo  {journal} {Phys.
  Rev. C}\ }\textbf {\bibinfo {volume} {85}},\ \bibinfo {pages} {014611}
  (\bibinfo {year} {2012})}\BibitemShut {NoStop}%
\bibitem [{\citenamefont {du~Rietz}\ \emph {et~al.}(2013)\citenamefont
  {du~Rietz}, \citenamefont {Williams}, \citenamefont {Hinde}, \citenamefont
  {Dasgupta}, \citenamefont {Evers}, \citenamefont {Lin}, \citenamefont
  {Luong}, \citenamefont {Simenel},\ and\ \citenamefont
  {Wakhle}}]{durietz2013}%
  \BibitemOpen
  \bibfield  {author} {\bibinfo {author} {\bibfnamefont {R.}~\bibnamefont
  {du~Rietz}}, \bibinfo {author} {\bibfnamefont {E.}~\bibnamefont {Williams}},
  \bibinfo {author} {\bibfnamefont {D.~J.}\ \bibnamefont {Hinde}}, \bibinfo
  {author} {\bibfnamefont {M.}~\bibnamefont {Dasgupta}}, \bibinfo {author}
  {\bibfnamefont {M.}~\bibnamefont {Evers}}, \bibinfo {author} {\bibfnamefont
  {C.~J.}\ \bibnamefont {Lin}}, \bibinfo {author} {\bibfnamefont {D.~H.}\
  \bibnamefont {Luong}}, \bibinfo {author} {\bibfnamefont {C.}~\bibnamefont
  {Simenel}}, \ and\ \bibinfo {author} {\bibfnamefont {A.}~\bibnamefont
  {Wakhle}},\ }\bibfield  {title} {\enquote {\bibinfo {title} {Mapping
  quasifission characteristics and timescales in heavy element formation
  reactions},}\ }\href {\doibase 10.1103/PhysRevC.88.054618} {\bibfield
  {journal} {\bibinfo  {journal} {Phys. Rev. C}\ }\textbf {\bibinfo {volume}
  {88}},\ \bibinfo {pages} {054618} (\bibinfo {year} {2013})}\BibitemShut
  {NoStop}%
\bibitem [{\citenamefont {Hinde}\ \emph {et~al.}(1995)\citenamefont {Hinde},
  \citenamefont {Dasgupta}, \citenamefont {Leigh}, \citenamefont {Lestone},
  \citenamefont {Mein}, \citenamefont {Morton}, \citenamefont {Newton},\ and\
  \citenamefont {Timmers}}]{hinde1995}%
  \BibitemOpen
  \bibfield  {author} {\bibinfo {author} {\bibfnamefont {D.~J.}\ \bibnamefont
  {Hinde}}, \bibinfo {author} {\bibfnamefont {M.}~\bibnamefont {Dasgupta}},
  \bibinfo {author} {\bibfnamefont {J.~R.}\ \bibnamefont {Leigh}}, \bibinfo
  {author} {\bibfnamefont {J.~P.}\ \bibnamefont {Lestone}}, \bibinfo {author}
  {\bibfnamefont {J.~C.}\ \bibnamefont {Mein}}, \bibinfo {author}
  {\bibfnamefont {C.~R.}\ \bibnamefont {Morton}}, \bibinfo {author}
  {\bibfnamefont {J.~O.}\ \bibnamefont {Newton}}, \ and\ \bibinfo {author}
  {\bibfnamefont {H.}~\bibnamefont {Timmers}},\ }\bibfield  {title} {\enquote
  {\bibinfo {title} {{F}usion-{F}ission versus {Q}uasifission: {E}ffect of
  {N}uclear {O}rientation},}\ }\href {\doibase 10.1103/PhysRevLett.74.1295}
  {\bibfield  {journal} {\bibinfo  {journal} {Phys. Rev. Lett.}\ }\textbf
  {\bibinfo {volume} {74}},\ \bibinfo {pages} {1295--1298} (\bibinfo {year}
  {1995})}\BibitemShut {NoStop}%
\bibitem [{\citenamefont {Hinde}\ \emph {et~al.}(1996)\citenamefont {Hinde},
  \citenamefont {Dasgupta}, \citenamefont {Leigh}, \citenamefont {Mein},
  \citenamefont {Morton}, \citenamefont {Newton},\ and\ \citenamefont
  {Timmers}}]{hinde1996}%
  \BibitemOpen
  \bibfield  {author} {\bibinfo {author} {\bibfnamefont {D.~J.}\ \bibnamefont
  {Hinde}}, \bibinfo {author} {\bibfnamefont {M.}~\bibnamefont {Dasgupta}},
  \bibinfo {author} {\bibfnamefont {J.~R.}\ \bibnamefont {Leigh}}, \bibinfo
  {author} {\bibfnamefont {J.~C.}\ \bibnamefont {Mein}}, \bibinfo {author}
  {\bibfnamefont {C.~R.}\ \bibnamefont {Morton}}, \bibinfo {author}
  {\bibfnamefont {J.~O.}\ \bibnamefont {Newton}}, \ and\ \bibinfo {author}
  {\bibfnamefont {H.}~\bibnamefont {Timmers}},\ }\bibfield  {title} {\enquote
  {\bibinfo {title} {{C}onclusive evidence for the influence of nuclear
  orientation on quasifission},}\ }\href {\doibase 10.1103/PhysRevC.53.1290}
  {\bibfield  {journal} {\bibinfo  {journal} {Phys. Rev. C}\ }\textbf {\bibinfo
  {volume} {53}},\ \bibinfo {pages} {1290--1300} (\bibinfo {year}
  {1996})}\BibitemShut {NoStop}%
\bibitem [{\citenamefont {Knyazheva}\ \emph {et~al.}(2007)\citenamefont
  {Knyazheva}, \citenamefont {Kozulin}, \citenamefont {Sagaidak}, \citenamefont
  {{A. Yu. Chizhov}}, \citenamefont {Itkis}, \citenamefont {Kondratiev},
  \citenamefont {Voskressensky}, \citenamefont {Stefanini}, \citenamefont
  {Behera}, \citenamefont {Corradi}, \citenamefont {Fioretto}, \citenamefont
  {Gadea}, \citenamefont {Latina}, \citenamefont {Szilner}, \citenamefont
  {Trotta}, \citenamefont {Beghini}, \citenamefont {Montagnoli}, \citenamefont
  {Scarlassara}, \citenamefont {Haas}, \citenamefont {Rowley}, \citenamefont
  {Gomes},\ and\ \citenamefont {{Szanto de Toledo}}}]{knyazheva2007}%
  \BibitemOpen
  \bibfield  {author} {\bibinfo {author} {\bibfnamefont {G.~N.}\ \bibnamefont
  {Knyazheva}}, \bibinfo {author} {\bibfnamefont {E.~M.}\ \bibnamefont
  {Kozulin}}, \bibinfo {author} {\bibfnamefont {R.~N.}\ \bibnamefont
  {Sagaidak}}, \bibinfo {author} {\bibnamefont {{A. Yu. Chizhov}}}, \bibinfo
  {author} {\bibfnamefont {M.~G.}\ \bibnamefont {Itkis}}, \bibinfo {author}
  {\bibfnamefont {N.~A.}\ \bibnamefont {Kondratiev}}, \bibinfo {author}
  {\bibfnamefont {V.~M.}\ \bibnamefont {Voskressensky}}, \bibinfo {author}
  {\bibfnamefont {A.~M.}\ \bibnamefont {Stefanini}}, \bibinfo {author}
  {\bibfnamefont {B.~R.}\ \bibnamefont {Behera}}, \bibinfo {author}
  {\bibfnamefont {L.}~\bibnamefont {Corradi}}, \bibinfo {author} {\bibfnamefont
  {E.}~\bibnamefont {Fioretto}}, \bibinfo {author} {\bibfnamefont
  {A.}~\bibnamefont {Gadea}}, \bibinfo {author} {\bibfnamefont
  {A.}~\bibnamefont {Latina}}, \bibinfo {author} {\bibfnamefont
  {S.}~\bibnamefont {Szilner}}, \bibinfo {author} {\bibfnamefont
  {M.}~\bibnamefont {Trotta}}, \bibinfo {author} {\bibfnamefont
  {S.}~\bibnamefont {Beghini}}, \bibinfo {author} {\bibfnamefont
  {G.}~\bibnamefont {Montagnoli}}, \bibinfo {author} {\bibfnamefont
  {F.}~\bibnamefont {Scarlassara}}, \bibinfo {author} {\bibfnamefont
  {F.}~\bibnamefont {Haas}}, \bibinfo {author} {\bibfnamefont {N.}~\bibnamefont
  {Rowley}}, \bibinfo {author} {\bibfnamefont {P.~R.~S.}\ \bibnamefont
  {Gomes}}, \ and\ \bibinfo {author} {\bibfnamefont {A.}~\bibnamefont {{Szanto
  de Toledo}}},\ }\bibfield  {title} {\enquote {\bibinfo {title}
  {{Q}uasifission processes in $^{40,48}\mathrm{Ca}+{}^{144,154}\mathrm{Sm}$
  reactions},}\ }\href {\doibase 10.1103/PhysRevC.75.064602} {\bibfield
  {journal} {\bibinfo  {journal} {Phys. Rev. C}\ }\textbf {\bibinfo {volume}
  {75}},\ \bibinfo {pages} {064602} (\bibinfo {year} {2007})}\BibitemShut
  {NoStop}%
\bibitem [{\citenamefont {Hinde}\ \emph {et~al.}(2008)\citenamefont {Hinde},
  \citenamefont {Thomas}, \citenamefont {du~Rietz}, \citenamefont
  {Diaz-Torres}, \citenamefont {Dasgupta}, \citenamefont {Brown}, \citenamefont
  {Evers}, \citenamefont {Gasques}, \citenamefont {Rafiei},\ and\ \citenamefont
  {Rodriguez}}]{hinde2008}%
  \BibitemOpen
  \bibfield  {author} {\bibinfo {author} {\bibfnamefont {D.~J.}\ \bibnamefont
  {Hinde}}, \bibinfo {author} {\bibfnamefont {R.~G.}\ \bibnamefont {Thomas}},
  \bibinfo {author} {\bibfnamefont {R.}~\bibnamefont {du~Rietz}}, \bibinfo
  {author} {\bibfnamefont {A.}~\bibnamefont {Diaz-Torres}}, \bibinfo {author}
  {\bibfnamefont {M.}~\bibnamefont {Dasgupta}}, \bibinfo {author}
  {\bibfnamefont {M.~L.}\ \bibnamefont {Brown}}, \bibinfo {author}
  {\bibfnamefont {M.}~\bibnamefont {Evers}}, \bibinfo {author} {\bibfnamefont
  {L.~R.}\ \bibnamefont {Gasques}}, \bibinfo {author} {\bibfnamefont
  {R.}~\bibnamefont {Rafiei}}, \ and\ \bibinfo {author} {\bibfnamefont {M.~D.}\
  \bibnamefont {Rodriguez}},\ }\bibfield  {title} {\enquote {\bibinfo {title}
  {{D}isentangling {E}ffects of {N}uclear {S}tructure in {H}eavy {E}lement
  {F}ormation},}\ }\href {\doibase 10.1103/PhysRevLett.100.202701} {\bibfield
  {journal} {\bibinfo  {journal} {Phys. Rev. Lett.}\ }\textbf {\bibinfo
  {volume} {100}},\ \bibinfo {pages} {202701} (\bibinfo {year}
  {2008})}\BibitemShut {NoStop}%
\bibitem [{\citenamefont {Simenel}\ \emph {et~al.}(2012)\citenamefont
  {Simenel}, \citenamefont {Hinde}, \citenamefont {{du Rietz}}, \citenamefont
  {Dasgupta}, \citenamefont {Evers}, \citenamefont {Lin}, \citenamefont
  {Luong},\ and\ \citenamefont {Wakhle}}]{simenel2012b}%
  \BibitemOpen
  \bibfield  {author} {\bibinfo {author} {\bibfnamefont {C.}~\bibnamefont
  {Simenel}}, \bibinfo {author} {\bibfnamefont {D.~J.}\ \bibnamefont {Hinde}},
  \bibinfo {author} {\bibfnamefont {R.}~\bibnamefont {{du Rietz}}}, \bibinfo
  {author} {\bibfnamefont {M.}~\bibnamefont {Dasgupta}}, \bibinfo {author}
  {\bibfnamefont {M.}~\bibnamefont {Evers}}, \bibinfo {author} {\bibfnamefont
  {C.~J.}\ \bibnamefont {Lin}}, \bibinfo {author} {\bibfnamefont {D.~H.}\
  \bibnamefont {Luong}}, \ and\ \bibinfo {author} {\bibfnamefont
  {A.}~\bibnamefont {Wakhle}},\ }\bibfield  {title} {\enquote {\bibinfo {title}
  {{I}nfluence of entrance-channel magicity and isospin on quasi-fission},}\
  }\href {\doibase 10.1016/j.physletb.2012.03.063} {\bibfield  {journal}
  {\bibinfo  {journal} {Phys. Lett. B}\ }\textbf {\bibinfo {volume} {710}},\
  \bibinfo {pages} {607--611} (\bibinfo {year} {2012})}\BibitemShut {NoStop}%
\bibitem [{\citenamefont {Mohanto}\ \emph {et~al.}(2018)\citenamefont
  {Mohanto}, \citenamefont {Hinde}, \citenamefont {Banerjee}, \citenamefont
  {Dasgupta}, \citenamefont {Jeung}, \citenamefont {Simenel}, \citenamefont
  {Simpson}, \citenamefont {Wakhle}, \citenamefont {Williams}, \citenamefont
  {Carter}, \citenamefont {Cook}, \citenamefont {Luong}, \citenamefont
  {Palshetkar},\ and\ \citenamefont {Rafferty}}]{mohanto2018}%
  \BibitemOpen
  \bibfield  {author} {\bibinfo {author} {\bibfnamefont {G.}~\bibnamefont
  {Mohanto}}, \bibinfo {author} {\bibfnamefont {D.~J.}\ \bibnamefont {Hinde}},
  \bibinfo {author} {\bibfnamefont {K.}~\bibnamefont {Banerjee}}, \bibinfo
  {author} {\bibfnamefont {M.}~\bibnamefont {Dasgupta}}, \bibinfo {author}
  {\bibfnamefont {D.~Y.}\ \bibnamefont {Jeung}}, \bibinfo {author}
  {\bibfnamefont {C.}~\bibnamefont {Simenel}}, \bibinfo {author} {\bibfnamefont
  {E.~C.}\ \bibnamefont {Simpson}}, \bibinfo {author} {\bibfnamefont
  {A.}~\bibnamefont {Wakhle}}, \bibinfo {author} {\bibfnamefont
  {E.}~\bibnamefont {Williams}}, \bibinfo {author} {\bibfnamefont {I.~P.}\
  \bibnamefont {Carter}}, \bibinfo {author} {\bibfnamefont {K.~J.}\
  \bibnamefont {Cook}}, \bibinfo {author} {\bibfnamefont {D.~H.}\ \bibnamefont
  {Luong}}, \bibinfo {author} {\bibfnamefont {C.~S.}\ \bibnamefont
  {Palshetkar}}, \ and\ \bibinfo {author} {\bibfnamefont {D.~C.}\ \bibnamefont
  {Rafferty}},\ }\bibfield  {title} {\enquote {\bibinfo {title} {Interplay of
  spherical closed shells and {$N/Z$} asymmetry in quasifission dynamics},}\
  }\href {\doibase 10.1103/PhysRevC.97.054603} {\bibfield  {journal} {\bibinfo
  {journal} {Phys. Rev. C}\ }\textbf {\bibinfo {volume} {97}},\ \bibinfo
  {pages} {054603} (\bibinfo {year} {2018})}\BibitemShut {NoStop}%
\bibitem [{\citenamefont {Hammerton}\ \emph {et~al.}(2015)\citenamefont
  {Hammerton}, \citenamefont {Kohley}, \citenamefont {Hinde}, \citenamefont
  {Dasgupta}, \citenamefont {Wakhle}, \citenamefont {Williams}, \citenamefont
  {Oberacker}, \citenamefont {Umar}, \citenamefont {Carter}, \citenamefont
  {Cook}, \citenamefont {Greene}, \citenamefont {Jeung}, \citenamefont {Luong},
  \citenamefont {{McNeil}}, \citenamefont {Palshetkar}, \citenamefont
  {Rafferty}, \citenamefont {Simenel},\ and\ \citenamefont
  {Stiefel}}]{hammerton2015}%
  \BibitemOpen
  \bibfield  {author} {\bibinfo {author} {\bibfnamefont {K.}~\bibnamefont
  {Hammerton}}, \bibinfo {author} {\bibfnamefont {Z.}~\bibnamefont {Kohley}},
  \bibinfo {author} {\bibfnamefont {D.~J.}\ \bibnamefont {Hinde}}, \bibinfo
  {author} {\bibfnamefont {M.}~\bibnamefont {Dasgupta}}, \bibinfo {author}
  {\bibfnamefont {A.}~\bibnamefont {Wakhle}}, \bibinfo {author} {\bibfnamefont
  {E.}~\bibnamefont {Williams}}, \bibinfo {author} {\bibfnamefont {V.~E.}\
  \bibnamefont {Oberacker}}, \bibinfo {author} {\bibfnamefont {A.~S.}\
  \bibnamefont {Umar}}, \bibinfo {author} {\bibfnamefont {I.~P.}\ \bibnamefont
  {Carter}}, \bibinfo {author} {\bibfnamefont {K.~J.}\ \bibnamefont {Cook}},
  \bibinfo {author} {\bibfnamefont {J.}~\bibnamefont {Greene}}, \bibinfo
  {author} {\bibfnamefont {D.~Y.}\ \bibnamefont {Jeung}}, \bibinfo {author}
  {\bibfnamefont {D.~H.}\ \bibnamefont {Luong}}, \bibinfo {author}
  {\bibfnamefont {S.~D.}\ \bibnamefont {{McNeil}}}, \bibinfo {author}
  {\bibfnamefont {C.~S.}\ \bibnamefont {Palshetkar}}, \bibinfo {author}
  {\bibfnamefont {D.~C.}\ \bibnamefont {Rafferty}}, \bibinfo {author}
  {\bibfnamefont {C.}~\bibnamefont {Simenel}}, \ and\ \bibinfo {author}
  {\bibfnamefont {K.}~\bibnamefont {Stiefel}},\ }\bibfield  {title} {\enquote
  {\bibinfo {title} {{R}educed quasifission competition in fusion reactions
  forming neutron-rich heavy elements},}\ }\href {\doibase
  10.1103/PhysRevC.91.041602} {\bibfield  {journal} {\bibinfo  {journal} {Phys.
  Rev. C}\ }\textbf {\bibinfo {volume} {91}},\ \bibinfo {pages} {041602(R)}
  (\bibinfo {year} {2015})}\BibitemShut {NoStop}%
\bibitem [{\citenamefont {Hammerton}\ \emph {et~al.}(2019)\citenamefont
  {Hammerton}, \citenamefont {Morrissey}, \citenamefont {Kohley}, \citenamefont
  {Hinde}, \citenamefont {Dasgupta}, \citenamefont {Wakhle}, \citenamefont
  {Williams}, \citenamefont {Carter}, \citenamefont {Cook}, \citenamefont
  {Greene}, \citenamefont {Jeung}, \citenamefont {Luong}, \citenamefont
  {McNeil}, \citenamefont {Palshetkar}, \citenamefont {Rafferty}, \citenamefont
  {Simenel},\ and\ \citenamefont {Stiefel}}]{hammerton2019}%
  \BibitemOpen
  \bibfield  {author} {\bibinfo {author} {\bibfnamefont {K.}~\bibnamefont
  {Hammerton}}, \bibinfo {author} {\bibfnamefont {D.~J.}\ \bibnamefont
  {Morrissey}}, \bibinfo {author} {\bibfnamefont {Z.}~\bibnamefont {Kohley}},
  \bibinfo {author} {\bibfnamefont {D.~J.}\ \bibnamefont {Hinde}}, \bibinfo
  {author} {\bibfnamefont {M.}~\bibnamefont {Dasgupta}}, \bibinfo {author}
  {\bibfnamefont {A.}~\bibnamefont {Wakhle}}, \bibinfo {author} {\bibfnamefont
  {E.}~\bibnamefont {Williams}}, \bibinfo {author} {\bibfnamefont {I.~P.}\
  \bibnamefont {Carter}}, \bibinfo {author} {\bibfnamefont {K.~J.}\
  \bibnamefont {Cook}}, \bibinfo {author} {\bibfnamefont {J.}~\bibnamefont
  {Greene}}, \bibinfo {author} {\bibfnamefont {D.~Y.}\ \bibnamefont {Jeung}},
  \bibinfo {author} {\bibfnamefont {D.~H.}\ \bibnamefont {Luong}}, \bibinfo
  {author} {\bibfnamefont {S.~D.}\ \bibnamefont {McNeil}}, \bibinfo {author}
  {\bibfnamefont {C.}~\bibnamefont {Palshetkar}}, \bibinfo {author}
  {\bibfnamefont {D.~C.}\ \bibnamefont {Rafferty}}, \bibinfo {author}
  {\bibfnamefont {C.}~\bibnamefont {Simenel}}, \ and\ \bibinfo {author}
  {\bibfnamefont {K.}~\bibnamefont {Stiefel}},\ }\bibfield  {title} {\enquote
  {\bibinfo {title} {Entrance channel effects on the quasifission reaction
  channel in {C}r+{W} systems},}\ }\href {\doibase 10.1103/PhysRevC.99.054621}
  {\bibfield  {journal} {\bibinfo  {journal} {Phys. Rev. C}\ }\textbf {\bibinfo
  {volume} {99}},\ \bibinfo {pages} {054621} (\bibinfo {year}
  {2019})}\BibitemShut {NoStop}%
\bibitem [{\citenamefont {Diaz-Torres}\ \emph {et~al.}(2001)\citenamefont
  {Diaz-Torres}, \citenamefont {Adamian}, \citenamefont {Antonenko},\ and\
  \citenamefont {Scheid}}]{diaz-torres2001}%
  \BibitemOpen
  \bibfield  {author} {\bibinfo {author} {\bibfnamefont {A.}~\bibnamefont
  {Diaz-Torres}}, \bibinfo {author} {\bibfnamefont {G.~G.}\ \bibnamefont
  {Adamian}}, \bibinfo {author} {\bibfnamefont {N.~V.}\ \bibnamefont
  {Antonenko}}, \ and\ \bibinfo {author} {\bibfnamefont {W.}~\bibnamefont
  {Scheid}},\ }\bibfield  {title} {\enquote {\bibinfo {title} {{Q}uasifission
  process in a transport model for a dinuclear system},}\ }\href {\doibase
  10.1103/PhysRevC.64.024604} {\bibfield  {journal} {\bibinfo  {journal} {Phys.
  Rev. C}\ }\textbf {\bibinfo {volume} {64}},\ \bibinfo {pages} {024604}
  (\bibinfo {year} {2001})}\BibitemShut {NoStop}%
\bibitem [{\citenamefont {Adamian}\ \emph {et~al.}(2003)\citenamefont
  {Adamian}, \citenamefont {Antonenko},\ and\ \citenamefont
  {Scheid}}]{adamian2003}%
  \BibitemOpen
  \bibfield  {author} {\bibinfo {author} {\bibfnamefont {G.~G.}\ \bibnamefont
  {Adamian}}, \bibinfo {author} {\bibfnamefont {N.~V.}\ \bibnamefont
  {Antonenko}}, \ and\ \bibinfo {author} {\bibfnamefont {W.}~\bibnamefont
  {Scheid}},\ }\bibfield  {title} {\enquote {\bibinfo {title}
  {{C}haracteristics of quasifission products within the dinuclear system
  model},}\ }\href {\doibase 10.1103/PhysRevC.68.034601} {\bibfield  {journal}
  {\bibinfo  {journal} {Phys. Rev. C}\ }\textbf {\bibinfo {volume} {68}},\
  \bibinfo {pages} {034601} (\bibinfo {year} {2003})}\BibitemShut {NoStop}%
\bibitem [{\citenamefont {Huang}\ \emph {et~al.}(2010)\citenamefont {Huang},
  \citenamefont {Gan}, \citenamefont {Zhou}, \citenamefont {Li},\ and\
  \citenamefont {Scheid}}]{huang2010}%
  \BibitemOpen
  \bibfield  {author} {\bibinfo {author} {\bibfnamefont {Minghui}\ \bibnamefont
  {Huang}}, \bibinfo {author} {\bibfnamefont {Zaiguo}\ \bibnamefont {Gan}},
  \bibinfo {author} {\bibfnamefont {Xiaohong}\ \bibnamefont {Zhou}}, \bibinfo
  {author} {\bibfnamefont {Junqing}\ \bibnamefont {Li}}, \ and\ \bibinfo
  {author} {\bibfnamefont {W.}~\bibnamefont {Scheid}},\ }\bibfield  {title}
  {\enquote {\bibinfo {title} {Competing fusion and quasifission reaction
  mechanisms in the production of superheavy nuclei},}\ }\href {\doibase
  10.1103/PhysRevC.82.044614} {\bibfield  {journal} {\bibinfo  {journal} {Phys.
  Rev. C}\ }\textbf {\bibinfo {volume} {82}},\ \bibinfo {pages} {044614}
  (\bibinfo {year} {2010})}\BibitemShut {NoStop}%
\bibitem [{\citenamefont {Bao}\ \emph {et~al.}(2015)\citenamefont {Bao},
  \citenamefont {Gao}, \citenamefont {Li},\ and\ \citenamefont
  {Zhang}}]{bao2015}%
  \BibitemOpen
  \bibfield  {author} {\bibinfo {author} {\bibfnamefont {X.~J.}\ \bibnamefont
  {Bao}}, \bibinfo {author} {\bibfnamefont {Y.}~\bibnamefont {Gao}}, \bibinfo
  {author} {\bibfnamefont {J.~Q.}\ \bibnamefont {Li}}, \ and\ \bibinfo {author}
  {\bibfnamefont {H.~F.}\ \bibnamefont {Zhang}},\ }\bibfield  {title} {\enquote
  {\bibinfo {title} {Theoretical study of the synthesis of superheavy nuclei
  using radioactive beams},}\ }\href {\doibase 10.1103/PhysRevC.91.064612}
  {\bibfield  {journal} {\bibinfo  {journal} {Phys. Rev. C}\ }\textbf {\bibinfo
  {volume} {91}},\ \bibinfo {pages} {064612} (\bibinfo {year}
  {2015})}\BibitemShut {NoStop}%
\bibitem [{\citenamefont {Guo}\ \emph {et~al.}(2017)\citenamefont {Guo},
  \citenamefont {Gao}, \citenamefont {Li},\ and\ \citenamefont
  {Zhang}}]{guo2018c}%
  \BibitemOpen
  \bibfield  {author} {\bibinfo {author} {\bibfnamefont {S.~Q.}\ \bibnamefont
  {Guo}}, \bibinfo {author} {\bibfnamefont {Y.}~\bibnamefont {Gao}}, \bibinfo
  {author} {\bibfnamefont {J.~Q.}\ \bibnamefont {Li}}, \ and\ \bibinfo {author}
  {\bibfnamefont {H.~F.}\ \bibnamefont {Zhang}},\ }\bibfield  {title} {\enquote
  {\bibinfo {title} {Dynamical deformation in heavy ion reactions and the
  characteristics of quasifission products},}\ }\href {\doibase
  10.1103/PhysRevC.96.044622} {\bibfield  {journal} {\bibinfo  {journal} {Phys.
  Rev. C}\ }\textbf {\bibinfo {volume} {96}},\ \bibinfo {pages} {044622}
  (\bibinfo {year} {2017})}\BibitemShut {NoStop}%
\bibitem [{\citenamefont {Zagrebaev}\ and\ \citenamefont
  {Greiner}(2005)}]{zagrebaev2005}%
  \BibitemOpen
  \bibfield  {author} {\bibinfo {author} {\bibfnamefont {Valery}\ \bibnamefont
  {Zagrebaev}}\ and\ \bibinfo {author} {\bibfnamefont {Walter}\ \bibnamefont
  {Greiner}},\ }\bibfield  {title} {\enquote {\bibinfo {title} {Unified
  consideration of deep inelastic, quasi-fission and fusion-fission
  phenomena},}\ }\href {\doibase 10.1088/0954-3899/31/7/024} {\bibfield
  {journal} {\bibinfo  {journal} {J. Phys. G}\ }\textbf {\bibinfo {volume}
  {31}},\ \bibinfo {pages} {825} (\bibinfo {year} {2005})}\BibitemShut
  {NoStop}%
\bibitem [{\citenamefont {Aritomo}(2009)}]{aritomo2009}%
  \BibitemOpen
  \bibfield  {author} {\bibinfo {author} {\bibfnamefont {Y.}~\bibnamefont
  {Aritomo}},\ }\bibfield  {title} {\enquote {\bibinfo {title} {{A}nalysis of
  dynamical processes using the mass distribution of fission fragments in
  heavy-ion reactions},}\ }\href {\doibase 10.1103/PhysRevC.80.064604}
  {\bibfield  {journal} {\bibinfo  {journal} {Phys. Rev. C}\ }\textbf {\bibinfo
  {volume} {80}},\ \bibinfo {pages} {064604} (\bibinfo {year}
  {2009})}\BibitemShut {NoStop}%
\bibitem [{\citenamefont {Aritomo}\ \emph {et~al.}(2012)\citenamefont
  {Aritomo}, \citenamefont {Hagino}, \citenamefont {Nishio},\ and\
  \citenamefont {Chiba}}]{aritomo2012}%
  \BibitemOpen
  \bibfield  {author} {\bibinfo {author} {\bibfnamefont {Y.}~\bibnamefont
  {Aritomo}}, \bibinfo {author} {\bibfnamefont {K.}~\bibnamefont {Hagino}},
  \bibinfo {author} {\bibfnamefont {K.}~\bibnamefont {Nishio}}, \ and\ \bibinfo
  {author} {\bibfnamefont {S.}~\bibnamefont {Chiba}},\ }\bibfield  {title}
  {\enquote {\bibinfo {title} {Dynamical approach to heavy-ion induced fission
  using actinide target nuclei at energies around the {C}oulomb barrier},}\
  }\href {\doibase 10.1103/PhysRevC.85.044614} {\bibfield  {journal} {\bibinfo
  {journal} {Phys. Rev. C}\ }\textbf {\bibinfo {volume} {85}},\ \bibinfo
  {pages} {044614} (\bibinfo {year} {2012})}\BibitemShut {NoStop}%
\bibitem [{\citenamefont {Karpov}\ and\ \citenamefont
  {Saiko}(2017)}]{karpov2017}%
  \BibitemOpen
  \bibfield  {author} {\bibinfo {author} {\bibfnamefont {A.~V.}\ \bibnamefont
  {Karpov}}\ and\ \bibinfo {author} {\bibfnamefont {V.~V.}\ \bibnamefont
  {Saiko}},\ }\bibfield  {title} {\enquote {\bibinfo {title} {Modeling
  near-barrier collisions of heavy ions based on a {L}angevin-type approach},}\
  }\href {\doibase 10.1103/PhysRevC.96.024618} {\bibfield  {journal} {\bibinfo
  {journal} {Phys. Rev. C}\ }\textbf {\bibinfo {volume} {96}},\ \bibinfo
  {pages} {024618} (\bibinfo {year} {2017})}\BibitemShut {NoStop}%
\bibitem [{\citenamefont {Sekizawa}\ and\ \citenamefont
  {Hagino}(2019)}]{sekizawa2019b}%
  \BibitemOpen
  \bibfield  {author} {\bibinfo {author} {\bibfnamefont {K.}~\bibnamefont
  {Sekizawa}}\ and\ \bibinfo {author} {\bibfnamefont {K.}~\bibnamefont
  {Hagino}},\ }\bibfield  {title} {\enquote {\bibinfo {title} {Time-dependent
  {H}artree-{F}ock plus {L}angevin approach for hot fusion reactions to
  synthesize the ${Z}=120$ superheavy element},}\ }\href {\doibase
  10.1103/PhysRevC.99.051602} {\bibfield  {journal} {\bibinfo  {journal} {Phys.
  Rev. C}\ }\textbf {\bibinfo {volume} {99}},\ \bibinfo {pages} {051602}
  (\bibinfo {year} {2019})}\BibitemShut {NoStop}%
\bibitem [{\citenamefont {Wen}\ \emph {et~al.}(2013)\citenamefont {Wen},
  \citenamefont {Sakata}, \citenamefont {Li}, \citenamefont {Wu}, \citenamefont
  {Zhang},\ and\ \citenamefont {Zhou}}]{wen2013}%
  \BibitemOpen
  \bibfield  {author} {\bibinfo {author} {\bibfnamefont {Kai}\ \bibnamefont
  {Wen}}, \bibinfo {author} {\bibfnamefont {Fumihiko}\ \bibnamefont {Sakata}},
  \bibinfo {author} {\bibfnamefont {Zhu-Xia}\ \bibnamefont {Li}}, \bibinfo
  {author} {\bibfnamefont {Xi-Zhen}\ \bibnamefont {Wu}}, \bibinfo {author}
  {\bibfnamefont {Ying-Xun}\ \bibnamefont {Zhang}}, \ and\ \bibinfo {author}
  {\bibfnamefont {Shan-Gui}\ \bibnamefont {Zhou}},\ }\bibfield  {title}
  {\enquote {\bibinfo {title} {{N}on-{G}aussian {F}luctuations and
  {N}on-{M}arkovian {E}ffects in the {N}uclear {F}usion {P}rocess: {L}angevin
  {D}ynamics {E}merging from {Q}uantum {M}olecular {D}ynamics {S}imulations},}\
  }\href {\doibase 10.1103/PhysRevLett.111.012501} {\bibfield  {journal}
  {\bibinfo  {journal} {Phys. Rev. Lett.}\ }\textbf {\bibinfo {volume} {111}},\
  \bibinfo {pages} {012501} (\bibinfo {year} {2013})}\BibitemShut {NoStop}%
\bibitem [{\citenamefont {Wang}\ and\ \citenamefont {Guo}(2016)}]{wang2016}%
  \BibitemOpen
  \bibfield  {author} {\bibinfo {author} {\bibfnamefont {Ning}\ \bibnamefont
  {Wang}}\ and\ \bibinfo {author} {\bibfnamefont {Lu}~\bibnamefont {Guo}},\
  }\bibfield  {title} {\enquote {\bibinfo {title} {New neutron-rich isotope
  production in $^{154}${S}m+$^{160}${G}d},}\ }\href {\doibase
  10.1016/j.physletb.2016.06.073} {\bibfield  {journal} {\bibinfo  {journal}
  {Phys. Lett. B}\ }\textbf {\bibinfo {volume} {760}},\ \bibinfo {pages}
  {236--241} (\bibinfo {year} {2016})}\BibitemShut {NoStop}%
\bibitem [{\citenamefont {Zhao}\ \emph {et~al.}(2016)\citenamefont {Zhao},
  \citenamefont {Li}, \citenamefont {Zhang}, \citenamefont {Wang},
  \citenamefont {Li}, \citenamefont {Shen}, \citenamefont {Wang},\ and\
  \citenamefont {Wu}}]{zhao2016}%
  \BibitemOpen
  \bibfield  {author} {\bibinfo {author} {\bibfnamefont {Kai}\ \bibnamefont
  {Zhao}}, \bibinfo {author} {\bibfnamefont {Zhuxia}\ \bibnamefont {Li}},
  \bibinfo {author} {\bibfnamefont {Yingxun}\ \bibnamefont {Zhang}}, \bibinfo
  {author} {\bibfnamefont {Ning}\ \bibnamefont {Wang}}, \bibinfo {author}
  {\bibfnamefont {Qingfeng}\ \bibnamefont {Li}}, \bibinfo {author}
  {\bibfnamefont {Caiwan}\ \bibnamefont {Shen}}, \bibinfo {author}
  {\bibfnamefont {Yongjia}\ \bibnamefont {Wang}}, \ and\ \bibinfo {author}
  {\bibfnamefont {Xizhen}\ \bibnamefont {Wu}},\ }\bibfield  {title} {\enquote
  {\bibinfo {title} {Production of unknown neutron--rich isotopes in
  $^{238}${U}+$^{238}${U} collisions at near--barrier energy},}\ }\href
  {\doibase 10.1103/PhysRevC.94.024601} {\bibfield  {journal} {\bibinfo
  {journal} {Phys. Rev. C}\ }\textbf {\bibinfo {volume} {94}},\ \bibinfo
  {pages} {024601} (\bibinfo {year} {2016})}\BibitemShut {NoStop}%
\bibitem [{\citenamefont {{C\'edric Golabek}}\ and\ \citenamefont {{C\'edric
  Simenel}}(2009)}]{golabek2009}%
  \BibitemOpen
  \bibfield  {author} {\bibinfo {author} {\bibnamefont {{C\'edric Golabek}}}\
  and\ \bibinfo {author} {\bibnamefont {{C\'edric Simenel}}},\ }\bibfield
  {title} {\enquote {\bibinfo {title} {{C}ollision {D}ynamics of {T}wo
  $^{238}${U A}tomic {N}uclei},}\ }\href {\doibase
  10.1103/PhysRevLett.103.042701} {\bibfield  {journal} {\bibinfo  {journal}
  {Phys. Rev. Lett.}\ }\textbf {\bibinfo {volume} {103}},\ \bibinfo {pages}
  {042701} (\bibinfo {year} {2009})}\BibitemShut {NoStop}%
\bibitem [{\citenamefont {{David J. Kedziora}}\ and\ \citenamefont {{C\'edric
  Simenel}}(2010)}]{kedziora2010}%
  \BibitemOpen
  \bibfield  {author} {\bibinfo {author} {\bibnamefont {{David J. Kedziora}}}\
  and\ \bibinfo {author} {\bibnamefont {{C\'edric Simenel}}},\ }\bibfield
  {title} {\enquote {\bibinfo {title} {{N}ew inverse quasifission mechanism to
  produce neutron-rich transfermium nuclei},}\ }\href {\doibase
  10.1103/PhysRevC.81.044613} {\bibfield  {journal} {\bibinfo  {journal} {Phys.
  Rev. C}\ }\textbf {\bibinfo {volume} {81}},\ \bibinfo {pages} {044613}
  (\bibinfo {year} {2010})}\BibitemShut {NoStop}%
\bibitem [{\citenamefont {Oberacker}\ \emph {et~al.}(2014)\citenamefont
  {Oberacker}, \citenamefont {Umar},\ and\ \citenamefont
  {Simenel}}]{oberacker2014}%
  \BibitemOpen
  \bibfield  {author} {\bibinfo {author} {\bibfnamefont {V.~E.}\ \bibnamefont
  {Oberacker}}, \bibinfo {author} {\bibfnamefont {A.~S.}\ \bibnamefont {Umar}},
  \ and\ \bibinfo {author} {\bibfnamefont {C.}~\bibnamefont {Simenel}},\
  }\bibfield  {title} {\enquote {\bibinfo {title} {{D}issipative dynamics in
  quasifission},}\ }\href {\doibase 10.1103/PhysRevC.90.054605} {\bibfield
  {journal} {\bibinfo  {journal} {Phys. Rev. C}\ }\textbf {\bibinfo {volume}
  {90}},\ \bibinfo {pages} {054605} (\bibinfo {year} {2014})}\BibitemShut
  {NoStop}%
\bibitem [{\citenamefont {Umar}\ \emph {et~al.}(2015)\citenamefont {Umar},
  \citenamefont {Oberacker},\ and\ \citenamefont {Simenel}}]{umar2015a}%
  \BibitemOpen
  \bibfield  {author} {\bibinfo {author} {\bibfnamefont {A.~S.}\ \bibnamefont
  {Umar}}, \bibinfo {author} {\bibfnamefont {V.~E.}\ \bibnamefont {Oberacker}},
  \ and\ \bibinfo {author} {\bibfnamefont {C.}~\bibnamefont {Simenel}},\
  }\bibfield  {title} {\enquote {\bibinfo {title} {{S}hape evolution and
  collective dynamics of quasifission in the time-dependent {H}artree-{F}ock
  approach},}\ }\href {\doibase 10.1103/PhysRevC.92.024621} {\bibfield
  {journal} {\bibinfo  {journal} {Phys. Rev. C}\ }\textbf {\bibinfo {volume}
  {92}},\ \bibinfo {pages} {024621} (\bibinfo {year} {2015})}\BibitemShut
  {NoStop}%
\bibitem [{\citenamefont {Umar}\ \emph {et~al.}(2016)\citenamefont {Umar},
  \citenamefont {Oberacker},\ and\ \citenamefont {Simenel}}]{umar2016}%
  \BibitemOpen
  \bibfield  {author} {\bibinfo {author} {\bibfnamefont {A.~S.}\ \bibnamefont
  {Umar}}, \bibinfo {author} {\bibfnamefont {V.~E.}\ \bibnamefont {Oberacker}},
  \ and\ \bibinfo {author} {\bibfnamefont {C.}~\bibnamefont {Simenel}},\
  }\bibfield  {title} {\enquote {\bibinfo {title} {Fusion and quasifission
  dynamics in the reactions $^{48}\mathrm{Ca}+{}^{249}\mathrm{Bk}$ and
  $^{50}\mathrm{Ti}+{}^{249}\mathrm{Bk}$ using a time-dependent
  {H}artree-{F}ock approach},}\ }\href {\doibase 10.1103/PhysRevC.94.024605}
  {\bibfield  {journal} {\bibinfo  {journal} {Phys. Rev. C}\ }\textbf {\bibinfo
  {volume} {94}},\ \bibinfo {pages} {024605} (\bibinfo {year}
  {2016})}\BibitemShut {NoStop}%
\bibitem [{\citenamefont {Sekizawa}\ and\ \citenamefont
  {Yabana}(2016)}]{sekizawa2016}%
  \BibitemOpen
  \bibfield  {author} {\bibinfo {author} {\bibfnamefont {Kazuyuki}\
  \bibnamefont {Sekizawa}}\ and\ \bibinfo {author} {\bibfnamefont {Kazuhiro}\
  \bibnamefont {Yabana}},\ }\bibfield  {title} {\enquote {\bibinfo {title}
  {{T}ime-dependent {H}artree-{F}ock calculations for multinucleon transfer and
  quasifission processes in the $^{64}\text{Ni}+{}^{238}\text{U}$ reaction},}\
  }\href {\doibase 10.1103/PhysRevC.93.054616} {\bibfield  {journal} {\bibinfo
  {journal} {Phys. Rev. C}\ }\textbf {\bibinfo {volume} {93}},\ \bibinfo
  {pages} {054616} (\bibinfo {year} {2016})}\BibitemShut {NoStop}%
\bibitem [{\citenamefont {{Chong Yu}}\ and\ \citenamefont {{Lu
  Guo}}(2017)}]{yu2017}%
  \BibitemOpen
  \bibfield  {author} {\bibinfo {author} {\bibnamefont {{Chong Yu}}}\ and\
  \bibinfo {author} {\bibnamefont {{Lu Guo}}},\ }\bibfield  {title} {\enquote
  {\bibinfo {title} {Angular momentum dependence of quasifission dynamics in
  the reaction $^{48}${C}a+$^{244}${P}u},}\ }\href {\doibase
  10.1007/s11433-017-9063-3} {\bibfield  {journal} {\bibinfo  {journal} {Sci.
  China Phys.}\ }\textbf {\bibinfo {volume} {60}},\ \bibinfo {pages} {092011}
  (\bibinfo {year} {2017})}\BibitemShut {NoStop}%
\bibitem [{\citenamefont {Ayik}\ \emph {et~al.}(2017)\citenamefont {Ayik},
  \citenamefont {Yilmaz}, \citenamefont {Yilmaz}, \citenamefont {Umar},\ and\
  \citenamefont {Turan}}]{ayik2017}%
  \BibitemOpen
  \bibfield  {author} {\bibinfo {author} {\bibfnamefont {S.}~\bibnamefont
  {Ayik}}, \bibinfo {author} {\bibfnamefont {B.}~\bibnamefont {Yilmaz}},
  \bibinfo {author} {\bibfnamefont {O.}~\bibnamefont {Yilmaz}}, \bibinfo
  {author} {\bibfnamefont {A.~S.}\ \bibnamefont {Umar}}, \ and\ \bibinfo
  {author} {\bibfnamefont {G.}~\bibnamefont {Turan}},\ }\bibfield  {title}
  {\enquote {\bibinfo {title} {Multinucleon transfer in central collisions of
  $^{238}\mathrm{U}+{}^{238}\mathrm{U}$},}\ }\href {\doibase
  10.1103/PhysRevC.96.024611} {\bibfield  {journal} {\bibinfo  {journal} {Phys.
  Rev. C}\ }\textbf {\bibinfo {volume} {96}},\ \bibinfo {pages} {024611}
  (\bibinfo {year} {2017})}\BibitemShut {NoStop}%
\bibitem [{\citenamefont {Ayik}\ \emph {et~al.}(2018)\citenamefont {Ayik},
  \citenamefont {Yilmaz}, \citenamefont {Yilmaz},\ and\ \citenamefont
  {Umar}}]{ayik2018}%
  \BibitemOpen
  \bibfield  {author} {\bibinfo {author} {\bibfnamefont {S.}~\bibnamefont
  {Ayik}}, \bibinfo {author} {\bibfnamefont {B.}~\bibnamefont {Yilmaz}},
  \bibinfo {author} {\bibfnamefont {O.}~\bibnamefont {Yilmaz}}, \ and\ \bibinfo
  {author} {\bibfnamefont {A.~S.}\ \bibnamefont {Umar}},\ }\bibfield  {title}
  {\enquote {\bibinfo {title} {Quantal diffusion description of multinucleon
  transfers in heavy--ion collisions},}\ }\href {\doibase
  10.1103/PhysRevC.97.054618} {\bibfield  {journal} {\bibinfo  {journal} {Phys.
  Rev. C}\ }\textbf {\bibinfo {volume} {97}},\ \bibinfo {pages} {054618}
  (\bibinfo {year} {2018})}\BibitemShut {NoStop}%
\bibitem [{\citenamefont {Sekizawa}(2017)}]{sekizawa2017a}%
  \BibitemOpen
  \bibfield  {author} {\bibinfo {author} {\bibfnamefont {Kazuyuki}\
  \bibnamefont {Sekizawa}},\ }\bibfield  {title} {\enquote {\bibinfo {title}
  {Enhanced nucleon transfer in tip collisions of
  $^{238}\mathrm{U}+{}^{124}\mathrm{Sn}$},}\ }\href {\doibase
  10.1103/PhysRevC.96.041601} {\bibfield  {journal} {\bibinfo  {journal} {Phys.
  Rev. C}\ }\textbf {\bibinfo {volume} {96}},\ \bibinfo {pages} {041601(R)}
  (\bibinfo {year} {2017})}\BibitemShut {NoStop}%
\bibitem [{\citenamefont {Wakhle}\ \emph {et~al.}(2018)\citenamefont {Wakhle},
  \citenamefont {Hammerton}, \citenamefont {Kohley}, \citenamefont {Morrissey},
  \citenamefont {Stiefel}, \citenamefont {Yurkon}, \citenamefont {Walshe},
  \citenamefont {Cook}, \citenamefont {Dasgupta}, \citenamefont {Hinde},
  \citenamefont {Jeung}, \citenamefont {Prasad}, \citenamefont {Rafferty},
  \citenamefont {Simenel}, \citenamefont {Simpson}, \citenamefont {Vo-Phuoc},
  \citenamefont {King}, \citenamefont {Loveland},\ and\ \citenamefont
  {Yanez}}]{wakhle2018}%
  \BibitemOpen
  \bibfield  {author} {\bibinfo {author} {\bibfnamefont {A.}~\bibnamefont
  {Wakhle}}, \bibinfo {author} {\bibfnamefont {K.}~\bibnamefont {Hammerton}},
  \bibinfo {author} {\bibfnamefont {Z.}~\bibnamefont {Kohley}}, \bibinfo
  {author} {\bibfnamefont {D.~J.}\ \bibnamefont {Morrissey}}, \bibinfo {author}
  {\bibfnamefont {K.}~\bibnamefont {Stiefel}}, \bibinfo {author} {\bibfnamefont
  {J.}~\bibnamefont {Yurkon}}, \bibinfo {author} {\bibfnamefont
  {J.}~\bibnamefont {Walshe}}, \bibinfo {author} {\bibfnamefont {K.~J.}\
  \bibnamefont {Cook}}, \bibinfo {author} {\bibfnamefont {M.}~\bibnamefont
  {Dasgupta}}, \bibinfo {author} {\bibfnamefont {D.~J.}\ \bibnamefont {Hinde}},
  \bibinfo {author} {\bibfnamefont {D.~J.}\ \bibnamefont {Jeung}}, \bibinfo
  {author} {\bibfnamefont {E.}~\bibnamefont {Prasad}}, \bibinfo {author}
  {\bibfnamefont {D.~C.}\ \bibnamefont {Rafferty}}, \bibinfo {author}
  {\bibfnamefont {C.}~\bibnamefont {Simenel}}, \bibinfo {author} {\bibfnamefont
  {E.~C.}\ \bibnamefont {Simpson}}, \bibinfo {author} {\bibfnamefont
  {K.}~\bibnamefont {Vo-Phuoc}}, \bibinfo {author} {\bibfnamefont
  {J.}~\bibnamefont {King}}, \bibinfo {author} {\bibfnamefont {W.}~\bibnamefont
  {Loveland}}, \ and\ \bibinfo {author} {\bibfnamefont {R.}~\bibnamefont
  {Yanez}},\ }\bibfield  {title} {\enquote {\bibinfo {title} {Capture cross
  sections for the synthesis of new heavy nuclei using radioactive beams},}\
  }\href {\doibase 10.1103/PhysRevC.97.021602} {\bibfield  {journal} {\bibinfo
  {journal} {Phys. Rev. C}\ }\textbf {\bibinfo {volume} {97}},\ \bibinfo
  {pages} {021602} (\bibinfo {year} {2018})}\BibitemShut {NoStop}%
\bibitem [{\citenamefont {Simenel}(2012)}]{simenel2012}%
  \BibitemOpen
  \bibfield  {author} {\bibinfo {author} {\bibfnamefont {C\'edric}\
  \bibnamefont {Simenel}},\ }\bibfield  {title} {\enquote {\bibinfo {title}
  {{N}uclear quantum many-body dynamics},}\ }\href {\doibase
  10.1140/epja/i2012-12152-0} {\bibfield  {journal} {\bibinfo  {journal} {Eur.
  Phys. J. A}\ }\textbf {\bibinfo {volume} {48}},\ \bibinfo {pages} {152}
  (\bibinfo {year} {2012})}\BibitemShut {NoStop}%
\bibitem [{\citenamefont {Simenel}\ and\ \citenamefont
  {Umar}(2018)}]{simenel2018}%
  \BibitemOpen
  \bibfield  {author} {\bibinfo {author} {\bibfnamefont {C.}~\bibnamefont
  {Simenel}}\ and\ \bibinfo {author} {\bibfnamefont {A.~S.}\ \bibnamefont
  {Umar}},\ }\bibfield  {title} {\enquote {\bibinfo {title} {Heavy--ion
  collisions and fission dynamics with the time--dependent {H}artree-{F}ock
  theory and its extensions},}\ }\href {\doibase 10.1016/j.ppnp.2018.07.002}
  {\bibfield  {journal} {\bibinfo  {journal} {Prog. Part. Nucl. Phys.}\
  }\textbf {\bibinfo {volume} {103}},\ \bibinfo {pages} {19--66} (\bibinfo
  {year} {2018})}\BibitemShut {NoStop}%
\bibitem [{\citenamefont {{Kazuyuki Sekizawa}}(2019)}]{sekizawa2019}%
  \BibitemOpen
  \bibfield  {author} {\bibinfo {author} {\bibnamefont {{Kazuyuki Sekizawa}}},\
  }\bibfield  {title} {\enquote {\bibinfo {title} {{TDHF} {T}heory and {I}ts
  {E}xtensions for the {M}ultinucleon {T}ransfer {R}eaction: {A} {M}ini
  {R}eview},}\ }\href {\doibase 10.3389/fphy.2019.00020} {\bibfield  {journal}
  {\bibinfo  {journal} {Front. Phys.}\ }\textbf {\bibinfo {volume} {7}},\
  \bibinfo {pages} {20} (\bibinfo {year} {2019})}\BibitemShut {NoStop}%
\bibitem [{\citenamefont {Stevenson}\ and\ \citenamefont
  {Barton}(2019)}]{stevenson2019}%
  \BibitemOpen
  \bibfield  {author} {\bibinfo {author} {\bibfnamefont {P.~D.}\ \bibnamefont
  {Stevenson}}\ and\ \bibinfo {author} {\bibfnamefont {M.~C.}\ \bibnamefont
  {Barton}},\ }\bibfield  {title} {\enquote {\bibinfo {title} {Low--energy
  heavy-ion reactions and the {S}kyrme effective interaction},}\ }\href
  {\doibase 10.1016/j.ppnp.2018.09.002} {\bibfield  {journal} {\bibinfo
  {journal} {Prog. Part. Nucl. Phys.}\ }\textbf {\bibinfo {volume} {104}},\
  \bibinfo {pages} {142--164} (\bibinfo {year} {2019})}\BibitemShut {NoStop}%
\bibitem [{\citenamefont {Scamps}\ and\ \citenamefont
  {Simenel}(2018)}]{scamps2018}%
  \BibitemOpen
  \bibfield  {author} {\bibinfo {author} {\bibfnamefont {Guillaume}\
  \bibnamefont {Scamps}}\ and\ \bibinfo {author} {\bibfnamefont {C{\'e}dric}\
  \bibnamefont {Simenel}},\ }\bibfield  {title} {\enquote {\bibinfo {title}
  {Impact of pear-shaped fission fragments on mass-asymmetric fission in
  actinides},}\ }\href {\doibase 10.1038/s41586-018-0780-0} {\bibfield
  {journal} {\bibinfo  {journal} {Nature}\ }\textbf {\bibinfo {volume} {564}},\
  \bibinfo {pages} {382--385} (\bibinfo {year} {2018})}\BibitemShut {NoStop}%
\bibitem [{\citenamefont {Scamps}\ and\ \citenamefont
  {Simenel}(2019)}]{scamps2019}%
  \BibitemOpen
  \bibfield  {author} {\bibinfo {author} {\bibfnamefont {G.}~\bibnamefont
  {Scamps}}\ and\ \bibinfo {author} {\bibfnamefont {C.}~\bibnamefont
  {Simenel}},\ }\bibfield  {title} {\enquote {\bibinfo {title} {Effect of shell
  structure on the fission of sub-lead nuclei},}\ }\href
  {https://arxiv.org/abs/1904.01275} {\bibfield  {journal} {\bibinfo  {journal}
  {Arxiv:1904.012705}\ } (\bibinfo {year} {2019})}\BibitemShut {NoStop}%
\bibitem [{\citenamefont {Poenaru}\ and\ \citenamefont
  {Gherghescu}(2018)}]{poenaru2018}%
  \BibitemOpen
  \bibfield  {author} {\bibinfo {author} {\bibfnamefont {D.~N.}\ \bibnamefont
  {Poenaru}}\ and\ \bibinfo {author} {\bibfnamefont {R.~A.}\ \bibnamefont
  {Gherghescu}},\ }\bibfield  {title} {\enquote {\bibinfo {title} {$\alpha$
  decay and cluster radioactivity of nuclei of interest to the synthesis of
  ${Z}=119$, 120 isotopes},}\ }\href {\doibase 10.1103/PhysRevC.97.044621}
  {\bibfield  {journal} {\bibinfo  {journal} {Phys. Rev. C}\ }\textbf {\bibinfo
  {volume} {97}},\ \bibinfo {pages} {044621} (\bibinfo {year}
  {2018})}\BibitemShut {NoStop}%
\bibitem [{\citenamefont {Warda}\ \emph {et~al.}(2018)\citenamefont {Warda},
  \citenamefont {Zdeb},\ and\ \citenamefont {Robledo}}]{warda2018}%
  \BibitemOpen
  \bibfield  {author} {\bibinfo {author} {\bibfnamefont {M.}~\bibnamefont
  {Warda}}, \bibinfo {author} {\bibfnamefont {A.}~\bibnamefont {Zdeb}}, \ and\
  \bibinfo {author} {\bibfnamefont {L.~M.}\ \bibnamefont {Robledo}},\
  }\bibfield  {title} {\enquote {\bibinfo {title} {Cluster radioactivity in
  superheavy nuclei},}\ }\href {\doibase 10.1103/PhysRevC.98.041602} {\bibfield
   {journal} {\bibinfo  {journal} {Phys. Rev. C}\ }\textbf {\bibinfo {volume}
  {98}},\ \bibinfo {pages} {041602} (\bibinfo {year} {2018})}\BibitemShut
  {NoStop}%
\bibitem [{\citenamefont {Matheson}\ \emph {et~al.}(2019)\citenamefont
  {Matheson}, \citenamefont {Giuliani}, \citenamefont {Nazarewicz},
  \citenamefont {Sadhukhan},\ and\ \citenamefont {Schunck}}]{matheson2019}%
  \BibitemOpen
  \bibfield  {author} {\bibinfo {author} {\bibfnamefont {Zachary}\ \bibnamefont
  {Matheson}}, \bibinfo {author} {\bibfnamefont {Samuel~A.}\ \bibnamefont
  {Giuliani}}, \bibinfo {author} {\bibfnamefont {Witold}\ \bibnamefont
  {Nazarewicz}}, \bibinfo {author} {\bibfnamefont {Jhilam}\ \bibnamefont
  {Sadhukhan}}, \ and\ \bibinfo {author} {\bibfnamefont {Nicolas}\ \bibnamefont
  {Schunck}},\ }\bibfield  {title} {\enquote {\bibinfo {title} {Cluster
  radioactivity of $_{118}^{294}\mathrm{Og}_{176}$},}\ }\href {\doibase
  10.1103/PhysRevC.99.041304} {\bibfield  {journal} {\bibinfo  {journal} {Phys.
  Rev. C}\ }\textbf {\bibinfo {volume} {99}},\ \bibinfo {pages} {041304}
  (\bibinfo {year} {2019})}\BibitemShut {NoStop}%
\bibitem [{\citenamefont {Zhang}\ and\ \citenamefont
  {Wang}(2018)}]{zhang2018b}%
  \BibitemOpen
  \bibfield  {author} {\bibinfo {author} {\bibfnamefont {Y.~L.}\ \bibnamefont
  {Zhang}}\ and\ \bibinfo {author} {\bibfnamefont {Y.~Z.}\ \bibnamefont
  {Wang}},\ }\bibfield  {title} {\enquote {\bibinfo {title} {Systematic study
  of cluster radioactivity of superheavy nuclei},}\ }\href {\doibase
  10.1103/PhysRevC.97.014318} {\bibfield  {journal} {\bibinfo  {journal} {Phys.
  Rev. C}\ }\textbf {\bibinfo {volume} {97}},\ \bibinfo {pages} {014318}
  (\bibinfo {year} {2018})}\BibitemShut {NoStop}%
\bibitem [{\citenamefont {{Yu. Ts. Oganessian}}\ \emph
  {et~al.}(2010)\citenamefont {{Yu. Ts. Oganessian}}, \citenamefont {{F. Sh.
  Abdullin}}, \citenamefont {Bailey}, \citenamefont {Benker}, \citenamefont
  {Bennett}, \citenamefont {Dmitriev}, \citenamefont {Ezold}, \citenamefont
  {Hamilton}, \citenamefont {Henderson}, \citenamefont {Itkis}, \citenamefont
  {{Yu. V. Lobanov}}, \citenamefont {Mezentsev}, \citenamefont {Moody},
  \citenamefont {Nelson}, \citenamefont {Polyakov}, \citenamefont {Porter},
  \citenamefont {Ramayya}, \citenamefont {Riley}, \citenamefont {Roberto},
  \citenamefont {Ryabinin}, \citenamefont {Rykaczewski}, \citenamefont
  {Sagaidak}, \citenamefont {Shaughnessy}, \citenamefont {Shirokovsky},
  \citenamefont {Stoyer}, \citenamefont {Subbotin}, \citenamefont {Sudowe},
  \citenamefont {Sukhov}, \citenamefont {{Yu. S. Tsyganov}}, \citenamefont
  {Utyonkov}, \citenamefont {Voinov}, \citenamefont {Vostokin},\ and\
  \citenamefont {Wilk}}]{oganessian2010}%
  \BibitemOpen
  \bibfield  {author} {\bibinfo {author} {\bibnamefont {{Yu. Ts. Oganessian}}},
  \bibinfo {author} {\bibnamefont {{F. Sh. Abdullin}}}, \bibinfo {author}
  {\bibfnamefont {P.~D.}\ \bibnamefont {Bailey}}, \bibinfo {author}
  {\bibfnamefont {D.~E.}\ \bibnamefont {Benker}}, \bibinfo {author}
  {\bibfnamefont {M.~E.}\ \bibnamefont {Bennett}}, \bibinfo {author}
  {\bibfnamefont {S.~N.}\ \bibnamefont {Dmitriev}}, \bibinfo {author}
  {\bibfnamefont {J.~G.}\ \bibnamefont {Ezold}}, \bibinfo {author}
  {\bibfnamefont {J.~H.}\ \bibnamefont {Hamilton}}, \bibinfo {author}
  {\bibfnamefont {R.~A.}\ \bibnamefont {Henderson}}, \bibinfo {author}
  {\bibfnamefont {M.~G.}\ \bibnamefont {Itkis}}, \bibinfo {author}
  {\bibnamefont {{Yu. V. Lobanov}}}, \bibinfo {author} {\bibfnamefont {A.~N.}\
  \bibnamefont {Mezentsev}}, \bibinfo {author} {\bibfnamefont {K.~J.}\
  \bibnamefont {Moody}}, \bibinfo {author} {\bibfnamefont {S.~L.}\ \bibnamefont
  {Nelson}}, \bibinfo {author} {\bibfnamefont {A.~N.}\ \bibnamefont
  {Polyakov}}, \bibinfo {author} {\bibfnamefont {C.~E.}\ \bibnamefont
  {Porter}}, \bibinfo {author} {\bibfnamefont {A.~V.}\ \bibnamefont {Ramayya}},
  \bibinfo {author} {\bibfnamefont {F.~D.}\ \bibnamefont {Riley}}, \bibinfo
  {author} {\bibfnamefont {J.~B.}\ \bibnamefont {Roberto}}, \bibinfo {author}
  {\bibfnamefont {M.~A.}\ \bibnamefont {Ryabinin}}, \bibinfo {author}
  {\bibfnamefont {K.~P.}\ \bibnamefont {Rykaczewski}}, \bibinfo {author}
  {\bibfnamefont {R.~N.}\ \bibnamefont {Sagaidak}}, \bibinfo {author}
  {\bibfnamefont {D.~A.}\ \bibnamefont {Shaughnessy}}, \bibinfo {author}
  {\bibfnamefont {I.~V.}\ \bibnamefont {Shirokovsky}}, \bibinfo {author}
  {\bibfnamefont {M.~A.}\ \bibnamefont {Stoyer}}, \bibinfo {author}
  {\bibfnamefont {V.~G.}\ \bibnamefont {Subbotin}}, \bibinfo {author}
  {\bibfnamefont {R.}~\bibnamefont {Sudowe}}, \bibinfo {author} {\bibfnamefont
  {A.~M.}\ \bibnamefont {Sukhov}}, \bibinfo {author} {\bibnamefont {{Yu. S.
  Tsyganov}}}, \bibinfo {author} {\bibfnamefont {V.~K.}\ \bibnamefont
  {Utyonkov}}, \bibinfo {author} {\bibfnamefont {A.~A.}\ \bibnamefont
  {Voinov}}, \bibinfo {author} {\bibfnamefont {G.~K.}\ \bibnamefont
  {Vostokin}}, \ and\ \bibinfo {author} {\bibfnamefont {P.~A.}\ \bibnamefont
  {Wilk}},\ }\bibfield  {title} {\enquote {\bibinfo {title} {{S}ynthesis of a
  {N}ew {E}lement with {A}tomic {N}umber ${Z}=117$},}\ }\href {\doibase
  10.1103/PhysRevLett.104.142502} {\bibfield  {journal} {\bibinfo  {journal}
  {Phys. Rev. Lett.}\ }\textbf {\bibinfo {volume} {104}},\ \bibinfo {pages}
  {142502} (\bibinfo {year} {2010})}\BibitemShut {NoStop}%
\bibitem [{\citenamefont {{Yu. Ts. Oganessian}}\ \emph
  {et~al.}(2011)\citenamefont {{Yu. Ts. Oganessian}}, \citenamefont {{F. Sh.
  Abdullin}}, \citenamefont {Bailey}, \citenamefont {Benker}, \citenamefont
  {Bennett}, \citenamefont {Dmitriev}, \citenamefont {Ezold}, \citenamefont
  {Hamilton}, \citenamefont {Henderson}, \citenamefont {Itkis}, \citenamefont
  {{Yu. V. Lobanov}}, \citenamefont {Mezentsev}, \citenamefont {Moody},
  \citenamefont {Nelson}, \citenamefont {Polyakov}, \citenamefont {Porter},
  \citenamefont {Ramayya}, \citenamefont {Riley}, \citenamefont {Roberto},
  \citenamefont {Ryabinin}, \citenamefont {Rykaczewski}, \citenamefont
  {Sagaidak}, \citenamefont {Shaughnessy}, \citenamefont {Shirokovsky},
  \citenamefont {Stoyer}, \citenamefont {Subbotin}, \citenamefont {Sudowe},
  \citenamefont {Sukhov}, \citenamefont {Taylor}, \citenamefont {{Yu. S.
  Tsyganov}}, \citenamefont {Utyonkov}, \citenamefont {Voinov}, \citenamefont
  {Vostokin},\ and\ \citenamefont {Wilk}}]{oganessian2011}%
  \BibitemOpen
  \bibfield  {author} {\bibinfo {author} {\bibnamefont {{Yu. Ts. Oganessian}}},
  \bibinfo {author} {\bibnamefont {{F. Sh. Abdullin}}}, \bibinfo {author}
  {\bibfnamefont {P.~D.}\ \bibnamefont {Bailey}}, \bibinfo {author}
  {\bibfnamefont {D.~E.}\ \bibnamefont {Benker}}, \bibinfo {author}
  {\bibfnamefont {M.~E.}\ \bibnamefont {Bennett}}, \bibinfo {author}
  {\bibfnamefont {S.~N.}\ \bibnamefont {Dmitriev}}, \bibinfo {author}
  {\bibfnamefont {J.~G.}\ \bibnamefont {Ezold}}, \bibinfo {author}
  {\bibfnamefont {J.~H.}\ \bibnamefont {Hamilton}}, \bibinfo {author}
  {\bibfnamefont {R.~A.}\ \bibnamefont {Henderson}}, \bibinfo {author}
  {\bibfnamefont {M.~G.}\ \bibnamefont {Itkis}}, \bibinfo {author}
  {\bibnamefont {{Yu. V. Lobanov}}}, \bibinfo {author} {\bibfnamefont {A.~N.}\
  \bibnamefont {Mezentsev}}, \bibinfo {author} {\bibfnamefont {K.~J.}\
  \bibnamefont {Moody}}, \bibinfo {author} {\bibfnamefont {S.~L.}\ \bibnamefont
  {Nelson}}, \bibinfo {author} {\bibfnamefont {A.~N.}\ \bibnamefont
  {Polyakov}}, \bibinfo {author} {\bibfnamefont {C.~E.}\ \bibnamefont
  {Porter}}, \bibinfo {author} {\bibfnamefont {A.~V.}\ \bibnamefont {Ramayya}},
  \bibinfo {author} {\bibfnamefont {F.~D.}\ \bibnamefont {Riley}}, \bibinfo
  {author} {\bibfnamefont {J.~B.}\ \bibnamefont {Roberto}}, \bibinfo {author}
  {\bibfnamefont {M.~A.}\ \bibnamefont {Ryabinin}}, \bibinfo {author}
  {\bibfnamefont {K.~P.}\ \bibnamefont {Rykaczewski}}, \bibinfo {author}
  {\bibfnamefont {R.~N.}\ \bibnamefont {Sagaidak}}, \bibinfo {author}
  {\bibfnamefont {D.~A.}\ \bibnamefont {Shaughnessy}}, \bibinfo {author}
  {\bibfnamefont {I.~V.}\ \bibnamefont {Shirokovsky}}, \bibinfo {author}
  {\bibfnamefont {M.~A.}\ \bibnamefont {Stoyer}}, \bibinfo {author}
  {\bibfnamefont {V.~G.}\ \bibnamefont {Subbotin}}, \bibinfo {author}
  {\bibfnamefont {R.}~\bibnamefont {Sudowe}}, \bibinfo {author} {\bibfnamefont
  {A.~M.}\ \bibnamefont {Sukhov}}, \bibinfo {author} {\bibfnamefont
  {R.}~\bibnamefont {Taylor}}, \bibinfo {author} {\bibnamefont {{Yu. S.
  Tsyganov}}}, \bibinfo {author} {\bibfnamefont {V.~K.}\ \bibnamefont
  {Utyonkov}}, \bibinfo {author} {\bibfnamefont {A.~A.}\ \bibnamefont
  {Voinov}}, \bibinfo {author} {\bibfnamefont {G.~K.}\ \bibnamefont
  {Vostokin}}, \ and\ \bibinfo {author} {\bibfnamefont {P.~A.}\ \bibnamefont
  {Wilk}},\ }\bibfield  {title} {\enquote {\bibinfo {title} {{E}leven new
  heaviest isotopes of elements {Z} = 105 to {Z} = 117 identified among the
  products of $^{249}${B}k+$^{48}${C}a reactions},}\ }\href {\doibase
  10.1103/physrevc.83.054315} {\bibfield  {journal} {\bibinfo  {journal} {Phys.
  Rev. C}\ }\textbf {\bibinfo {volume} {83}},\ \bibinfo {pages} {054315}
  (\bibinfo {year} {2011})}\BibitemShut {NoStop}%
\bibitem [{\citenamefont {Oganessian}\ \emph {et~al.}(2012)\citenamefont
  {Oganessian}, \citenamefont {Abdullin}, \citenamefont {Alexander},
  \citenamefont {Binder}, \citenamefont {Boll}, \citenamefont {Dmitriev},
  \citenamefont {Ezold}, \citenamefont {Felker}, \citenamefont {Gostic},
  \citenamefont {Grzywacz}, \citenamefont {Hamilton}, \citenamefont
  {Henderson}, \citenamefont {Itkis}, \citenamefont {Miernik}, \citenamefont
  {Miller}, \citenamefont {Moody}, \citenamefont {Polyakov}, \citenamefont
  {Ramayya}, \citenamefont {Roberto}, \citenamefont {Ryabinin}, \citenamefont
  {Rykaczewski}, \citenamefont {Sagaidak}, \citenamefont {Shaughnessy},
  \citenamefont {Shirokovsky}, \citenamefont {Shumeiko}, \citenamefont
  {Stoyer}, \citenamefont {Stoyer}, \citenamefont {Subbotin}, \citenamefont
  {Sukhov}, \citenamefont {Tsyganov}, \citenamefont {Utyonkov}, \citenamefont
  {Voinov},\ and\ \citenamefont {Vostokin}}]{oganessian2012}%
  \BibitemOpen
  \bibfield  {author} {\bibinfo {author} {\bibfnamefont {{\relax Yu.
  Ts}.}~\bibnamefont {Oganessian}}, \bibinfo {author} {\bibfnamefont {{\relax
  F. Sh}.}~\bibnamefont {Abdullin}}, \bibinfo {author} {\bibfnamefont
  {C.}~\bibnamefont {Alexander}}, \bibinfo {author} {\bibfnamefont
  {J.}~\bibnamefont {Binder}}, \bibinfo {author} {\bibfnamefont {R.~A.}\
  \bibnamefont {Boll}}, \bibinfo {author} {\bibfnamefont {S.~N.}\ \bibnamefont
  {Dmitriev}}, \bibinfo {author} {\bibfnamefont {J.}~\bibnamefont {Ezold}},
  \bibinfo {author} {\bibfnamefont {K.}~\bibnamefont {Felker}}, \bibinfo
  {author} {\bibfnamefont {J.~M.}\ \bibnamefont {Gostic}}, \bibinfo {author}
  {\bibfnamefont {R.~K.}\ \bibnamefont {Grzywacz}}, \bibinfo {author}
  {\bibfnamefont {J.~H.}\ \bibnamefont {Hamilton}}, \bibinfo {author}
  {\bibfnamefont {R.~A.}\ \bibnamefont {Henderson}}, \bibinfo {author}
  {\bibfnamefont {M.~G.}\ \bibnamefont {Itkis}}, \bibinfo {author}
  {\bibfnamefont {K.}~\bibnamefont {Miernik}}, \bibinfo {author} {\bibfnamefont
  {D.}~\bibnamefont {Miller}}, \bibinfo {author} {\bibfnamefont {K.~J.}\
  \bibnamefont {Moody}}, \bibinfo {author} {\bibfnamefont {A.~N.}\ \bibnamefont
  {Polyakov}}, \bibinfo {author} {\bibfnamefont {A.~V.}\ \bibnamefont
  {Ramayya}}, \bibinfo {author} {\bibfnamefont {J.~B.}\ \bibnamefont
  {Roberto}}, \bibinfo {author} {\bibfnamefont {M.~A.}\ \bibnamefont
  {Ryabinin}}, \bibinfo {author} {\bibfnamefont {K.~P.}\ \bibnamefont
  {Rykaczewski}}, \bibinfo {author} {\bibfnamefont {R.~N.}\ \bibnamefont
  {Sagaidak}}, \bibinfo {author} {\bibfnamefont {D.~A.}\ \bibnamefont
  {Shaughnessy}}, \bibinfo {author} {\bibfnamefont {I.~V.}\ \bibnamefont
  {Shirokovsky}}, \bibinfo {author} {\bibfnamefont {M.~V.}\ \bibnamefont
  {Shumeiko}}, \bibinfo {author} {\bibfnamefont {M.~A.}\ \bibnamefont
  {Stoyer}}, \bibinfo {author} {\bibfnamefont {N.~J.}\ \bibnamefont {Stoyer}},
  \bibinfo {author} {\bibfnamefont {V.~G.}\ \bibnamefont {Subbotin}}, \bibinfo
  {author} {\bibfnamefont {A.~M.}\ \bibnamefont {Sukhov}}, \bibinfo {author}
  {\bibfnamefont {{\relax Yu. S}.}~\bibnamefont {Tsyganov}}, \bibinfo {author}
  {\bibfnamefont {V.~K.}\ \bibnamefont {Utyonkov}}, \bibinfo {author}
  {\bibfnamefont {A.~A.}\ \bibnamefont {Voinov}}, \ and\ \bibinfo {author}
  {\bibfnamefont {G.~K.}\ \bibnamefont {Vostokin}},\ }\bibfield  {title}
  {\enquote {\bibinfo {title} {{P}roduction and {D}ecay of the {H}eaviest
  {N}uclei $^{293,294}117$ and $^{294}118$},}\ }\href {\doibase
  10.1103/PhysRevLett.109.162501} {\bibfield  {journal} {\bibinfo  {journal}
  {Phys. Rev. Lett.}\ }\textbf {\bibinfo {volume} {109}},\ \bibinfo {pages}
  {162501} (\bibinfo {year} {2012})}\BibitemShut {NoStop}%
\bibitem [{\citenamefont {Oganessian}\ \emph {et~al.}(2013)\citenamefont
  {Oganessian}, \citenamefont {Abdullin}, \citenamefont {Alexander},
  \citenamefont {Binder}, \citenamefont {Boll}, \citenamefont {Dmitriev},
  \citenamefont {Ezold}, \citenamefont {Felker}, \citenamefont {Gostic},
  \citenamefont {Grzywacz}, \citenamefont {Hamilton}, \citenamefont
  {Henderson}, \citenamefont {Itkis}, \citenamefont {Miernik}, \citenamefont
  {Miller}, \citenamefont {Moody}, \citenamefont {Polyakov}, \citenamefont
  {Ramayya}, \citenamefont {Roberto}, \citenamefont {Ryabinin}, \citenamefont
  {Rykaczewski}, \citenamefont {Sagaidak}, \citenamefont {Shaughnessy},
  \citenamefont {Shirokovsky}, \citenamefont {Shumeiko}, \citenamefont
  {Stoyer}, \citenamefont {Stoyer}, \citenamefont {Subbotin}, \citenamefont
  {Sukhov}, \citenamefont {Tsyganov}, \citenamefont {Utyonkov}, \citenamefont
  {Voinov},\ and\ \citenamefont {Vostokin}}]{oganessian2013}%
  \BibitemOpen
  \bibfield  {author} {\bibinfo {author} {\bibfnamefont {{\relax Yu.
  Ts}.}~\bibnamefont {Oganessian}}, \bibinfo {author} {\bibfnamefont {{\relax
  F. Sh}.}~\bibnamefont {Abdullin}}, \bibinfo {author} {\bibfnamefont
  {C.}~\bibnamefont {Alexander}}, \bibinfo {author} {\bibfnamefont
  {J.}~\bibnamefont {Binder}}, \bibinfo {author} {\bibfnamefont {R.~A.}\
  \bibnamefont {Boll}}, \bibinfo {author} {\bibfnamefont {S.~N.}\ \bibnamefont
  {Dmitriev}}, \bibinfo {author} {\bibfnamefont {J.}~\bibnamefont {Ezold}},
  \bibinfo {author} {\bibfnamefont {K.}~\bibnamefont {Felker}}, \bibinfo
  {author} {\bibfnamefont {J.~M.}\ \bibnamefont {Gostic}}, \bibinfo {author}
  {\bibfnamefont {R.~K.}\ \bibnamefont {Grzywacz}}, \bibinfo {author}
  {\bibfnamefont {J.~H.}\ \bibnamefont {Hamilton}}, \bibinfo {author}
  {\bibfnamefont {R.~A.}\ \bibnamefont {Henderson}}, \bibinfo {author}
  {\bibfnamefont {M.~G.}\ \bibnamefont {Itkis}}, \bibinfo {author}
  {\bibfnamefont {K.}~\bibnamefont {Miernik}}, \bibinfo {author} {\bibfnamefont
  {D.}~\bibnamefont {Miller}}, \bibinfo {author} {\bibfnamefont {K.~J.}\
  \bibnamefont {Moody}}, \bibinfo {author} {\bibfnamefont {A.~N.}\ \bibnamefont
  {Polyakov}}, \bibinfo {author} {\bibfnamefont {A.~V.}\ \bibnamefont
  {Ramayya}}, \bibinfo {author} {\bibfnamefont {J.~B.}\ \bibnamefont
  {Roberto}}, \bibinfo {author} {\bibfnamefont {M.~A.}\ \bibnamefont
  {Ryabinin}}, \bibinfo {author} {\bibfnamefont {K.~P.}\ \bibnamefont
  {Rykaczewski}}, \bibinfo {author} {\bibfnamefont {R.~N.}\ \bibnamefont
  {Sagaidak}}, \bibinfo {author} {\bibfnamefont {D.~A.}\ \bibnamefont
  {Shaughnessy}}, \bibinfo {author} {\bibfnamefont {I.~V.}\ \bibnamefont
  {Shirokovsky}}, \bibinfo {author} {\bibfnamefont {M.~V.}\ \bibnamefont
  {Shumeiko}}, \bibinfo {author} {\bibfnamefont {M.~A.}\ \bibnamefont
  {Stoyer}}, \bibinfo {author} {\bibfnamefont {N.~J.}\ \bibnamefont {Stoyer}},
  \bibinfo {author} {\bibfnamefont {V.~G.}\ \bibnamefont {Subbotin}}, \bibinfo
  {author} {\bibfnamefont {A.~M.}\ \bibnamefont {Sukhov}}, \bibinfo {author}
  {\bibfnamefont {{\relax Yu. S}.}~\bibnamefont {Tsyganov}}, \bibinfo {author}
  {\bibfnamefont {V.~K.}\ \bibnamefont {Utyonkov}}, \bibinfo {author}
  {\bibfnamefont {A.~A.}\ \bibnamefont {Voinov}}, \ and\ \bibinfo {author}
  {\bibfnamefont {G.~K.}\ \bibnamefont {Vostokin}},\ }\bibfield  {title}
  {\enquote {\bibinfo {title} {{E}xperimental studies of the ${}^{249}${B}k +
  ${}^{48}${C}a reaction including decay properties and excitation function for
  isotopes of element 117, and discovery of the new isotope ${}^{277}${M}t},}\
  }\href {\doibase 10.1103/PhysRevC.87.054621} {\bibfield  {journal} {\bibinfo
  {journal} {Phys. Rev. C}\ }\textbf {\bibinfo {volume} {87}},\ \bibinfo
  {pages} {054621} (\bibinfo {year} {2013})}\BibitemShut {NoStop}%
\bibitem [{\citenamefont {Khuyagbaatar}\ \emph {et~al.}(2014)\citenamefont
  {Khuyagbaatar}, \citenamefont {Yakushev}, \citenamefont {D\"ullmann},
  \citenamefont {Ackermann}, \citenamefont {Andersson}, \citenamefont {Asai},
  \citenamefont {Block}, \citenamefont {Boll}, \citenamefont {Brand},
  \citenamefont {Cox}, \citenamefont {Dasgupta}, \citenamefont {Derkx},
  \citenamefont {{Di Nitto}}, \citenamefont {Eberhardt}, \citenamefont {Even},
  \citenamefont {Evers}, \citenamefont {Fahlander}, \citenamefont {Forsberg},
  \citenamefont {Gates}, \citenamefont {Gharibyan}, \citenamefont {Golubev},
  \citenamefont {Gregorich}, \citenamefont {Hamilton}, \citenamefont
  {Hartmann}, \citenamefont {Herzberg}, \citenamefont {He\ss{}berger},
  \citenamefont {Hinde}, \citenamefont {Hoffmann}, \citenamefont {Hollinger},
  \citenamefont {H\"ubner}, \citenamefont {J\"ager}, \citenamefont {Kindler},
  \citenamefont {Kratz}, \citenamefont {Krier}, \citenamefont {Kurz},
  \citenamefont {Laatiaoui}, \citenamefont {Lahiri}, \citenamefont {Lang},
  \citenamefont {Lommel}, \citenamefont {Maiti}, \citenamefont {Miernik},
  \citenamefont {Minami}, \citenamefont {Mistry}, \citenamefont {Mokry},
  \citenamefont {Nitsche}, \citenamefont {Omtvedt}, \citenamefont {Pang},
  \citenamefont {Papadakis}, \citenamefont {Renisch}, \citenamefont {Roberto},
  \citenamefont {Rudolph}, \citenamefont {Runke}, \citenamefont {Rykaczewski},
  \citenamefont {Sarmiento}, \citenamefont {Sch\"adel}, \citenamefont
  {Schausten}, \citenamefont {Semchenkov}, \citenamefont {Shaughnessy},
  \citenamefont {Steinegger}, \citenamefont {Steiner}, \citenamefont
  {Tereshatov}, \citenamefont {Th\"orle-Pospiech}, \citenamefont {Tinschert},
  \citenamefont {Torres De~Heidenreich}, \citenamefont {Trautmann},
  \citenamefont {T\"urler}, \citenamefont {Uusitalo}, \citenamefont {Ward},
  \citenamefont {Wegrzecki}, \citenamefont {Wiehl}, \citenamefont {Van~Cleve},\
  and\ \citenamefont {Yakusheva}}]{khuyagbaatar2014}%
  \BibitemOpen
  \bibfield  {author} {\bibinfo {author} {\bibfnamefont {J.}~\bibnamefont
  {Khuyagbaatar}}, \bibinfo {author} {\bibfnamefont {A.}~\bibnamefont
  {Yakushev}}, \bibinfo {author} {\bibfnamefont {{\relax Ch. E}.}~\bibnamefont
  {D\"ullmann}}, \bibinfo {author} {\bibfnamefont {D.}~\bibnamefont
  {Ackermann}}, \bibinfo {author} {\bibfnamefont {L.-L.}\ \bibnamefont
  {Andersson}}, \bibinfo {author} {\bibfnamefont {M.}~\bibnamefont {Asai}},
  \bibinfo {author} {\bibfnamefont {M.}~\bibnamefont {Block}}, \bibinfo
  {author} {\bibfnamefont {R.~A.}\ \bibnamefont {Boll}}, \bibinfo {author}
  {\bibfnamefont {H.}~\bibnamefont {Brand}}, \bibinfo {author} {\bibfnamefont
  {D.~M.}\ \bibnamefont {Cox}}, \bibinfo {author} {\bibfnamefont
  {M.}~\bibnamefont {Dasgupta}}, \bibinfo {author} {\bibfnamefont
  {X.}~\bibnamefont {Derkx}}, \bibinfo {author} {\bibfnamefont
  {A.}~\bibnamefont {{Di Nitto}}}, \bibinfo {author} {\bibfnamefont
  {K.}~\bibnamefont {Eberhardt}}, \bibinfo {author} {\bibfnamefont
  {J.}~\bibnamefont {Even}}, \bibinfo {author} {\bibfnamefont {M.}~\bibnamefont
  {Evers}}, \bibinfo {author} {\bibfnamefont {C.}~\bibnamefont {Fahlander}},
  \bibinfo {author} {\bibfnamefont {U.}~\bibnamefont {Forsberg}}, \bibinfo
  {author} {\bibfnamefont {J.~M.}\ \bibnamefont {Gates}}, \bibinfo {author}
  {\bibfnamefont {N.}~\bibnamefont {Gharibyan}}, \bibinfo {author}
  {\bibfnamefont {P.}~\bibnamefont {Golubev}}, \bibinfo {author} {\bibfnamefont
  {K.~E.}\ \bibnamefont {Gregorich}}, \bibinfo {author} {\bibfnamefont {J.~H.}\
  \bibnamefont {Hamilton}}, \bibinfo {author} {\bibfnamefont {W.}~\bibnamefont
  {Hartmann}}, \bibinfo {author} {\bibfnamefont {R.-D.}\ \bibnamefont
  {Herzberg}}, \bibinfo {author} {\bibfnamefont {F.~P.}\ \bibnamefont
  {He\ss{}berger}}, \bibinfo {author} {\bibfnamefont {D.~J.}\ \bibnamefont
  {Hinde}}, \bibinfo {author} {\bibfnamefont {J.}~\bibnamefont {Hoffmann}},
  \bibinfo {author} {\bibfnamefont {R.}~\bibnamefont {Hollinger}}, \bibinfo
  {author} {\bibfnamefont {A.}~\bibnamefont {H\"ubner}}, \bibinfo {author}
  {\bibfnamefont {E.}~\bibnamefont {J\"ager}}, \bibinfo {author} {\bibfnamefont
  {B.}~\bibnamefont {Kindler}}, \bibinfo {author} {\bibfnamefont {J.~V.}\
  \bibnamefont {Kratz}}, \bibinfo {author} {\bibfnamefont {J.}~\bibnamefont
  {Krier}}, \bibinfo {author} {\bibfnamefont {N.}~\bibnamefont {Kurz}},
  \bibinfo {author} {\bibfnamefont {M.}~\bibnamefont {Laatiaoui}}, \bibinfo
  {author} {\bibfnamefont {S.}~\bibnamefont {Lahiri}}, \bibinfo {author}
  {\bibfnamefont {R.}~\bibnamefont {Lang}}, \bibinfo {author} {\bibfnamefont
  {B.}~\bibnamefont {Lommel}}, \bibinfo {author} {\bibfnamefont
  {M.}~\bibnamefont {Maiti}}, \bibinfo {author} {\bibfnamefont
  {K.}~\bibnamefont {Miernik}}, \bibinfo {author} {\bibfnamefont
  {S.}~\bibnamefont {Minami}}, \bibinfo {author} {\bibfnamefont
  {A.}~\bibnamefont {Mistry}}, \bibinfo {author} {\bibfnamefont
  {C.}~\bibnamefont {Mokry}}, \bibinfo {author} {\bibfnamefont
  {H.}~\bibnamefont {Nitsche}}, \bibinfo {author} {\bibfnamefont {J.~P.}\
  \bibnamefont {Omtvedt}}, \bibinfo {author} {\bibfnamefont {G.~K.}\
  \bibnamefont {Pang}}, \bibinfo {author} {\bibfnamefont {P.}~\bibnamefont
  {Papadakis}}, \bibinfo {author} {\bibfnamefont {D.}~\bibnamefont {Renisch}},
  \bibinfo {author} {\bibfnamefont {J.}~\bibnamefont {Roberto}}, \bibinfo
  {author} {\bibfnamefont {D.}~\bibnamefont {Rudolph}}, \bibinfo {author}
  {\bibfnamefont {J.}~\bibnamefont {Runke}}, \bibinfo {author} {\bibfnamefont
  {K.~P.}\ \bibnamefont {Rykaczewski}}, \bibinfo {author} {\bibfnamefont
  {L.~G.}\ \bibnamefont {Sarmiento}}, \bibinfo {author} {\bibfnamefont
  {M.}~\bibnamefont {Sch\"adel}}, \bibinfo {author} {\bibfnamefont
  {B.}~\bibnamefont {Schausten}}, \bibinfo {author} {\bibfnamefont
  {A.}~\bibnamefont {Semchenkov}}, \bibinfo {author} {\bibfnamefont {D.~A.}\
  \bibnamefont {Shaughnessy}}, \bibinfo {author} {\bibfnamefont
  {P.}~\bibnamefont {Steinegger}}, \bibinfo {author} {\bibfnamefont
  {J.}~\bibnamefont {Steiner}}, \bibinfo {author} {\bibfnamefont {E.~E.}\
  \bibnamefont {Tereshatov}}, \bibinfo {author} {\bibfnamefont
  {P.}~\bibnamefont {Th\"orle-Pospiech}}, \bibinfo {author} {\bibfnamefont
  {K.}~\bibnamefont {Tinschert}}, \bibinfo {author} {\bibfnamefont
  {T.}~\bibnamefont {Torres De~Heidenreich}}, \bibinfo {author} {\bibfnamefont
  {N.}~\bibnamefont {Trautmann}}, \bibinfo {author} {\bibfnamefont
  {A.}~\bibnamefont {T\"urler}}, \bibinfo {author} {\bibfnamefont
  {J.}~\bibnamefont {Uusitalo}}, \bibinfo {author} {\bibfnamefont {D.~E.}\
  \bibnamefont {Ward}}, \bibinfo {author} {\bibfnamefont {M.}~\bibnamefont
  {Wegrzecki}}, \bibinfo {author} {\bibfnamefont {N.}~\bibnamefont {Wiehl}},
  \bibinfo {author} {\bibfnamefont {S.~M.}\ \bibnamefont {Van~Cleve}}, \ and\
  \bibinfo {author} {\bibfnamefont {V.}~\bibnamefont {Yakusheva}},\ }\bibfield
  {title} {\enquote {\bibinfo {title} {$^{48}\mathrm{Ca}+{}^{249}\mathrm{Bk}$
  {F}usion {R}eaction {L}eading to {E}lement ${Z}=117$: {L}ong-{L}ived
  $\alpha$-{D}ecaying $^{270}\mathrm{Db}$ and {D}iscovery of
  $^{266}\mathrm{Lr}$},}\ }\href {\doibase 10.1103/PhysRevLett.112.172501}
  {\bibfield  {journal} {\bibinfo  {journal} {Phys. Rev. Lett.}\ }\textbf
  {\bibinfo {volume} {112}},\ \bibinfo {pages} {172501} (\bibinfo {year}
  {2014})}\BibitemShut {NoStop}%
\bibitem [{\citenamefont {Negele}(1982)}]{negele1982}%
  \BibitemOpen
  \bibfield  {author} {\bibinfo {author} {\bibfnamefont {J.~W.}\ \bibnamefont
  {Negele}},\ }\bibfield  {title} {\enquote {\bibinfo {title} {{T}he mean-field
  theory of nuclear-structure and dynamics},}\ }\href {\doibase
  10.1103/RevModPhys.54.913} {\bibfield  {journal} {\bibinfo  {journal} {Rev.
  Mod. Phys.}\ }\textbf {\bibinfo {volume} {54}},\ \bibinfo {pages} {913--1015}
  (\bibinfo {year} {1982})}\BibitemShut {NoStop}%
\bibitem [{\citenamefont {Bonche}\ \emph {et~al.}(1978)\citenamefont {Bonche},
  \citenamefont {Grammaticos},\ and\ \citenamefont {Koonin}}]{bonche1978}%
  \BibitemOpen
  \bibfield  {author} {\bibinfo {author} {\bibfnamefont {P.}~\bibnamefont
  {Bonche}}, \bibinfo {author} {\bibfnamefont {B.}~\bibnamefont {Grammaticos}},
  \ and\ \bibinfo {author} {\bibfnamefont {S.}~\bibnamefont {Koonin}},\
  }\bibfield  {title} {\enquote {\bibinfo {title} {{T}hree-dimensional
  time-dependent {H}artree-{F}ock calculations of $^{16}${O}+$^{16}${O} and
  $^{40}${C}a+$^{40}${C}a fusion cross sections},}\ }\href {\doibase
  10.1103/PhysRevC.17.1700} {\bibfield  {journal} {\bibinfo  {journal} {Phys.
  Rev. C}\ }\textbf {\bibinfo {volume} {17}},\ \bibinfo {pages} {1700--1705}
  (\bibinfo {year} {1978})}\BibitemShut {NoStop}%
\bibitem [{\citenamefont {Flocard}\ \emph {et~al.}(1978)\citenamefont
  {Flocard}, \citenamefont {Koonin},\ and\ \citenamefont
  {Weiss}}]{flocard1978}%
  \BibitemOpen
  \bibfield  {author} {\bibinfo {author} {\bibfnamefont {H.}~\bibnamefont
  {Flocard}}, \bibinfo {author} {\bibfnamefont {S.~E.}\ \bibnamefont {Koonin}},
  \ and\ \bibinfo {author} {\bibfnamefont {M.~S.}\ \bibnamefont {Weiss}},\
  }\bibfield  {title} {\enquote {\bibinfo {title} {{T}hree-dimensional
  time-dependent {H}artree-{F}ock calculations: {A}pplication to
  $^{16}${O}+$^{16}${O} collisions},}\ }\href {\doibase
  10.1103/PhysRevC.17.1682} {\bibfield  {journal} {\bibinfo  {journal} {Phys.
  Rev. C}\ }\textbf {\bibinfo {volume} {17}},\ \bibinfo {pages} {1682--1699}
  (\bibinfo {year} {1978})}\BibitemShut {NoStop}%
\bibitem [{\citenamefont {Simenel}\ \emph {et~al.}(2001)\citenamefont
  {Simenel}, \citenamefont {Chomaz},\ and\ \citenamefont {{de
  France}}}]{simenel2001}%
  \BibitemOpen
  \bibfield  {author} {\bibinfo {author} {\bibfnamefont {C.}~\bibnamefont
  {Simenel}}, \bibinfo {author} {\bibfnamefont {Ph.}\ \bibnamefont {Chomaz}}, \
  and\ \bibinfo {author} {\bibfnamefont {G.}~\bibnamefont {{de France}}},\
  }\bibfield  {title} {\enquote {\bibinfo {title} {{Q}uantum {C}alculation of
  the {D}ipole {E}xcitation in {F}usion {R}eactions},}\ }\href {\doibase
  10.1103/PhysRevLett.86.2971} {\bibfield  {journal} {\bibinfo  {journal}
  {Phys. Rev. Lett.}\ }\textbf {\bibinfo {volume} {86}},\ \bibinfo {pages}
  {2971--2974} (\bibinfo {year} {2001})}\BibitemShut {NoStop}%
\bibitem [{\citenamefont {Umar}\ \emph {et~al.}(2008)\citenamefont {Umar},
  \citenamefont {Oberacker},\ and\ \citenamefont {Maruhn}}]{umar2008a}%
  \BibitemOpen
  \bibfield  {author} {\bibinfo {author} {\bibfnamefont {A.~S.}\ \bibnamefont
  {Umar}}, \bibinfo {author} {\bibfnamefont {V.~E.}\ \bibnamefont {Oberacker}},
  \ and\ \bibinfo {author} {\bibfnamefont {J.~A.}\ \bibnamefont {Maruhn}},\
  }\bibfield  {title} {\enquote {\bibinfo {title} {{N}eutron transfer dynamics
  and doorway to fusion in time-dependent {H}artree-{F}ock theory},}\ }\href
  {\doibase 10.1140/epja/i2008-10614-6} {\bibfield  {journal} {\bibinfo
  {journal} {Eur. Phys. J. A}\ }\textbf {\bibinfo {volume} {37}},\ \bibinfo
  {pages} {245--250} (\bibinfo {year} {2008})}\BibitemShut {NoStop}%
\bibitem [{\citenamefont {Umar}\ and\ \citenamefont
  {Oberacker}(2006{\natexlab{a}})}]{umar2006d}%
  \BibitemOpen
  \bibfield  {author} {\bibinfo {author} {\bibfnamefont {A.~S.}\ \bibnamefont
  {Umar}}\ and\ \bibinfo {author} {\bibfnamefont {V.~E.}\ \bibnamefont
  {Oberacker}},\ }\bibfield  {title} {\enquote {\bibinfo {title} {{T}ime
  dependent {H}artree-{F}ock fusion calculations for spherical, deformed
  systems},}\ }\href {\doibase 10.1103/PhysRevC.74.024606} {\bibfield
  {journal} {\bibinfo  {journal} {Phys. Rev. C}\ }\textbf {\bibinfo {volume}
  {74}},\ \bibinfo {pages} {024606} (\bibinfo {year}
  {2006}{\natexlab{a}})}\BibitemShut {NoStop}%
\bibitem [{\citenamefont {{Kouhei Washiyama}}\ and\ \citenamefont {{Denis
  Lacroix}}(2008)}]{washiyama2008}%
  \BibitemOpen
  \bibfield  {author} {\bibinfo {author} {\bibnamefont {{Kouhei Washiyama}}}\
  and\ \bibinfo {author} {\bibnamefont {{Denis Lacroix}}},\ }\bibfield  {title}
  {\enquote {\bibinfo {title} {{E}nergy dependence of the nucleus-nucleus
  potential close to the {C}oulomb barrier},}\ }\href {\doibase
  10.1103/PhysRevC.78.024610} {\bibfield  {journal} {\bibinfo  {journal} {Phys.
  Rev. C}\ }\textbf {\bibinfo {volume} {78}},\ \bibinfo {pages} {024610}
  (\bibinfo {year} {2008})}\BibitemShut {NoStop}%
\bibitem [{\citenamefont {Umar}\ \emph {et~al.}(2010)\citenamefont {Umar},
  \citenamefont {Oberacker}, \citenamefont {Maruhn},\ and\ \citenamefont
  {Reinhard}}]{umar2010a}%
  \BibitemOpen
  \bibfield  {author} {\bibinfo {author} {\bibfnamefont {A.~S.}\ \bibnamefont
  {Umar}}, \bibinfo {author} {\bibfnamefont {V.~E.}\ \bibnamefont {Oberacker}},
  \bibinfo {author} {\bibfnamefont {J.~A.}\ \bibnamefont {Maruhn}}, \ and\
  \bibinfo {author} {\bibfnamefont {P.-G.}\ \bibnamefont {Reinhard}},\
  }\bibfield  {title} {\enquote {\bibinfo {title} {{E}ntrance channel dynamics
  of hot and cold fusion reactions leading to superheavy elements},}\ }\href
  {\doibase 10.1103/PhysRevC.81.064607} {\bibfield  {journal} {\bibinfo
  {journal} {Phys. Rev. C}\ }\textbf {\bibinfo {volume} {81}},\ \bibinfo
  {pages} {064607} (\bibinfo {year} {2010})}\BibitemShut {NoStop}%
\bibitem [{\citenamefont {{Lu Guo}}\ and\ \citenamefont {{Takashi
  Nakatsukasa}}(2012)}]{guo2012}%
  \BibitemOpen
  \bibfield  {author} {\bibinfo {author} {\bibnamefont {{Lu Guo}}}\ and\
  \bibinfo {author} {\bibnamefont {{Takashi Nakatsukasa}}},\ }\bibfield
  {title} {\enquote {\bibinfo {title} {{T}ime-dependent {H}artree-{F}ock
  studies of the dynamical fusion threshold},}\ }\href {\doibase
  10.1051/epjconf/20123809003} {\bibfield  {journal} {\bibinfo  {journal} {EPJ
  Web Conf.}\ }\textbf {\bibinfo {volume} {38}},\ \bibinfo {pages} {09003}
  (\bibinfo {year} {2012})}\BibitemShut {NoStop}%
\bibitem [{\citenamefont {Keser}\ \emph {et~al.}(2012)\citenamefont {Keser},
  \citenamefont {Umar},\ and\ \citenamefont {Oberacker}}]{keser2012}%
  \BibitemOpen
  \bibfield  {author} {\bibinfo {author} {\bibfnamefont {R.}~\bibnamefont
  {Keser}}, \bibinfo {author} {\bibfnamefont {A.~S.}\ \bibnamefont {Umar}}, \
  and\ \bibinfo {author} {\bibfnamefont {V.~E.}\ \bibnamefont {Oberacker}},\
  }\bibfield  {title} {\enquote {\bibinfo {title} {{M}icroscopic study of
  {C}a$+${C}a fusion},}\ }\href {\doibase 10.1103/PhysRevC.85.044606}
  {\bibfield  {journal} {\bibinfo  {journal} {Phys. Rev. C}\ }\textbf {\bibinfo
  {volume} {85}},\ \bibinfo {pages} {044606} (\bibinfo {year}
  {2012})}\BibitemShut {NoStop}%
\bibitem [{\citenamefont {Simenel}\ \emph
  {et~al.}(2013{\natexlab{a}})\citenamefont {Simenel}, \citenamefont {Keser},
  \citenamefont {Umar},\ and\ \citenamefont {Oberacker}}]{simenel2013a}%
  \BibitemOpen
  \bibfield  {author} {\bibinfo {author} {\bibfnamefont {C.}~\bibnamefont
  {Simenel}}, \bibinfo {author} {\bibfnamefont {R.}~\bibnamefont {Keser}},
  \bibinfo {author} {\bibfnamefont {A.~S.}\ \bibnamefont {Umar}}, \ and\
  \bibinfo {author} {\bibfnamefont {V.~E.}\ \bibnamefont {Oberacker}},\
  }\bibfield  {title} {\enquote {\bibinfo {title} {Microscopic study of
  ${}^{16}\mathrm{O}+{}^{16}\mathrm{O}$ fusion},}\ }\href {\doibase
  10.1103/PhysRevC.88.024617} {\bibfield  {journal} {\bibinfo  {journal} {Phys.
  Rev. C}\ }\textbf {\bibinfo {volume} {88}},\ \bibinfo {pages} {024617}
  (\bibinfo {year} {2013}{\natexlab{a}})}\BibitemShut {NoStop}%
\bibitem [{\citenamefont {Oberacker}\ \emph {et~al.}(2012)\citenamefont
  {Oberacker}, \citenamefont {Umar}, \citenamefont {Maruhn},\ and\
  \citenamefont {Reinhard}}]{oberacker2012}%
  \BibitemOpen
  \bibfield  {author} {\bibinfo {author} {\bibfnamefont {V.~E.}\ \bibnamefont
  {Oberacker}}, \bibinfo {author} {\bibfnamefont {A.~S.}\ \bibnamefont {Umar}},
  \bibinfo {author} {\bibfnamefont {J.~A.}\ \bibnamefont {Maruhn}}, \ and\
  \bibinfo {author} {\bibfnamefont {P.-G.}\ \bibnamefont {Reinhard}},\
  }\bibfield  {title} {\enquote {\bibinfo {title} {{D}ynamic microscopic study
  of pre-equilibrium giant resonance excitation and fusion in the reactions
  ${}^{132}${S}n $+$ ${}^{48}${C}a and ${}^{124}${S}n $+$ ${}^{40}${C}a},}\
  }\href {\doibase 10.1103/PhysRevC.85.034609} {\bibfield  {journal} {\bibinfo
  {journal} {Phys. Rev. C}\ }\textbf {\bibinfo {volume} {85}},\ \bibinfo
  {pages} {034609} (\bibinfo {year} {2012})}\BibitemShut {NoStop}%
\bibitem [{\citenamefont {Oberacker}\ \emph {et~al.}(2010)\citenamefont
  {Oberacker}, \citenamefont {Umar}, \citenamefont {Maruhn},\ and\
  \citenamefont {Reinhard}}]{oberacker2010}%
  \BibitemOpen
  \bibfield  {author} {\bibinfo {author} {\bibfnamefont {V.~E.}\ \bibnamefont
  {Oberacker}}, \bibinfo {author} {\bibfnamefont {A.~S.}\ \bibnamefont {Umar}},
  \bibinfo {author} {\bibfnamefont {J.~A.}\ \bibnamefont {Maruhn}}, \ and\
  \bibinfo {author} {\bibfnamefont {{P.--G.}}\ \bibnamefont {Reinhard}},\
  }\bibfield  {title} {\enquote {\bibinfo {title} {{M}icroscopic study of the
  $^{132,124}\mathrm{Sn}+{}^{96}\mathrm{Zr}$ reactions: {D}ynamic excitation
  energy, energy-dependent heavy-ion potential, and capture cross section},}\
  }\href {\doibase 10.1103/PhysRevC.82.034603} {\bibfield  {journal} {\bibinfo
  {journal} {Phys. Rev. C}\ }\textbf {\bibinfo {volume} {82}},\ \bibinfo
  {pages} {034603} (\bibinfo {year} {2010})}\BibitemShut {NoStop}%
\bibitem [{\citenamefont {Umar}\ \emph {et~al.}(2012)\citenamefont {Umar},
  \citenamefont {Oberacker},\ and\ \citenamefont {Horowitz}}]{umar2012a}%
  \BibitemOpen
  \bibfield  {author} {\bibinfo {author} {\bibfnamefont {A.~S.}\ \bibnamefont
  {Umar}}, \bibinfo {author} {\bibfnamefont {V.~E.}\ \bibnamefont {Oberacker}},
  \ and\ \bibinfo {author} {\bibfnamefont {C.~J.}\ \bibnamefont {Horowitz}},\
  }\bibfield  {title} {\enquote {\bibinfo {title} {{M}icroscopic sub-barrier
  fusion calculations for the neutron star crust},}\ }\href {\doibase
  10.1103/PhysRevC.85.055801} {\bibfield  {journal} {\bibinfo  {journal} {Phys.
  Rev. C}\ }\textbf {\bibinfo {volume} {85}},\ \bibinfo {pages} {055801}
  (\bibinfo {year} {2012})}\BibitemShut {NoStop}%
\bibitem [{\citenamefont {Simenel}\ \emph
  {et~al.}(2013{\natexlab{b}})\citenamefont {Simenel}, \citenamefont
  {Dasgupta}, \citenamefont {Hinde},\ and\ \citenamefont
  {Williams}}]{simenel2013b}%
  \BibitemOpen
  \bibfield  {author} {\bibinfo {author} {\bibfnamefont {C.}~\bibnamefont
  {Simenel}}, \bibinfo {author} {\bibfnamefont {M.}~\bibnamefont {Dasgupta}},
  \bibinfo {author} {\bibfnamefont {D.~J.}\ \bibnamefont {Hinde}}, \ and\
  \bibinfo {author} {\bibfnamefont {E.}~\bibnamefont {Williams}},\ }\bibfield
  {title} {\enquote {\bibinfo {title} {{M}icroscopic approach to
  coupled-channels effects on fusion},}\ }\href {\doibase
  10.1103/PhysRevC.88.064604} {\bibfield  {journal} {\bibinfo  {journal} {Phys.
  Rev. C}\ }\textbf {\bibinfo {volume} {88}},\ \bibinfo {pages} {064604}
  (\bibinfo {year} {2013}{\natexlab{b}})}\BibitemShut {NoStop}%
\bibitem [{\citenamefont {Umar}\ \emph {et~al.}(2014)\citenamefont {Umar},
  \citenamefont {Simenel},\ and\ \citenamefont {Oberacker}}]{umar2014a}%
  \BibitemOpen
  \bibfield  {author} {\bibinfo {author} {\bibfnamefont {A.~S.}\ \bibnamefont
  {Umar}}, \bibinfo {author} {\bibfnamefont {C.}~\bibnamefont {Simenel}}, \
  and\ \bibinfo {author} {\bibfnamefont {V.~E.}\ \bibnamefont {Oberacker}},\
  }\bibfield  {title} {\enquote {\bibinfo {title} {{E}nergy dependence of
  potential barriers and its effect on fusion cross sections},}\ }\href
  {\doibase 10.1103/PhysRevC.89.034611} {\bibfield  {journal} {\bibinfo
  {journal} {Phys. Rev. C}\ }\textbf {\bibinfo {volume} {89}},\ \bibinfo
  {pages} {034611} (\bibinfo {year} {2014})}\BibitemShut {NoStop}%
\bibitem [{\citenamefont {Jiang}\ \emph {et~al.}(2014)\citenamefont {Jiang},
  \citenamefont {Maruhn},\ and\ \citenamefont {Yan}}]{jiang2014}%
  \BibitemOpen
  \bibfield  {author} {\bibinfo {author} {\bibfnamefont {Xiang}\ \bibnamefont
  {Jiang}}, \bibinfo {author} {\bibfnamefont {Joachim~A.}\ \bibnamefont
  {Maruhn}}, \ and\ \bibinfo {author} {\bibfnamefont {Shiwei}\ \bibnamefont
  {Yan}},\ }\bibfield  {title} {\enquote {\bibinfo {title} {{M}icroscopic study
  of noncentral effects in heavy-ion fusion reactions with spherical nuclei},}\
  }\href {\doibase 10.1103/PhysRevC.90.064618} {\bibfield  {journal} {\bibinfo
  {journal} {Phys. Rev. C}\ }\textbf {\bibinfo {volume} {90}},\ \bibinfo
  {pages} {064618} (\bibinfo {year} {2014})}\BibitemShut {NoStop}%
\bibitem [{\citenamefont {Koonin}\ \emph {et~al.}(1977)\citenamefont {Koonin},
  \citenamefont {Davies}, \citenamefont {Maruhn-Rezwani}, \citenamefont
  {Feldmeier}, \citenamefont {Krieger},\ and\ \citenamefont
  {Negele}}]{koonin1977}%
  \BibitemOpen
  \bibfield  {author} {\bibinfo {author} {\bibfnamefont {S.~E.}\ \bibnamefont
  {Koonin}}, \bibinfo {author} {\bibfnamefont {K.~T.~R.}\ \bibnamefont
  {Davies}}, \bibinfo {author} {\bibfnamefont {V.}~\bibnamefont
  {Maruhn-Rezwani}}, \bibinfo {author} {\bibfnamefont {H.}~\bibnamefont
  {Feldmeier}}, \bibinfo {author} {\bibfnamefont {S.~J.}\ \bibnamefont
  {Krieger}}, \ and\ \bibinfo {author} {\bibfnamefont {J.~W.}\ \bibnamefont
  {Negele}},\ }\bibfield  {title} {\enquote {\bibinfo {title} {{T}ime-dependent
  {H}artree-{F}ock calculations for $^{16}${O} $+$ $^{16}${O} and $^{40}${C}a
  $+$ $^{40}${C}a reactions},}\ }\href {\doibase 10.1103/PhysRevC.15.1359}
  {\bibfield  {journal} {\bibinfo  {journal} {Phys. Rev. C}\ }\textbf {\bibinfo
  {volume} {15}},\ \bibinfo {pages} {1359--1374} (\bibinfo {year}
  {1977})}\BibitemShut {NoStop}%
\bibitem [{\citenamefont {Simenel}(2010)}]{simenel2010}%
  \BibitemOpen
  \bibfield  {author} {\bibinfo {author} {\bibfnamefont {C\'edric}\
  \bibnamefont {Simenel}},\ }\bibfield  {title} {\enquote {\bibinfo {title}
  {{P}article {T}ransfer {R}eactions with the {T}ime-{D}ependent
  {H}artree-{F}ock {T}heory {U}sing a {P}article {N}umber {P}rojection
  {T}echnique},}\ }\href {\doibase 10.1103/PhysRevLett.105.192701} {\bibfield
  {journal} {\bibinfo  {journal} {Phys. Rev. Lett.}\ }\textbf {\bibinfo
  {volume} {105}},\ \bibinfo {pages} {192701} (\bibinfo {year}
  {2010})}\BibitemShut {NoStop}%
\bibitem [{\citenamefont {Simenel}(2011)}]{simenel2011}%
  \BibitemOpen
  \bibfield  {author} {\bibinfo {author} {\bibfnamefont {C{\'e}dric}\
  \bibnamefont {Simenel}},\ }\bibfield  {title} {\enquote {\bibinfo {title}
  {{P}article-{N}umber {F}luctuations and {C}orrelations in {T}ransfer
  {R}eactions {O}btained {U}sing the {B}alian-{V}\'en\'eroni {V}ariational
  {P}rinciple},}\ }\href {\doibase 10.1103/PhysRevLett.106.112502} {\bibfield
  {journal} {\bibinfo  {journal} {Phys. Rev. Lett.}\ }\textbf {\bibinfo
  {volume} {106}},\ \bibinfo {pages} {112502} (\bibinfo {year}
  {2011})}\BibitemShut {NoStop}%
\bibitem [{\citenamefont {{Kazuyuki Sekizawa}}\ and\ \citenamefont {{Kazuhiro
  Yabana}}(2013)}]{sekizawa2013}%
  \BibitemOpen
  \bibfield  {author} {\bibinfo {author} {\bibnamefont {{Kazuyuki Sekizawa}}}\
  and\ \bibinfo {author} {\bibnamefont {{Kazuhiro Yabana}}},\ }\bibfield
  {title} {\enquote {\bibinfo {title} {{T}ime-dependent {H}artree-{F}ock
  calculations for multinucleon transfer processes in
  $^{40,48}${C}a+$^{124}${S}n, $^{40}${C}a+$^{208}${P}b, and
  $^{58}${N}i+$^{208}${P}b reactions},}\ }\href {\doibase
  10.1103/PhysRevC.88.014614} {\bibfield  {journal} {\bibinfo  {journal} {Phys.
  Rev. C}\ }\textbf {\bibinfo {volume} {88}},\ \bibinfo {pages} {014614}
  (\bibinfo {year} {2013})}\BibitemShut {NoStop}%
\bibitem [{\citenamefont {Scamps}\ and\ \citenamefont
  {Lacroix}(2013)}]{scamps2013a}%
  \BibitemOpen
  \bibfield  {author} {\bibinfo {author} {\bibfnamefont {Guillaume}\
  \bibnamefont {Scamps}}\ and\ \bibinfo {author} {\bibfnamefont {Denis}\
  \bibnamefont {Lacroix}},\ }\bibfield  {title} {\enquote {\bibinfo {title}
  {{E}ffect of pairing on one- and two-nucleon transfer below the {C}oulomb
  barrier: {A} time-dependent microscopic description},}\ }\href {\doibase
  10.1103/PhysRevC.87.014605} {\bibfield  {journal} {\bibinfo  {journal} {Phys.
  Rev. C}\ }\textbf {\bibinfo {volume} {87}},\ \bibinfo {pages} {014605}
  (\bibinfo {year} {2013})}\BibitemShut {NoStop}%
\bibitem [{\citenamefont {Sekizawa}\ and\ \citenamefont
  {Yabana}(2014)}]{sekizawa2014}%
  \BibitemOpen
  \bibfield  {author} {\bibinfo {author} {\bibfnamefont {Kazuyuki}\
  \bibnamefont {Sekizawa}}\ and\ \bibinfo {author} {\bibfnamefont {Kazuhiro}\
  \bibnamefont {Yabana}},\ }\bibfield  {title} {\enquote {\bibinfo {title}
  {{P}article-number projection method in time-dependent {H}artree-{F}ock
  theory: {P}roperties of reaction products},}\ }\href {\doibase
  10.1103/PhysRevC.90.064614} {\bibfield  {journal} {\bibinfo  {journal} {Phys.
  Rev. C}\ }\textbf {\bibinfo {volume} {90}},\ \bibinfo {pages} {064614}
  (\bibinfo {year} {2014})}\BibitemShut {NoStop}%
\bibitem [{\citenamefont {Bourgin}\ \emph {et~al.}(2016)\citenamefont
  {Bourgin}, \citenamefont {Simenel}, \citenamefont {Courtin},\ and\
  \citenamefont {Haas}}]{bourgin2016}%
  \BibitemOpen
  \bibfield  {author} {\bibinfo {author} {\bibfnamefont {D.}~\bibnamefont
  {Bourgin}}, \bibinfo {author} {\bibfnamefont {C.}~\bibnamefont {Simenel}},
  \bibinfo {author} {\bibfnamefont {S.}~\bibnamefont {Courtin}}, \ and\
  \bibinfo {author} {\bibfnamefont {F.}~\bibnamefont {Haas}},\ }\bibfield
  {title} {\enquote {\bibinfo {title} {{M}icroscopic study of $^{40}${C}a $+$
  $^{58,64}${N}i fusion reactions},}\ }\href {\doibase
  10.1103/PhysRevC.93.034604} {\bibfield  {journal} {\bibinfo  {journal} {Phys.
  Rev. C}\ }\textbf {\bibinfo {volume} {93}},\ \bibinfo {pages} {034604}
  (\bibinfo {year} {2016})}\BibitemShut {NoStop}%
\bibitem [{\citenamefont {Umar}\ \emph {et~al.}(2017)\citenamefont {Umar},
  \citenamefont {Simenel},\ and\ \citenamefont {Ye}}]{umar2017}%
  \BibitemOpen
  \bibfield  {author} {\bibinfo {author} {\bibfnamefont {A.~S.}\ \bibnamefont
  {Umar}}, \bibinfo {author} {\bibfnamefont {C.}~\bibnamefont {Simenel}}, \
  and\ \bibinfo {author} {\bibfnamefont {W.}~\bibnamefont {Ye}},\ }\bibfield
  {title} {\enquote {\bibinfo {title} {Transport properties of isospin
  asymmetric nuclear matter using the time-dependent {H}artree--{F}ock
  method},}\ }\href {\doibase 10.1103/PhysRevC.96.024625} {\bibfield  {journal}
  {\bibinfo  {journal} {Phys. Rev. C}\ }\textbf {\bibinfo {volume} {96}},\
  \bibinfo {pages} {024625} (\bibinfo {year} {2017})}\BibitemShut {NoStop}%
\bibitem [{\citenamefont {Simenel}\ and\ \citenamefont
  {Umar}(2014)}]{simenel2014a}%
  \BibitemOpen
  \bibfield  {author} {\bibinfo {author} {\bibfnamefont {C.}~\bibnamefont
  {Simenel}}\ and\ \bibinfo {author} {\bibfnamefont {A.~S.}\ \bibnamefont
  {Umar}},\ }\bibfield  {title} {\enquote {\bibinfo {title} {{F}ormation and
  dynamics of fission fragments},}\ }\href {\doibase
  10.1103/PhysRevC.89.031601} {\bibfield  {journal} {\bibinfo  {journal} {Phys.
  Rev. C}\ }\textbf {\bibinfo {volume} {89}},\ \bibinfo {pages} {031601(R)}
  (\bibinfo {year} {2014})}\BibitemShut {NoStop}%
\bibitem [{\citenamefont {Umar}\ and\ \citenamefont
  {Oberacker}(2015)}]{umar2015c}%
  \BibitemOpen
  \bibfield  {author} {\bibinfo {author} {\bibfnamefont {A.~S.}\ \bibnamefont
  {Umar}}\ and\ \bibinfo {author} {\bibfnamefont {V.~E.}\ \bibnamefont
  {Oberacker}},\ }\bibfield  {title} {\enquote {\bibinfo {title}
  {{T}ime-dependent {HF} approach to {SHE} dynamics},}\ }\href {\doibase
  10.1016/j.nuclphysa.2015.02.011} {\bibfield  {journal} {\bibinfo  {journal}
  {Nucl. Phys. A}\ }\textbf {\bibinfo {volume} {944}},\ \bibinfo {pages}
  {238--256} (\bibinfo {year} {2015})}\BibitemShut {NoStop}%
\bibitem [{\citenamefont {Scamps}\ \emph {et~al.}(2015)\citenamefont {Scamps},
  \citenamefont {Simenel},\ and\ \citenamefont {Lacroix}}]{scamps2015a}%
  \BibitemOpen
  \bibfield  {author} {\bibinfo {author} {\bibfnamefont {Guillaume}\
  \bibnamefont {Scamps}}, \bibinfo {author} {\bibfnamefont {C\'edric}\
  \bibnamefont {Simenel}}, \ and\ \bibinfo {author} {\bibfnamefont {Denis}\
  \bibnamefont {Lacroix}},\ }\bibfield  {title} {\enquote {\bibinfo {title}
  {{S}uperfluid dynamics of $^{258}\mathrm{Fm}$ fission},}\ }\href {\doibase
  10.1103/PhysRevC.92.011602} {\bibfield  {journal} {\bibinfo  {journal} {Phys.
  Rev. C}\ }\textbf {\bibinfo {volume} {92}},\ \bibinfo {pages} {011602}
  (\bibinfo {year} {2015})}\BibitemShut {NoStop}%
\bibitem [{\citenamefont {Goddard}\ \emph {et~al.}(2015)\citenamefont
  {Goddard}, \citenamefont {Stevenson},\ and\ \citenamefont
  {Rios}}]{goddard2015}%
  \BibitemOpen
  \bibfield  {author} {\bibinfo {author} {\bibfnamefont {P.~M.}\ \bibnamefont
  {Goddard}}, \bibinfo {author} {\bibfnamefont {P.~D.}\ \bibnamefont
  {Stevenson}}, \ and\ \bibinfo {author} {\bibfnamefont {A.}~\bibnamefont
  {Rios}},\ }\bibfield  {title} {\enquote {\bibinfo {title} {{F}ission dynamics
  within time-dependent {H}artree-{F}ock: deformation-induced fission},}\
  }\href {\doibase 10.1103/PhysRevC.92.054610} {\bibfield  {journal} {\bibinfo
  {journal} {Phys. Rev. C}\ }\textbf {\bibinfo {volume} {92}},\ \bibinfo
  {pages} {054610} (\bibinfo {year} {2015})}\BibitemShut {NoStop}%
\bibitem [{\citenamefont {{Aurel Bulgac}}\ \emph {et~al.}(2016)\citenamefont
  {{Aurel Bulgac}}, \citenamefont {{Piotr Magierski}}, \citenamefont {{Kenneth
  J. Roche}},\ and\ \citenamefont {{Ionel Stetcu}}}]{bulgac2016}%
  \BibitemOpen
  \bibfield  {author} {\bibinfo {author} {\bibnamefont {{Aurel Bulgac}}},
  \bibinfo {author} {\bibnamefont {{Piotr Magierski}}}, \bibinfo {author}
  {\bibnamefont {{Kenneth J. Roche}}}, \ and\ \bibinfo {author} {\bibnamefont
  {{Ionel Stetcu}}},\ }\bibfield  {title} {\enquote {\bibinfo {title}
  {{I}nduced {F}ission of $^{240}${P}u within a {R}eal-{T}ime {M}icroscopic
  {F}ramework},}\ }\href {\doibase 10.1103/physrevlett.116.122504} {\bibfield
  {journal} {\bibinfo  {journal} {Phys. Rev. Lett.}\ }\textbf {\bibinfo
  {volume} {116}},\ \bibinfo {pages} {122504} (\bibinfo {year}
  {2016})}\BibitemShut {NoStop}%
\bibitem [{\citenamefont {Bottcher}\ \emph {et~al.}(1989)\citenamefont
  {Bottcher}, \citenamefont {Strayer}, \citenamefont {Umar},\ and\
  \citenamefont {Reinhard}}]{bottcher1989}%
  \BibitemOpen
  \bibfield  {author} {\bibinfo {author} {\bibfnamefont {C.}~\bibnamefont
  {Bottcher}}, \bibinfo {author} {\bibfnamefont {M.~R.}\ \bibnamefont
  {Strayer}}, \bibinfo {author} {\bibfnamefont {A.~S.}\ \bibnamefont {Umar}}, \
  and\ \bibinfo {author} {\bibfnamefont {P.-G.}\ \bibnamefont {Reinhard}},\
  }\bibfield  {title} {\enquote {\bibinfo {title} {{D}amped relaxation
  techniques to calculate relativistic bound-states},}\ }\href {\doibase
  10.1103/PhysRevA.40.4182} {\bibfield  {journal} {\bibinfo  {journal} {Phys.
  Rev. A}\ }\textbf {\bibinfo {volume} {40}},\ \bibinfo {pages} {4182--4189}
  (\bibinfo {year} {1989})}\BibitemShut {NoStop}%
\bibitem [{\citenamefont {Umar}\ and\ \citenamefont
  {Oberacker}(2006{\natexlab{b}})}]{umar2006c}%
  \BibitemOpen
  \bibfield  {author} {\bibinfo {author} {\bibfnamefont {A.~S.}\ \bibnamefont
  {Umar}}\ and\ \bibinfo {author} {\bibfnamefont {V.~E.}\ \bibnamefont
  {Oberacker}},\ }\bibfield  {title} {\enquote {\bibinfo {title}
  {{T}hree-dimensional unrestricted time-dependent {H}artree-{F}ock fusion
  calculations using the full {S}kyrme interaction},}\ }\href {\doibase
  10.1103/PhysRevC.73.054607} {\bibfield  {journal} {\bibinfo  {journal} {Phys.
  Rev. C}\ }\textbf {\bibinfo {volume} {73}},\ \bibinfo {pages} {054607}
  (\bibinfo {year} {2006}{\natexlab{b}})}\BibitemShut {NoStop}%
\bibitem [{\citenamefont {Maruhn}\ \emph {et~al.}(2014)\citenamefont {Maruhn},
  \citenamefont {Reinhard}, \citenamefont {Stevenson},\ and\ \citenamefont
  {Umar}}]{maruhn2014}%
  \BibitemOpen
  \bibfield  {author} {\bibinfo {author} {\bibfnamefont {J.~A.}\ \bibnamefont
  {Maruhn}}, \bibinfo {author} {\bibfnamefont {P.-G.}\ \bibnamefont
  {Reinhard}}, \bibinfo {author} {\bibfnamefont {P.~D.}\ \bibnamefont
  {Stevenson}}, \ and\ \bibinfo {author} {\bibfnamefont {A.~S.}\ \bibnamefont
  {Umar}},\ }\bibfield  {title} {\enquote {\bibinfo {title} {{T}he {TDHF C}ode
  {S}ky3{D}},}\ }\href {\doibase 10.1016/j.cpc.2014.04.008} {\bibfield
  {journal} {\bibinfo  {journal} {Comput. Phys. Commun.}\ }\textbf {\bibinfo
  {volume} {185}},\ \bibinfo {pages} {2195--2216} (\bibinfo {year}
  {2014})}\BibitemShut {NoStop}%
\bibitem [{\citenamefont {{Ka--Hae Kim}}\ \emph {et~al.}(1997)\citenamefont
  {{Ka--Hae Kim}}, \citenamefont {{Takaharu Otsuka}},\ and\ \citenamefont
  {{Paul Bonche}}}]{kim1997}%
  \BibitemOpen
  \bibfield  {author} {\bibinfo {author} {\bibnamefont {{Ka--Hae Kim}}},
  \bibinfo {author} {\bibnamefont {{Takaharu Otsuka}}}, \ and\ \bibinfo
  {author} {\bibnamefont {{Paul Bonche}}},\ }\bibfield  {title} {\enquote
  {\bibinfo {title} {{T}hree-dimensional {TDHF} calculations for reactions of
  unstable nuclei},}\ }\href {\doibase 10.1088/0954-3899/23/10/014} {\bibfield
  {journal} {\bibinfo  {journal} {J. Phys. G}\ }\textbf {\bibinfo {volume}
  {23}},\ \bibinfo {pages} {1267} (\bibinfo {year} {1997})}\BibitemShut
  {NoStop}%
\bibitem [{\citenamefont {Pigg}\ \emph {et~al.}(2014)\citenamefont {Pigg},
  \citenamefont {Umar},\ and\ \citenamefont {Oberacker}}]{pigg2014}%
  \BibitemOpen
  \bibfield  {author} {\bibinfo {author} {\bibfnamefont {D.~A.}\ \bibnamefont
  {Pigg}}, \bibinfo {author} {\bibfnamefont {A.~S.}\ \bibnamefont {Umar}}, \
  and\ \bibinfo {author} {\bibfnamefont {V.~E.}\ \bibnamefont {Oberacker}},\
  }\bibfield  {title} {\enquote {\bibinfo {title} {{E}ulerian rotations of
  deformed nuclei for {TDDFT} calculations},}\ }\href {\doibase
  10.1016/j.cpc.2014.02.004} {\bibfield  {journal} {\bibinfo  {journal}
  {Comput. Phys. Commun.}\ }\textbf {\bibinfo {volume} {185}},\ \bibinfo
  {pages} {1410--1414} (\bibinfo {year} {2014})}\BibitemShut {NoStop}%
\bibitem [{\citenamefont {Viola}\ \emph {et~al.}(1985)\citenamefont {Viola},
  \citenamefont {Kwiatkowski},\ and\ \citenamefont {Walker}}]{viola1985}%
  \BibitemOpen
  \bibfield  {author} {\bibinfo {author} {\bibfnamefont {V.~E.}\ \bibnamefont
  {Viola}}, \bibinfo {author} {\bibfnamefont {K.}~\bibnamefont {Kwiatkowski}},
  \ and\ \bibinfo {author} {\bibfnamefont {M.}~\bibnamefont {Walker}},\
  }\bibfield  {title} {\enquote {\bibinfo {title} {{S}ystematics of fission
  fragment total kinetic-energy release},}\ }\href {\doibase
  10.1103/PhysRevC.31.1550} {\bibfield  {journal} {\bibinfo  {journal} {Phys.
  Rev. C}\ }\textbf {\bibinfo {volume} {31}},\ \bibinfo {pages} {1550--1552}
  (\bibinfo {year} {1985})}\BibitemShut {NoStop}%
\bibitem [{\citenamefont {Hinde}\ \emph {et~al.}(1987)\citenamefont {Hinde},
  \citenamefont {Leigh}, \citenamefont {Bokhorst}, \citenamefont {Newton},
  \citenamefont {Walsh},\ and\ \citenamefont {Boldeman}}]{hinde1987}%
  \BibitemOpen
  \bibfield  {author} {\bibinfo {author} {\bibfnamefont {D.~J.}\ \bibnamefont
  {Hinde}}, \bibinfo {author} {\bibfnamefont {J.~R.}\ \bibnamefont {Leigh}},
  \bibinfo {author} {\bibfnamefont {J.~J.~M.}\ \bibnamefont {Bokhorst}},
  \bibinfo {author} {\bibfnamefont {J.~O.}\ \bibnamefont {Newton}}, \bibinfo
  {author} {\bibfnamefont {R.~L.}\ \bibnamefont {Walsh}}, \ and\ \bibinfo
  {author} {\bibfnamefont {J.~W.}\ \bibnamefont {Boldeman}},\ }\bibfield
  {title} {\enquote {\bibinfo {title} {Mass-split dependence of the pre- and
  post-scission neutron multiplicities for fission of $^{251}${Es}},}\ }\href
  {\doibase 10.1016/0375-9474(87)90213-2} {\bibfield  {journal} {\bibinfo
  {journal} {Nucl. Phys. A}\ }\textbf {\bibinfo {volume} {472}},\ \bibinfo
  {pages} {318--332} (\bibinfo {year} {1987})}\BibitemShut {NoStop}%
\bibitem [{\citenamefont {Hinde}\ \emph {et~al.}(2018)\citenamefont {Hinde},
  \citenamefont {Jeung}, \citenamefont {Prasad}, \citenamefont {Wakhle},
  \citenamefont {Dasgupta}, \citenamefont {Evers}, \citenamefont {Luong},
  \citenamefont {du~Rietz}, \citenamefont {Simenel}, \citenamefont {Simpson},\
  and\ \citenamefont {Williams}}]{hinde2018}%
  \BibitemOpen
  \bibfield  {author} {\bibinfo {author} {\bibfnamefont {D.~J.}\ \bibnamefont
  {Hinde}}, \bibinfo {author} {\bibfnamefont {D.~Y.}\ \bibnamefont {Jeung}},
  \bibinfo {author} {\bibfnamefont {E.}~\bibnamefont {Prasad}}, \bibinfo
  {author} {\bibfnamefont {A.}~\bibnamefont {Wakhle}}, \bibinfo {author}
  {\bibfnamefont {M.}~\bibnamefont {Dasgupta}}, \bibinfo {author}
  {\bibfnamefont {M.}~\bibnamefont {Evers}}, \bibinfo {author} {\bibfnamefont
  {D.~H.}\ \bibnamefont {Luong}}, \bibinfo {author} {\bibfnamefont
  {R.}~\bibnamefont {du~Rietz}}, \bibinfo {author} {\bibfnamefont
  {C.}~\bibnamefont {Simenel}}, \bibinfo {author} {\bibfnamefont {E.~C.}\
  \bibnamefont {Simpson}}, \ and\ \bibinfo {author} {\bibfnamefont
  {E.}~\bibnamefont {Williams}},\ }\bibfield  {title} {\enquote {\bibinfo
  {title} {Sub--barrier quasifission in heavy element formation reactions with
  deformed actinide target nuclei},}\ }\href {\doibase
  10.1103/PhysRevC.97.024616} {\bibfield  {journal} {\bibinfo  {journal} {Phys.
  Rev. C}\ }\textbf {\bibinfo {volume} {97}},\ \bibinfo {pages} {024616}
  (\bibinfo {year} {2018})}\BibitemShut {NoStop}%
\bibitem [{\citenamefont {Scamps}\ and\ \citenamefont
  {Hashimoto}(2017)}]{scamps2017b}%
  \BibitemOpen
  \bibfield  {author} {\bibinfo {author} {\bibfnamefont {Guillaume}\
  \bibnamefont {Scamps}}\ and\ \bibinfo {author} {\bibfnamefont {Yukio}\
  \bibnamefont {Hashimoto}},\ }\bibfield  {title} {\enquote {\bibinfo {title}
  {Transfer probabilities for the reactions
  $^{14,20}\mathrm{O}+{}^{20}\mathrm{O}$ in terms of multiple time-dependent
  {H}artree-{F}ock-{B}ogoliubov trajectories},}\ }\href {\doibase
  10.1103/PhysRevC.96.031602} {\bibfield  {journal} {\bibinfo  {journal} {Phys.
  Rev. C}\ }\textbf {\bibinfo {volume} {96}},\ \bibinfo {pages} {031602}
  (\bibinfo {year} {2017})}\BibitemShut {NoStop}%
\bibitem [{\citenamefont {Dasso}\ \emph {et~al.}(1979)\citenamefont {Dasso},
  \citenamefont {Dossing},\ and\ \citenamefont {Pauli}}]{dasso1979}%
  \BibitemOpen
  \bibfield  {author} {\bibinfo {author} {\bibfnamefont {C.~H.}\ \bibnamefont
  {Dasso}}, \bibinfo {author} {\bibfnamefont {T.}~\bibnamefont {Dossing}}, \
  and\ \bibinfo {author} {\bibfnamefont {H.~C.}\ \bibnamefont {Pauli}},\
  }\bibfield  {title} {\enquote {\bibinfo {title} {{O}n the mass distribution
  in {T}ime-{D}ependent {H}artree-{F}ock calculations of heavy-ion
  collisions},}\ }\href {\doibase 10.1007/BF01409391} {\bibfield  {journal}
  {\bibinfo  {journal} {Z. Phys. A}\ }\textbf {\bibinfo {volume} {289}},\
  \bibinfo {pages} {395--398} (\bibinfo {year} {1979})}\BibitemShut {NoStop}%
\bibitem [{\citenamefont {Lacroix}\ and\ \citenamefont
  {Ayik}(2014)}]{lacroix2014}%
  \BibitemOpen
  \bibfield  {author} {\bibinfo {author} {\bibfnamefont {Denis}\ \bibnamefont
  {Lacroix}}\ and\ \bibinfo {author} {\bibfnamefont {Sakir}\ \bibnamefont
  {Ayik}},\ }\bibfield  {title} {\enquote {\bibinfo {title} {{S}tochastic
  quantum dynamics beyond mean field},}\ }\href {\doibase
  10.1140/epja/i2014-14095-8} {\bibfield  {journal} {\bibinfo  {journal} {Eur.
  Phys. J. A}\ }\textbf {\bibinfo {volume} {50}},\ \bibinfo {pages} {95}
  (\bibinfo {year} {2014})}\BibitemShut {NoStop}%
\bibitem [{\citenamefont {Ayik}\ \emph
  {et~al.}(2015{\natexlab{a}})\citenamefont {Ayik}, \citenamefont {Yilmaz},
  \citenamefont {Yilmaz}, \citenamefont {Umar}, \citenamefont {Gokalp},
  \citenamefont {Turan},\ and\ \citenamefont {Lacroix}}]{ayik2015}%
  \BibitemOpen
  \bibfield  {author} {\bibinfo {author} {\bibfnamefont {S.}~\bibnamefont
  {Ayik}}, \bibinfo {author} {\bibfnamefont {O.}~\bibnamefont {Yilmaz}},
  \bibinfo {author} {\bibfnamefont {B.}~\bibnamefont {Yilmaz}}, \bibinfo
  {author} {\bibfnamefont {A.~S.}\ \bibnamefont {Umar}}, \bibinfo {author}
  {\bibfnamefont {A.}~\bibnamefont {Gokalp}}, \bibinfo {author} {\bibfnamefont
  {G.}~\bibnamefont {Turan}}, \ and\ \bibinfo {author} {\bibfnamefont
  {D.}~\bibnamefont {Lacroix}},\ }\bibfield  {title} {\enquote {\bibinfo
  {title} {{Q}uantal description of nucleon exchange in a stochastic mean-field
  approach},}\ }\href {\doibase 10.1103/PhysRevC.91.054601} {\bibfield
  {journal} {\bibinfo  {journal} {Phys. Rev. C}\ }\textbf {\bibinfo {volume}
  {91}},\ \bibinfo {pages} {054601} (\bibinfo {year}
  {2015}{\natexlab{a}})}\BibitemShut {NoStop}%
\bibitem [{\citenamefont {Ayik}\ \emph
  {et~al.}(2015{\natexlab{b}})\citenamefont {Ayik}, \citenamefont {Yilmaz},\
  and\ \citenamefont {Yilmaz}}]{ayik2015a}%
  \BibitemOpen
  \bibfield  {author} {\bibinfo {author} {\bibfnamefont {S.}~\bibnamefont
  {Ayik}}, \bibinfo {author} {\bibfnamefont {B.}~\bibnamefont {Yilmaz}}, \ and\
  \bibinfo {author} {\bibfnamefont {O.}~\bibnamefont {Yilmaz}},\ }\bibfield
  {title} {\enquote {\bibinfo {title} {Multinucleon exchange in quasifission
  reactions},}\ }\href {\doibase 10.1103/physrevc.92.064615} {\bibfield
  {journal} {\bibinfo  {journal} {Phys. Rev. C}\ }\textbf {\bibinfo {volume}
  {92}},\ \bibinfo {pages} {064615} (\bibinfo {year}
  {2015}{\natexlab{b}})}\BibitemShut {NoStop}%
\bibitem [{\citenamefont {Ayik}\ \emph {et~al.}(2016)\citenamefont {Ayik},
  \citenamefont {Yilmaz}, \citenamefont {Yilmaz},\ and\ \citenamefont
  {Umar}}]{ayik2016}%
  \BibitemOpen
  \bibfield  {author} {\bibinfo {author} {\bibfnamefont {S.}~\bibnamefont
  {Ayik}}, \bibinfo {author} {\bibfnamefont {O.}~\bibnamefont {Yilmaz}},
  \bibinfo {author} {\bibfnamefont {B.}~\bibnamefont {Yilmaz}}, \ and\ \bibinfo
  {author} {\bibfnamefont {A.~S.}\ \bibnamefont {Umar}},\ }\bibfield  {title}
  {\enquote {\bibinfo {title} {Quantal nucleon diffusion: {C}entral collisions
  of symmetric nuclei},}\ }\href {\doibase 10.1103/PhysRevC.94.044624}
  {\bibfield  {journal} {\bibinfo  {journal} {Phys. Rev. C}\ }\textbf {\bibinfo
  {volume} {94}},\ \bibinfo {pages} {044624} (\bibinfo {year}
  {2016})}\BibitemShut {NoStop}%
\bibitem [{\citenamefont {Tanimura}\ \emph {et~al.}(2017)\citenamefont
  {Tanimura}, \citenamefont {Lacroix},\ and\ \citenamefont
  {Ayik}}]{tanimura2017}%
  \BibitemOpen
  \bibfield  {author} {\bibinfo {author} {\bibfnamefont {Yusuke}\ \bibnamefont
  {Tanimura}}, \bibinfo {author} {\bibfnamefont {Denis}\ \bibnamefont
  {Lacroix}}, \ and\ \bibinfo {author} {\bibfnamefont {Sakir}\ \bibnamefont
  {Ayik}},\ }\bibfield  {title} {\enquote {\bibinfo {title} {Microscopic
  {P}hase--{S}pace {E}xploration {M}odeling of $^{258}\mathrm{Fm}$
  {S}pontaneous {F}ission},}\ }\href {\doibase 10.1103/PhysRevLett.118.152501}
  {\bibfield  {journal} {\bibinfo  {journal} {Phys. Rev. Lett.}\ }\textbf
  {\bibinfo {volume} {118}},\ \bibinfo {pages} {152501} (\bibinfo {year}
  {2017})}\BibitemShut {NoStop}%
\bibitem [{\citenamefont {Lalazissis}\ \emph {et~al.}(1999)\citenamefont
  {Lalazissis}, \citenamefont {Raman},\ and\ \citenamefont
  {Ring}}]{lalazissis1999}%
  \BibitemOpen
  \bibfield  {author} {\bibinfo {author} {\bibfnamefont {G.~A.}\ \bibnamefont
  {Lalazissis}}, \bibinfo {author} {\bibfnamefont {S.}~\bibnamefont {Raman}}, \
  and\ \bibinfo {author} {\bibfnamefont {P.}~\bibnamefont {Ring}},\ }\bibfield
  {title} {\enquote {\bibinfo {title} {Ground--state properties of even--even
  nuclei in the relativistic mean--field theory},}\ }\href {\doibase
  10.1006/adnd.1998.0795} {\bibfield  {journal} {\bibinfo  {journal} {At. Data
  Nucl. Data Tables}\ }\textbf {\bibinfo {volume} {71}},\ \bibinfo {pages}
  {1--40} (\bibinfo {year} {1999})}\BibitemShut {NoStop}%
\bibitem [{\citenamefont {Blazkiewicz}\ \emph {et~al.}(2005)\citenamefont
  {Blazkiewicz}, \citenamefont {Oberacker}, \citenamefont {Umar},\ and\
  \citenamefont {Stoitsov}}]{blazkiewicz2005}%
  \BibitemOpen
  \bibfield  {author} {\bibinfo {author} {\bibfnamefont {A.}~\bibnamefont
  {Blazkiewicz}}, \bibinfo {author} {\bibfnamefont {V.~E.}\ \bibnamefont
  {Oberacker}}, \bibinfo {author} {\bibfnamefont {A.~S.}\ \bibnamefont {Umar}},
  \ and\ \bibinfo {author} {\bibfnamefont {M.}~\bibnamefont {Stoitsov}},\
  }\bibfield  {title} {\enquote {\bibinfo {title} {{C}oordinate space
  {H}artree-{F}ock-{B}ogoliubov calculations for the zirconium isotope chain up
  to the two-neutron drip line},}\ }\href {\doibase 10.1103/PhysRevC.71.054321}
  {\bibfield  {journal} {\bibinfo  {journal} {Phys. Rev. C}\ }\textbf {\bibinfo
  {volume} {71}},\ \bibinfo {pages} {054321} (\bibinfo {year}
  {2005})}\BibitemShut {NoStop}%
\bibitem [{\citenamefont {Hwang}\ \emph {et~al.}(2006)\citenamefont {Hwang},
  \citenamefont {Ramayya}, \citenamefont {Hamilton}, \citenamefont {Luo},
  \citenamefont {Daniel}, \citenamefont {Ter-Akopian}, \citenamefont {Cole},\
  and\ \citenamefont {Zhu}}]{hwang2006}%
  \BibitemOpen
  \bibfield  {author} {\bibinfo {author} {\bibfnamefont {J.~K.}\ \bibnamefont
  {Hwang}}, \bibinfo {author} {\bibfnamefont {A.~V.}\ \bibnamefont {Ramayya}},
  \bibinfo {author} {\bibfnamefont {J.~H.}\ \bibnamefont {Hamilton}}, \bibinfo
  {author} {\bibfnamefont {Y.~X.}\ \bibnamefont {Luo}}, \bibinfo {author}
  {\bibfnamefont {A.~V.}\ \bibnamefont {Daniel}}, \bibinfo {author}
  {\bibfnamefont {G.~M.}\ \bibnamefont {Ter-Akopian}}, \bibinfo {author}
  {\bibfnamefont {J.~D.}\ \bibnamefont {Cole}}, \ and\ \bibinfo {author}
  {\bibfnamefont {S.~J.}\ \bibnamefont {Zhu}},\ }\bibfield  {title} {\enquote
  {\bibinfo {title} {{H}alf-life measurements of several states in
  $^{95,97}\mathrm{Sr},^{97,100,104}\mathrm{Zr},^{106}\mathrm{Mo}$, and
  $^{148}\mathrm{Ce}$},}\ }\href {\doibase 10.1103/PhysRevC.73.044316}
  {\bibfield  {journal} {\bibinfo  {journal} {Phys. Rev. C}\ }\textbf {\bibinfo
  {volume} {73}},\ \bibinfo {pages} {044316} (\bibinfo {year}
  {2006})}\BibitemShut {NoStop}%
\end{thebibliography}%


\end{document}